\pdfoutput=1

\documentclass[prd,twocolumn,twoside,superscriptaddress,nofootinbib,preprintnumbers]{revtex4-1}

\usepackage{amsmath}
\usepackage{graphicx}
\usepackage[hyperfootnotes=false]{hyperref}
\usepackage{xspace}
\usepackage{xcolor}

\newcommand{\eq}[1]{Eq.~\eqref{eq:#1}}
\newcommand{\eqs}[2]{Eqs.~\eqref{eq:#1} and \eqref{eq:#2}}
\renewcommand{\sec}[1]{Sec.~\ref{sec:#1}}
\newcommand{\secs}[2]{Secs.~\ref{sec:#1} and \ref{sec:#2}}

\newcommand{\app}[1]{App.~\ref{app:#1}} 
\newcommand{\apps}[2]{Apps.~\ref{app:#1} and \ref{app:#2}}

\newcommand{\eg}{{\it e.g.~}}
\newcommand{\ie}{{\it i.e.~}}

\newcommand{\ord}[1]{{\mathcal O}(#1)}

\newcommand{\MAe}[3]{\Bigl\langle#1\Bigr\rvert#2\Bigr\rvert#3\Bigr\rangle}

\newcommand{\nn}{\nonumber}

\newcommand{\df}{\mathrm{d}}
\newcommand{\img}{\mathrm{i}}
\newcommand{\sdt}{\!\cdot\!}
\newcommand{\tr}{\mathrm{tr}}
\newcommand{\lra}{\leftrightarrow}

\newcommand{\al}{\alpha}

\newcommand{\ga}{\gamma}
\newcommand{\Ga}{\Gamma}
\newcommand{\de}{\delta}
\newcommand{\De}{\Delta}
\newcommand{\eps}{\epsilon}

\newcommand{\si}{\sigma}

\newcommand{\cG}{{\mathcal G}}
\newcommand{\cI}{{\mathcal I}}
\newcommand{\cJ}{{\mathcal J}}
\newcommand{\cL}{{\mathcal L}}

\newcommand{\cP}{{\mathcal P}}

\newcommand{\bn}{\bar{n}}
\newcommand{\bnP}{\overline {\mathcal P}}
\newcommand{\bnslash}{\bar{n}\!\!\!\slash}

\newcommand{\zero}{{(0)}}
\newcommand{\one}{{(1)}}
\newcommand{\two}{{(2)}}

\newcommand{\cGbar}{{g}}

\newcommand{\dPhiColl}[1]{\df \Phi_{#1}^\text{c}}
\newcommand{\dPhiCollISR}[1]{\df \Phi_{#1}^\text{c,ISR}}
\newcommand{\siColl}[1]{\si_{#1}^\text{c}}

\newcommand{\HypExp}{{\sc HypExp}\xspace}
\newcommand{\HPL}{{\sc HPL}\xspace}
\newcommand{\FIRE}{{\sc FIRE}\xspace}
\newcommand{\Reduze}{{\sc Reduze}\xspace}
\newcommand{\MT}{{\sc MT}\xspace}

\allowdisplaybreaks[2]

\begin{document}


\preprint{NIKHEF 2014-015}

\title{Fragmentation in Jets at NNLO}

\author{Mathias Ritzmann}
\affiliation{Nikhef, Theory Group, Amsterdam, The Netherlands}

\author{Wouter J.~Waalewijn}
\affiliation{Nikhef, Theory Group, Amsterdam, The Netherlands}
\affiliation{Institute for Theoretical Physics, University of Amsterdam, Amsterdam, The Netherlands}

\date{\today}

\begin{abstract}

Beam and jet functions in Soft-Collinear Effective Theory describe collinear initial- and final-state radiation (jets), and enter in factorization theorems for $N$-jet production, the Higgs $p_T$ spectrum, etc.
We show that they may directly be calculated as phase-space integrals of QCD splitting functions. At next-to-leading order (NLO) all computations are trivial, as we demonstrate explicitly for the beam function, the transverse-momentum-dependent beam function, the jet function and the fragmenting jet function. This approach also highlights the role of crossing symmetry in these calculations. At next-to-next-to leading order (NNLO) we reproduce the quark jet function and calculate the fragmenting quark jet function for the first time. Here we use two methods: a direct phase-space integration and a reduction to master integrals which are computed using differential equations. \\[1ex]

\end{abstract}

\maketitle

\section{Introduction}
\label{sec:intro}

All LHC processes involve QCD in some way: through the parton distribution functions (PDFs) describing the composition of the colliding protons in terms of quarks and gluons, through energetic collinear radiation (jet production), or through soft radiation effects, etc. 
Often there are hierarchies between scales of observables, \eg the jet mass $m_J$ is typically much smaller than the transverse momentum of the jet $p_T^J$. This leads to large logarithms (of \eg $m_J/p_T^J$) in the perturbative expansion of the cross section that require resummation. Soft-Collinear Effective Theory (SCET)~\cite{Bauer:2000ew,Bauer:2000yr, Bauer:2001ct, Bauer:2001yt} is a convenient framework for achieving higher-order logarithmic resummation and treating nonperturbative corrections, see \eg Refs.~\cite{Abbate:2010xh,Becher:2012qc,Stewart:2014nna}. In SCET, initial- and final-state collinear radiation is described by beam functions $B$ and jet functions $J$.

This paper focusses on the calculation of $B$ and $J$ by exploiting a new relationship with splitting functions. These calculations have many phenomenological applications, for example, the beam and jet function (schematically) enter in the factorization formula for the cross section of $pp \to X+N$ jets (with $X$ nonhadronic) as~\cite{Stewart:2010tn}
\begin{align} \label{eq:scet}
 \si = \sum_\kappa \int \df \Phi_{N+X} \tr [\widehat H_\kappa \widehat S_N^\kappa] \otimes \Big[B_{\kappa_a} B_{\kappa_b} \prod_{J=1}^N J_{\kappa_J} \Big]
\,.\end{align}
Here $\widehat H$ contains the tree-level partonic process plus short-distance virtual corrections and $\widehat S$ describes soft radiation effects. The phase-space is denoted by $\df \Phi_{N+X}$, the trace is over color configurations, and the dependence on the partonic process is labelled by $\kappa$.
Whereas $\widehat H$ and $\widehat S$ depend on the full partonic process (including color configuration), each beam function only depends on the flavor $\kappa_{a,b}$ of the colliding parton and each jet function only on the parton $\kappa_J$ that initiates the jet.
The convolution between the soft function and the beam and jet functions arises because measurements typically constrain the ``sum" of collinear and soft radiation.

\eq{scet} is valid for energetic well-separated beams and jets, receiving corrections that are suppressed by \eg $(m_J/p_T^J)^2$. It enables resummation by separating the cross section into contributions involving a single scale. This is accomplished by evaluating each object ($\widehat H$, $\widehat S$, $B$, $J$) at its natural scale and evolving it to some common scale using the renormalization group evolution. The order at which the resummation can be carried out is limited by the order at which each of the ingredients is known.

We can also analyze the process described by~\eq{scet} using the universality of collinear limits of QCD amplitudes~\cite{Berends:1987me,Mangano:1990by,Kosower:1999xi} (we will need the collinear limits of tree-level and one-loop~\cite{Bern:1993qk,Bern:1994zx,Bern:1995ix,Bern:1998sc,Kosower:1999rx,Bern:1999ry,Sborlini:2013jba} amplitudes as well as the triple-collinear limits of tree-level amplitudes~\cite{Campbell:1997hg,Catani:1998nv}). The contribution to the cross section in \eq{scet} from the tree-level process plus real and virtual corrections collinear to one specific jet $J$, can be written as
\begin{align} \label{eq:coll_fact}
 \si^\text{c} &= \sum_\kappa \int\! \df \Phi_{N+X} \si_\kappa^{(0)} f_{\kappa_a} f_{\kappa_b}
  \nn \\ & \quad \times
  \sum_m \sum_{\kappa^\text{c}} S_{\kappa^\text{c}} \int\! \dPhiColl{m} \siColl{m,\kappa^\text{c}}
\,.\end{align}
Here $\si_\kappa^{(0)}$ is the tree-level partonic cross section, $f$ is a PDF, $\siColl{m,\kappa^\text{c}}$ is the $\kappa_J \to \kappa^\text{c}$ splitting function (apart from an overall factor) where $\kappa^\text{c}$ consists of $m$ partons,  $\dPhiColl{m}$ is the $m$-body collinear phase-space and $S_{\kappa^\text{c}}$ is a symmetry factor. The first line of \eq{coll_fact} thus contains the tree-level cross section, producing the parton $\kappa_J$ that initiates a jet. The second line describes the collinear radiation produced by $\kappa_J$ that builds up this jet. A similar equation holds for collinear initial-state radiation.

By comparing the factorized form of the collinear radiation in \eq{coll_fact} to the SCET cross section in \eq{scet}, we establish a relationship between the jet function and the splitting functions. To this end, we need the tree-level results $\tr[\widehat H_\kappa^{(0)}\, \widehat S_N^{\kappa{(0)}}]= \si_\kappa^{(0)}\, \de(\dots)$, $J_{\kappa_J}^{(0)} \sim \delta(\dots)$ and $B_{\kappa_{a,b}}^{(0)} = f_{\kappa_{a,b}}\, \de(\dots)$. Each $\de(\dots)$ encodes the measurement on the soft radiation, or collinear final- or initial-state radiation, which is trivial at leading order. E.g.~$J_{\kappa_J}^{(0)}(s,\mu) = \delta(s)$ when the invariant mass $s$ of the jet is constrained. We conclude that
\begin{align} \label{eq:jet_split}
  J_{\kappa_J} &=  \sum_m \sum_{\kappa^\text{c}} S_{\kappa^\text{c}} \int\! \dPhiColl{m}\, \siColl{m,\kappa^\text{c}} \de(\dots)
\,.\end{align}
In this equation $\de(\dots)$ denotes the measurement imposed on the collinear final-state radiation in the jet, \eg$\de(\dots) \to \de(s-s_{\kappa^c})$ when the invariant mass $s$ is measured. There is an analogue of \eq{jet_split} for the beam function. The appropriate splitting function can be obtained from the one in \eq{jet_split} by crossing, but the collinear phase space needs to be replaced by its initial-state version.

It is instructive to compare \eq{jet_split} to the field-theoretic definition of the quark jet function in SCET~\cite{Bauer:2001yt}
\begin{widetext}
\begin{align} \label{eq:J_def}
J_q(s = p^- p^+, \mu)
= \frac{(2\pi)^2}{N_c} \!\int\! \frac{\df y^-}{2p^-}\, e^{\img p^+ y^-/2}\,
   \tr\MAe{0}{ \frac{\bnslash}{2} W_n^\dagger\Big(y^- \frac{n^\mu}{2}\Big) \xi_n \Big(y^- \frac{n^\mu}{2}\Big)
   \bigl[\delta(p^- + \bnP_n)\delta^2(\cP_{n\perp}) \bar \xi_n(0) W_n(0) \bigr]}{0}
\,,\end{align}
\end{widetext}
where the invariant mass $s$ of the collinear radiation is measured.
Here, $n=(1,\hat n)$, $\bn=(1,-\hat n)$ with $\hat n$ the jet direction, $\xi_n$ is the collinear quark field and $\cP_n$ picks out the large ``label" momentum. The Wilson line $W_n$ sums longitudinal gluon $\bn \cdot A$ emissions and is necessary to ensure gauge invariance.
In Eq.~\eqref{eq:jet_split}, we are explicitly integrating a gauge-invariant quantity (as long as we restrict ourselves to spin-averaged splitting functions).
However, the association of the splitting function with the process-independent diagrams describing one virtual parton splitting into several real ones is only valid in a gauge in which gluons are explicitly transverse (\eg$\bn \cdot A = 0$, for which $W_n=1$ in \eq{J_def}).

The argument underlying \eq{jet_split} only uses that the collinear approximation is valid for the collinear functions in the factorization theorem. It thus extends to arbitrary beam and jet functions.
One example we study in this paper is the fragmenting jet function $\cG_i^h$, which describes the momentum fraction $z$ of an energetic hadron $h$ \emph{in a jet}. This leads to an additional measurement delta function $\delta(z-z_h)$ in \eqs{jet_split}{J_def} compared to the regular jet function. Factorization theorems for processes involving jets can straightforwardly be extended to describe fragmentation as well, by replacing $J_i \to \cG_i^h$~\cite{Procura:2009vm}.

Even though normal QCD Feynman rules can be used to calculate the jet function from its definition in \eq{J_def}, it is not so easy for the uninitiated researcher. By contrast, \eq{jet_split} allows one to obtain the desired result by a straightforward phase-space integral. In practice the calculation involving \eq{jet_split} is also significantly easier at NLO, as we demonstrate explicitly in \sec{LO}. However, at NNLO the difficulty will strongly depend on the details of the measurement.
In addition to a direct phase-space integration, we also perform a reduction to master integrals which (in the case of the fragmenting jet function) are then computed using differential equations. It turns out that for the (fragmenting) jet function calculation the phase space restrictions are no impediment to the use of well-established techniques, and the fragmenting jet function can be expressed in terms of standard harmonic polylogarithms. 

In this paper we will compute the following:
\begin{itemize}
\item 
The jet function $J_q(s,\mu)$~\cite{Bauer:2001yt} where the invariant mass $s$ of a quark jet is constrained. We reproduce the known results at NLO~\cite{Bauer:2003pi, Fleming:2003gt} in \eq{J_q_1} and NNLO~\cite{Becher:2006qw} in \eq{J_q_2}.
\item
The fragmenting jet function $\cG_q^h(s,z,\mu)$~\cite{Procura:2009vm} where the momentum fraction $z$ of a hadron $h$ in the jet is also measured. The NLO results of Ref.~\cite{Jain:2011xz} are reproduced in \eq{J_1} and NNLO results are obtained for the first time in \sec{fragjet_nnlo_ren}. This agrees with the NNLO fragmentation of a light quark into heavy quarks calculated in Ref.~\cite{Bauer:2013bza}~\footnote{We thank E. Mereghetti for pointing out a contribution to the NNLO matching coefficient that was omitted in the original submission of this article.}. 
\item
The beam function $B_q(x,\vec k_\perp^{\,2},\mu)$~\cite{Becher:2010tm,GarciaEchevarria:2011rb,Chiu:2012ir,Collins:1984kg,Collins:2011zzd} describing the transverse momentum $\vec k_\perp$ of the colliding quark. This is essentially the transverse momentum dependent parton distribution function (TMD PDF) for $\vec k_\perp \gg \Lambda_{\text{QCD}}$. We reproduce the NLO results of Ref.~\cite{Becher:2010tm} in \eq{I_1_kt}. (The NNLO results have recently been calculated~\cite{Gehrmann:2012ze,Gehrmann:2014yya}.)
\item
The beam function $B_q(t,x,\mu)$~\cite{Fleming:2006cd,Stewart:2009yx} describing the dependence on the momentum fraction $x$ and transverse virtuality $t$ of the colliding parton. We rederive the NLO results of Ref.~\cite{Stewart:2010qs}, shown in \eq{I_1}. (The NNLO results are now known~\cite{Gaunt:2014xga,Gaunt:2014cfa}.)
\end{itemize}
Beam and jet functions involving more general phase space restrictions (such as jet algorithms) have been considered in phenomenological applications. We will briefly discuss some examples in the conclusions. At NLO their calculation will benefit from the method developed in this paper, though at NNLO it will depend on the details of the observable.

In \sec{ren_match}, we briefly discuss the renormalization of these objects, as well as their matching onto PDFs (for the beam functions) and fragmentation functions (for the fragmenting jet function). The LO splitting functions are used in \sec{LO} to perform the NLO calculations of these objects. In \sec{NLO} this is extended to NNLO for the jet function and fragmenting jet function. We conclude in \sec{discussion}. The definition and properties of plus distributions and harmonic polylogarithms can be found in \app{plus} and \app{hpl}, respectively. Intermediate results for the NNLO calculation of the jet function and fragmenting jet function using integral reduction are given in \app{jet_mi} and \ref{app:fragjet_mi} and in electronic form accompanying this paper. The IR divergences of the NNLO fragmenting jet function are given in \app{IR} and provide an important cross check.

\section{Renormalization and Matching}
\label{sec:ren_match}

We perform our calculations using dimensional regularization, removing UV divergences with the modified minimal subtraction scheme ($\overline{\text{MS}}$). The $J_q(s,\mu)$, $\cG_q^h(s,z,\mu)$ and $B_q(t,x,\mu)$ have the same renormalization $Z_{J_q}$~\cite{Procura:2009vm,Stewart:2010qs}
\begin{align} \label{eq:renormalization}
  J_q(s,\mu) &= \int_0^\infty\! \df s'\, Z_{J_q}(s',\mu)\, J_{q,\text{bare}}(s-s')
  \nn \,, \\ 
  \cG_q^h(s,z,\mu) &= \int_0^\infty\! \df s'\, Z_{J_q}(s',\mu)\, \cG_{q,\text{bare}}^h(s-s',z,\mu)
  \nn \,, \\ 
  B_q(t,x,\mu) &= \int_0^\infty\! \df t'\, Z_{J_q}(t',\mu)\, B_{q,\text{bare}}(t-t',x,\mu)
\,.\end{align}
The TMD beam function $B_q(x,\vec k_\perp^{\,2},\mu)$ is quite different as it has both UV and rapidity divergences. We will use the approach of Refs.~\cite{Chiu:2011qc,Chiu:2012ir} to perform the renormalization, to which we refer for further details.

The perturbative calculation of $\cG_q^h(s,z,\mu)$, $B_q(t,x,\mu)$ and $B_q(x,\vec k_\perp^{\,2},\mu)$ involves replacing the outgoing hadron $h$ or incoming proton by a parton. (For the beam functions we will denote the incoming parton $i$ by $B_{q/i}$ in our calculations.)
The corresponding IR divergences are removed by matching onto fragmentation functions and PDFs, 
\begin{align} \label{eq:matching}
  \cG_q^h(s,z,\mu) &= \sum_j \int_z^1\! \frac{\df z'}{z'}\, \cJ_{qj}\Big(s,\frac{z}{z'},\mu \Big) D_j^h(z',\mu)
\,,  \nn \\
  B_q(t,x,\mu) &= \sum_j \int_x^1\! \frac{\df x'}{x'}\, \cI_{qj}\Big(t,\frac{x}{x'},\mu \Big) f_j(x',\mu)
\,,  \nn \\
  B_q(x,\vec k_\perp^{\,2}, \mu, \nu) &= \sum_j \int_x^1\! \frac{\df x'}{x'}\, \cI_{qj}\Big(\frac{x}{x'},\vec k_\perp^{\,2},\mu, \nu \Big) f_j(x',\mu)
\,.\end{align}

An important cross check on our calculation is provided by the quark number and momentum sum rules~\cite{Procura:2011aq, Procura:2009vm,Jain:2011xz} which translate directly to the matching coefficients $\cJ_{ij}$ as
\begin{align} \label{eq:Jsumrules}
\int \text{d} z \big[ \cJ_{qq} (s, z, \mu) -  \cJ_{q\bar{q}}(s, z, \mu) \big] & = J_q(s,\mu) 
\,, \nn \\ 
\int \text{d} z \sum_i z\, \cJ_{qi}(s, z, \mu) &= J_q(s,\mu) 
\,.\end{align}
Here (and throughout this paper) we remove the spurious factor of $2(2\pi)^3$ in the definition of the fragmenting jet function and matching coefficients in Ref.~\cite{Procura:2009vm}.

\section{Beam and Jet Function at NLO}
\label{sec:LO}

\subsection{Splitting Function and Phase Space}

The real radiation $i^* \to jk$ in the collinear limit factors off the squared matrix element and is (in $\overline{\text{MS}}$) given by~\cite{Giele:1991vf}
\begin{equation} \label{eq:lo_pref}
  \siColl{2}(s,z) = \Big(\frac{\mu^2 e^{\ga_E}}{4\pi}\Big)^\eps\, \frac{2g^2}{s} \, P_{jk}(z)
\,,\end{equation}
where the LO splitting function is~\cite{Gross:1974cs,Altarelli:1977zs}
\begin{equation}
 P_{q^*\to qg}^\zero(s,z) \equiv \frac{1}{s}\, P_{qg}^\zero(z) = \frac{C_F}{s}\, \Big[ \frac{1+z^2}{1-z} - \eps (1-z) \Big]
\,.\end{equation}
Here $s\geq0$ is the time-like virtuality of the initial parton $i^*$ (the jet mass) and $0 \leq z \leq 1$ the momentum fraction of the final parton $j$. The corresponding collinear phase space for the final-state radiation is given by~\cite{Giele:1991vf}
\begin{equation} \label{eq:jet_ps}
 \dPhiColl{2}(s,z) = \df s\, \df z\, \frac{[z(1-z)s]^{-\eps}}{(4\pi)^{2-\eps} \Ga(1-\eps)}
\,.\end{equation}

\subsection{(Fragmenting) Jet Function}

We obtain the (fragmenting) jet function by combining these pieces and imposing the appropriate final-state measurement. In the fragmenting jet function both $s$ and $z$ are measured, so we just expand
\begin{widetext}
\begin{align} \label{eq:J_qq_1}
 \cG_{q,\text{bare}}^{q\one}(s,z) &= \int\! \dPhiColl{2}(s',z')\, \siColl{2}(s',z')\, \de(s-s')\, \de(z-z')
 \\ & 
 = \Big(\frac{\mu^2 e^{\ga_E}}{4\pi}\Big)^\eps\, \frac{[z(1-z)s]^{-\eps}}{(4\pi)^{2-\eps} \Ga(1-\eps)}\, \frac{2g^2}{s} P_{qg}^\zero(z)
  \nn \\ &
 = \frac{\al_s C_F}{2\pi} \frac{\mu^{2\eps}}{s^{1+\eps}} \bigg\{
\!-\!\frac{2}{\eps} \delta (1\!-\!z)
 \!+\!(1\!+\!z^2) \cL_0(1\!-\!z)
 \!+\!\eps \bigg[\!-\!(1\!+\!z^2) \Big(\cL_1(1\!-\!z)
 \!+\! \frac{\ln z}{1\!-\!z}\Big)
 \!+\!\frac{\pi^2}{6} \delta (1\!-\!z)
 \!+\!z\!-\!1 \bigg]
 \nn \\ & \quad
 \!+\!\eps^2 \bigg[
 \frac12 (1\!+\!z^2) \Big(\cL_2(1\!-\!z) \!+\! \frac{\ln^2[z(1\!-\!z)] - \ln^2(1-z)}{1\!-\!z} \!-\! \frac{\pi^2}{6}  \cL_0(1\!-\!z) \Big)
   \!+\!\frac{2}{3}\zeta_3 \delta (1\!-\!z)
 \!+\! (1\!-\!z) \ln [z(1-z)]
 \bigg]
 \nn \\ & \quad 
\!+\! \eps^3 \bigg[
\frac{1}{6}(1\!+\!z^2) \bigg(\!-\! \cL_3(1\!-\!z) 
\!+\! \frac{\ln^3(1-z) - \ln^3[z(1\!-\!z)]}{1-z} 
\!+\frac{\pi ^2}{2} \Big[ \cL_1(1\!-\!z) + \frac{\ln z}{1-z} \Big]
\!-\!2 \zeta_3 \cL_0(1\!-\!z)
\bigg)
 \nn \\ & \quad
\!-\!\frac{\pi^4}{720} \delta (1\!-\!z)
 \!-\! \frac12  (1\!-\!z) \Big(\ln^2[z(1-z)]
  \!-\! \frac{\pi^2}{6} \Big)
\bigg] + \ord{\eps^4} \bigg\}   
  \nn \\ &
  = \frac{\al_s C_F}{2\pi} \bigg\{\frac{2}{\eps^2} \de(s) \de(1-z) - \frac{1}{\eps} \bigg[\frac{2}{\mu^2} \cL_0\Big(\frac{s}{\mu^2}\Big) \de(1-z) 
  +\de(s) (1+z^2) \cL_0(1-z) \bigg] 
  + \frac{2}{\mu^2} \cL_1\Big(\frac{s}{\mu^2}\Big) \de(1-z) 
  \nn \\ & \quad
  +
  \frac{1}{\mu^2} \cL_0\Big(\frac{s}{\mu^2}\Big) (1\!+\!z^2) \cL_0(1\!-\!z)
  + \de(s) \bigg[(1\!+\!z^2) \cL_1(1\!-\!z) 
  + \frac{1+z^2}{1-z} \ln z + 1-z - \frac{\pi^2}{6} \de(1-z) \bigg] \bigg\} +\ord{\eps}
\,.\nn \end{align}
\end{widetext}
Here we used \eq{plus_exp} to perform the expansion in plus distributions $\cL_n$ [defined in \eq{plus_def}]. The virtual corrections are scaleless and vanish. We remind the reader that we have changed the normalization of the fragmenting jet function (and matching coefficients) with respect to Ref.~\cite{Procura:2009vm}, removing the spurious factor of $2(2\pi)^3$.

\eq{J_qq_1} contains UV divergences, which are removed by renormalization, and IR divergences, which cancel in the matching onto fragmentation functions in \eq{matching}. Because the one-loop renormalized fragmentation functions $D_i^{j\one}$ are pure IR divergences (in dimensional regularization), the finite part of \eq{J_qq_1} is the one-loop matching coefficient $\cJ_{qq}^\one$
\begin{align}
  \cG_q^{j\one}(s,z,\mu)\Big|_{\eps^0} &= \sum_i \int\! \df z'\,\cJ_{qi}^\one \Big(s,\frac{z}{z'},\mu\Big) D_i^{j\zero}(z',\mu) 
  \nn \\
  &= \cJ_{qj}^\one(s,z,\mu)
\,.\end{align}
We therefore find
\begin{align} \label{eq:J_1}
\cJ_{qq}^\one(s,z,\mu)
&= \frac{\al_s C_F}{2\pi} \bigg\{
  \frac{2}{\mu^2} \cL_1\Big(\frac{s}{\mu^2}\Big) \de(1\!-\!z) 
  \!+\! \frac{1}{\mu^2} \cL_0\Big(\frac{s}{\mu^2}\Big)
  \nn \\ & \quad \times  
  (1\!+\!z^2) \cL_0(1\!-\!z)
  \!+\! \de(s) \bigg[(1\!+\!z^2) \cL_1(1\!-\!z)
  \nn \\ & \quad
  \!+\! \frac{1\!+\!z^2}{1\!-\!z} \ln z \!+\! 1\!-\!z \!-\! \frac{\pi^2}{6} \de(1\!-\!z) \bigg] \bigg\}
\,,\end{align}
in agreement with Refs.~\cite{Jain:2011xz,Liu:2010ng}.
The other matching coefficient for quark jets follows from the symmetry relation $\cJ_{qg}^\one(s,z,\mu)=\cJ_{qq}^\one(s,1-z,\mu)$. Note that the limit $z \to 0$ does not require regularization, so the plus prescription may be dropped in this case.

The jet function only measures the invariant mass $s$. We can obtain the jet function $J_q$ by integrating the finite terms\footnote{Phase-space integration and operator renormalization do not commute in general. E.g.~the bare fragmenting jet function integrated over $s$ produces the partonic fragmentation function, which has a very different renormalization structure than \eq{renormalization}.}
 of \eq{J_qq_1} over the momentum fraction $z$,
\begin{align} \label{eq:J_q_1}
 \int\! \df z\, \cJ_{qq}^\one(s,z,\mu) &=  \frac{\al_s C_F}{2\pi} \Big[ \frac{2}{\mu^2} \cL_1\Big(\frac{s}{\mu^2}\Big) - \frac{3}{2 \mu^2} \cL_0\Big(\frac{s}{\mu^2}\Big) 
  \nn \\ & \quad
 +\Big(\frac{7}{2} - \frac{\pi^2}{2} \Big) \de(s) \Big]
 \nn \\ &
 = J_q^\one(s,\mu)
\,,\end{align}
which is the quark number sum rule in \eq{Jsumrules}.

\subsection{Beam Function}

In the beam function the initial parton taken out of the proton can be treated on-shell and instead the parton entering the hard interaction has a (space-like) virtuality. The splitting functions can be obtained from their all-outgoing counterparts by crossing symmetry.
If we denote the momentum fraction of the virtual parton entering the hard interaction as $x$, \ie the collinear limit is given by $p_j \rightarrow (1-x) p_i$ the crossing relation reads
\begin{align}
	P_{i \to k^* j} \left( 2 \, p_i \sdt p_j, x \right) &= (-1)^{\Delta_f} P_{k^* \to ij} \left( -2 \, p_i \sdt p_j , 1/x \right)
\,,\end{align}
where $\Delta_f$ is the difference in the number of incoming fermions and we use conventions in which both incoming and outgoing momenta have positive energy.
We will use the transverse virtuality of the colliding parton (with respect to the beam axis) $t = 2 \, x\, p_i \sdt p_j$ to parametrize the collinear phase space below.
We also have to keep track of the changes in the color and spin averaging factors, \eg for $g \to q^* \bar q$ we have an additional factor of 
\begin{align}
 \frac{2N}{(d-2)(N^2-1)} = \frac{1}{1-\eps}\,\frac{T_F}{C_F}
\,,\end{align}
since we now need to average over the colors and spins of an initial gluon rather than quark.

For the triple-collinear splitting functions, the crossing relation reads
\begin{widetext}
\begin{multline}
	P_{i \to l^* j k} \left( x=1-z_j-z_k, z_j, z_k, 2 \, p_i \sdt p_j, 2 \, p_i \sdt p_k, 2 \, p_j \sdt p_k \right) \\
	 = (-1)^{\Delta_f}  P_{l^* \to i j k} \left( \frac{1}{x}, \frac{-z_j}{1-z_j-z_k}, \frac{-z_k}{1-z_j-z_k}, 
	 	-2 \, p_i \sdt p_j, -2 \, p_i \sdt p_k, 2 \, p_j \sdt p_k \right)
\,,\end{multline}
with the collinear limit given by $p_j \to z_j p_i$, $p_k \to z_k p_i$.
The two-particle collinear phase space is
\begin{equation}
 \dPhiCollISR{2}(t,x) = \df t\, \df x\, \frac{[(1-x)t/x]^{-\eps}}{(4\pi)^{2-\eps} \Ga(1-\eps)}
\,.\end{equation}
Note that this cannot be obtained by crossing \eq{jet_ps}.

Combining these ingredients we find for $q \to q^* g$
\begin{align} \label{eq:I_qq_1}
 B_{q/q,\text{bare}}^{\one}(t,x) 
 &= \int\! \dPhiCollISR{2}(t',x')\, \siColl{2}\Big(\!-\!\frac{t'}{x'},\frac{1}{x'}\Big)\, \de(t-t')\, \de(x-x')
 \\ &
 = \Big(\frac{\mu^2 e^{\ga_E}}{4\pi}\Big)^\eps\, \frac{[(1-x)t/x]^{-\eps}}{(4\pi)^{2-\eps} \Ga(1-\eps)} \frac{2g^2}{-t/x} P_{qg}^\zero\Big(\frac{1}{x}\Big)
  \nn \\ & 
  = \frac{\al_s C_F}{2\pi} \Big(1 - \eps^2 \frac{\pi^2}{12} +\ord{\eps^3} \Big) \frac{\mu^{2\eps}}{t^{1+\eps}}\,[(1-x)/x]^{-\eps} \Big[\frac{1+x^2}{1-x} - \eps (1-x) \Big]
   \nn \\ &
  = \frac{\al_s C_F}{2\pi} \bigg\{\frac{2}{\eps^2} \de(t) \de(1-x) - \frac{1}{\eps} \bigg[\frac{2}{\mu^2} \cL_0\Big(\frac{t}{\mu^2}\Big) \de(1-x) 
  +\de(t) (1+x^2) \cL_0(1-x) \bigg] 
  + \frac{2}{\mu^2} \cL_1\Big(\frac{t}{\mu^2}\Big) \de(1-x)
  \nn \\ & \quad     
   + \frac{1}{\mu^2} \cL_0\Big(\frac{t}{\mu^2}\Big) (1+x^2) \cL_0(1-x)
  + \de(t) \bigg[(1+x^2) \cL_1(1-x) 
  - \frac{1+x^2}{1-x} \ln x + 1-x - \frac{\pi^2}{6} \de(1-x) \bigg] \bigg\}
\,.\nn \end{align}
Since the only change compared to $\cJ_{qq}$ is the phase space, only the sign of the $\ln x$ term is affected. For $g \to q^* \bar q$,
\begin{align} \label{eq:I_qg_1}
 B_{q/g,\text{bare}}^{\one}(t,x) 
 &= -\frac{1}{1-\eps} \frac{T_F}{C_F} \int\! \dPhiCollISR{2}(t',x')\, \siColl{2}\Big(\!-\!\frac{t'}{x'},\frac{x'-1}{x'}\Big)\, \de(t-t')\, \de(x-x')
 \nn \\ & 
 = -\frac{1}{1-\eps} \frac{T_F}{C_F} \Big(\frac{\mu^2 e^{\ga_E}}{4\pi}\Big)^\eps\, \frac{[(1-x)t/x]^{-\eps}}{(4\pi)^{2-\eps} \Ga(1-\eps)} \frac{2g^2}{-t/x} P_{qg}^\zero\Big(\frac{x-1}{x}\Big)
 \nn \\ &
 = \frac{\al_s T_F}{2\pi}\, \big(1+\eps + \ord{\eps^2}\big) \frac{\mu^{2\eps}}{t^{1+\eps}}\,[(1-x)/x]^{-\eps} [x^2+(1-x)^2 - \eps \big]
   \nn \\ & 
  = \frac{\al_s T_F}{2\pi} \bigg\{ \bigg[-\frac{1}{\eps} \de(t) + \frac{1}{\mu^2} \cL_0\Big(\frac{t}{\mu^2}\Big)\bigg] (x^2+(1-x)^2)
  + \de(t) \bigg[(x^2+(1-x)^2) \Big(\ln \frac{1-x}{x} -1\Big) + 1 \bigg] \bigg\}  
\,. \end{align}
The UV and IR divergences are again removed by renormalization and matching onto PDFs. The finite terms of \eqs{I_qq_1}{I_qg_1} reproduce the matching coefficients $\cI_{qq}^\one$ and $\cI_{qg}^\one$ calculated in Ref.~\cite{Stewart:2010qs}
\begin{align} \label{eq:I_1}
\cI_{qq}^\one(t,x,\mu) &= \frac{\al_s C_F}{2\pi} \bigg\{
  \frac{2}{\mu^2} \cL_1\Big(\frac{t}{\mu^2}\Big) \de(1-x) +
  \frac{1}{\mu^2} \cL_0\Big(\frac{t}{\mu^2}\Big) (1+x^2) \cL_0(1-x)
  \nn \\ & \quad  
  + \de(t) \bigg[(1+x^2) \cL_1(1-x) 
  - \frac{1+x^2}{1-x} \ln x + 1-x - \frac{\pi^2}{6} \de(1-x) \bigg] \bigg\}
\,,  \nn \\ 
  \cI_{qg}^\one(t,x,\mu)&= \frac{\al_s T_F}{2\pi} \bigg\{ \frac{1}{\mu^2} \cL_0\Big(\frac{t}{\mu^2}\Big) (x^2+(1-x)^2)
  + \de(t) \bigg[(x^2+(1-x)^2) \Big(\ln \frac{1-x}{x} -1\Big) + 1 \bigg] \bigg\}  
\,.\end{align}
\end{widetext}

\subsection{TMD Beam Function}

We now consider the beam function where instead of the transverse virtuality $t$, the transverse momentum $k_{\perp}$ of the colliding parton is measured. These beam functions have rapidity (light-cone) divergences which may be regularized using \eg\cite{Collins:1981uw,Ji:2004wu,Chiu:2009yx,Chiu:2011qc,Becher:2011dz}. The regulator in Ref.~\cite{Becher:2011dz} only affects the phase space and not the amplitude, making it the most suitable for our approach. 
We will use a slightly modified version of this regulator
\begin{equation}
  \int\! \df^d k\, \theta(k^0) \de(k^2) \to   \int\! \df^d k\, \theta(k^0) \de(k^2) \Big(\frac{\nu}{2k^z}\Big)^\eta
\,,\end{equation}
where $k^z$ is the momentum component along the energetic direction.
At one-loop the kinematics are fully constrained by $x$ and $t$, such that
\begin{align}
 \vec k_\perp^{\,2} = \frac{1-x}{x}\,t 
\,.\end{align}
The rapidity-regulated phase space for the initial state is then given by
\begin{align}
\dPhiCollISR{2}(\vec k_\perp^{\,2},x) &= 
\dPhiCollISR{2}\Big(t=\frac{x}{1-x}\, \vec k_\perp^{\,2},x\Big) 
 \Big(\frac{\nu}{(1-x)p^-}\Big)^\eta
\nn \\ &
=\Big(\frac{\nu}{p^-}\Big)^{\eta}\, \df \vec k_\perp^{\,2}\, \df x\,
 \frac{x(1-x)^{-1-\eta} (\vec k_\perp^{\,2})^{-\eps}}{(4\pi)^{2-\eps} \Ga(1-\eps)}
\,,\end{align}
where we used that in the collinear limit $2k^z = k^+ + k^- = k^- +$ power corrections, and $p^-$ is the large light-cone component of the incoming quark.
(The rapidity divergence occurs for $x\to1$, which is unregulated when $\eta=0$.)
This leads to
\begin{widetext}
\begin{align} \label{eq:I_qq_1_kt}
 B_{q/q,\text{bare}}^{\one}(x,\vec k_\perp^{\,2}) 
 &= \int\! \dPhiCollISR{2}(\vec k_\perp'\!\!{}^{2},x')\, \siColl{2}\Big(\!-\frac{\vec k_\perp'\!\!{}^{2}}{(1 -x')},\frac{1}{x'}\Big)\, \de(\vec k_\perp^{\,2}-\vec k_\perp'\!\!{}^{2})\, \de(x-x')
 \\ &
 = \Big(\frac{\nu}{p^-}\Big)^{\eta}\,\Big(\frac{\mu^2 e^{\ga_E}}{4\pi}\Big)^\eps\, 
\frac{x(1-x)^{-1-\eta} (\vec k_\perp^{\,2})^{-\eps}}{(4\pi)^{2-\eps} \Ga(1-\eps)} 
 \frac{2g^2}{-\vec k_\perp^{\,2}/(1 -x)} P_{qg}^\zero\Big(\frac{1}{x}\Big)
  \nn \\ & 
  = \frac{\al_s C_F}{2\pi}\, \frac{e^{\eps \ga_E}}{\Gamma(1-\eps)}\,
  \Big(\frac{\nu}{p^-}\Big)^{\eta}\,
  \frac{\mu^{2\eps}}{(\vec k_\perp^{\,2})^{1+\eps}}\,(1-x)^{-\eta} \Big[\frac{1+x^2}{1-x} - \eps (1-x) \Big]
   \nn \\ &
  = \frac{\al_s C_F}{2\pi} \bigg[\!-\!\frac{1}{\eps} \de(\vec k_\perp^{\,2}) + \frac{1}{\mu^2} \cL_0\Big(\frac{\vec k_\perp^{\,2}}{\mu^2}\Big) \bigg]\,\bigg[ \de(1\!-\!x) \bigg(\!-\!\frac{2}{\eta} + 2 \ln \frac{p^-}{\nu} \bigg) +(1\!+\!x^2) \cL_0(1\!-\!x) - \eps (1\!-\!x) \bigg] + \ord{\eta,\eps}
\,.\nn\end{align}
Similarly,
\begin{align} \label{eq:I_qg_1_kt}
 B_{q/g,\text{bare}}^{\one}(x,\vec k_\perp^{\,2}) 
 &= -\frac{1}{1-\eps} \frac{T_F}{C_F} \int\! \dPhiCollISR{2}(\vec k_\perp'\!\!{}^{2},x')\, \siColl{2}\Big(\!-\frac{\vec k_\perp'\!\!{}^{2}}{(1 -x')},\frac{x'-1}{x'}\Big)\, \de(\vec k_\perp^{\,2}-\vec k_\perp'\!\!{}^{2})\, \de(x-x')
 \nn \\ & 
 = -\frac{1}{1-\eps} \frac{T_F}{C_F} \Big(\frac{\nu}{p^-}\Big)^{\eta}\,\Big(\frac{\mu^2 e^{\ga_E}}{4\pi}\Big)^\eps\, 
\frac{x (1-x)^{-1-\eta} (\vec k_\perp^{\,2})^{-\eps}}{(4\pi)^{2-\eps} \Ga(1-\eps)} 
 \frac{2g^2}{-\vec k_\perp^{\,2}/(1 -x)}  P_{qg}^\zero\Big(\frac{x-1}{x}\Big)
 \nn \\ &
 = \frac{\al_s T_F}{2\pi}\, \frac{e^{\eps \ga_E}}{\Gamma(2-\eps)}\,
  \Big(\frac{\nu}{p^-}\Big)^{\eta}\,
  \frac{\mu^{2\eps}}{(\vec k_\perp^{\,2})^{1+\eps}}\,(1-x)^{-\eta} [x^2+(1-x)^2 - \eps \big]
   \nn \\ & 
  = \frac{\al_s T_F}{2\pi} \bigg[\!-\!\frac{1}{\eps} \de(\vec k_\perp^{\,2}) + \frac{1}{\mu^2} \cL_0\Big(\frac{\vec k_\perp^{\,2}}{\mu^2}\Big) \bigg]\,[x^2+(1-x)^2 - 2\eps x(1-x) \big]
 + \ord{\eta,\eps}
\,. \end{align}
\end{widetext}
Following the prescription in Ref.~\cite{Chiu:2011qc}, the $1/\eta$ and $1/\eps_\text{UV}$ get removed by the (rapidity) renormalization. Subsequently, the $1/\eps_\text{IR}$ cancels in the matching onto PDFs, leaving as matching coefficient
\begin{align} \label{eq:I_1_kt}
\cI_{qq}^\one(x,\vec k_\perp^{\,2},\mu,\nu) &= 
\frac{\al_s C_F}{2\pi} \bigg\{ \frac{1}{\mu^2} \cL_0\Big(\frac{\vec k_\perp^{\,2}}{\mu^2}\Big) \bigg[(1+x^2) \cL_0(1-x) 
\nn \\ & \quad
+ 2\, \de(1-x) \ln \frac{p^-}{\nu} \bigg] +
\de(\vec k_\perp^{\,2}) (1-x) \bigg\}
\,,\nn \\
\cI_{qg}^\one(x,\vec k_\perp^{\,2},\mu,\nu) &= 
\frac{\al_s T_F}{2\pi} \bigg\{ \frac{1}{\mu^2} \cL_0\Big(\frac{\vec k_\perp^{\,2}}{\mu^2}\Big) \big[x^2+(1-x)^2\big]
\nn \\ & \quad
+ 2\,\de(\vec k_\perp^{\,2}) x(1-x) \bigg\}
\,.\end{align}
Using the Fourier transforms in \eq{plus_fourier} and adding the contribution of the one-loop soft function, one finds agreement with Eqs.~(38) and (39) of Ref.~\cite{Becher:2010tm}.
We note that the soft function $S$ vanishes for the regulator chosen in Ref.~\cite{Becher:2010tm}. Its contribution is $\sqrt{S}$ for each beam function (in impact-parameter space) and can be obtained from Eq.~(5.62) of Ref.~\cite{Chiu:2012ir} by replacing $C_A \to C_F$,
\begin{align}
S(\vec k_\perp^{\,2},\mu,\nu) &= \de(\vec k_\perp^{\,2}) + \frac{\al_s C_F}{\pi} \bigg[- \frac{1}{\mu^2} \cL_1\Big(\frac{\vec k_\perp^{\,2}}{\mu^2}\Big) 
 \\ & \quad
+ \frac{1}{\mu^2} \cL_0\Big(\frac{\vec k_\perp^{\,2}}{\mu^2}\Big) \ln \frac{\nu^2}{\mu^2}
 - \frac{\pi^2}{12} \de(\vec k_\perp^{\,2})\bigg] + \ord{\al_s^2}
\,.\nn\end{align}

\section{(Fragmenting) Jet Function at NNLO}
\label{sec:NLO}

\subsection{Splitting Functions and Phase Space}

At two-loop order we have contributions with two real emissions, a real-virtual correction and a purely virtual correction. The latter vanishes again in dimensional regularization. 
Starting with two real emissions, the collinear phase space for \emph{nonidentical} particles is given by~\cite{GehrmannDeRidder:1997gf}
\begin{align}
 \dPhiColl{3} &= \df s_{123}\, \df s_{12}\,\df s_{13}\, \df s_{23}\,  \de(s_{123}-s_{12}-s_{13}-s_{23})
 \nn \\ & \quad \times
 \df z_1\, \df z_2\, \df z_3\,   \de(1-z_1-z_2-z_3)
 \nn \\ & \quad \times 
 \frac{4\Theta(-\De)(-\De)^{-\frac12-\eps}}{(4\pi)^{5-2\eps} \Ga(1-2\eps)}
\,,\end{align}
where 
\begin{align}
 \De &=  (z_3 s_{12} - z_1 s_{23}- z_2 s_{13})^2 - 4z_1z_2s_{13}s_{23}
\,,\end{align}
with $s_{123}\geq0$ the total invariant mass, $s_{ij}\geq 0$ the invariant mass of partons $i$ and $j$ and $0 \leq z_i \leq 1$ the momentum fraction of parton $i$. 
The collinear part of a squared matrix element factors off and is given by
\begin{equation}
  \siColl{3,ijk} = \Big(\frac{\mu^2 e^{\ga_E}}{4\pi}\Big)^{2\eps}\, \frac{4g^4}{s_{123}^2} P_{ijk}
\,,\end{equation}
where the LO splitting functions for $q^* \to ijk$ are~\cite{Campbell:1997hg,Catani:1998nv} 
\begin{widetext}
\begin{align}
  P_{\bar q' q' q} &= C_F T_F\, \frac{s_{123}}{2s_{12}} \bigg[-
  \frac{[z_1 (s_{12}+2s_{23}) - z_2 (s_{12}+2s_{13})]^2}{(z_1 + z_2)^2 s_{12}s_{123}}
 + \frac{4 z_3 + (z_1-z_2)^2}{z_1+z_2}+(1-2\eps)\bigg(z_1+z_2-\frac{s_{12}}{s_{123}}\bigg)\bigg]
 \nn \\
 P_{\bar q q q} & = (P_{\bar q' q' q} + 2 \lra 3) + P_{\bar q q q}^{(\text{id})}
 \nn \\
 P_{\bar q q q}^{(\text{id})} &= C_F\bigg(C_F - \frac12 C_A\bigg) \bigg\{(1-\eps)\bigg(\frac{2s_{23}}{s_{12}} - \eps\bigg) 
 +  \frac{s_{123}}{s_{12}} \bigg[\frac{1+z_1^2}{1-z_2} - \frac{2z_2}{1-z_3} - \eps \bigg( \frac{(1-z_3)^2}{1-z_2} + 1 + z_1 - \frac{2z_2}{1-z_3}\bigg) - \eps^2 (1-z_3)\bigg] 
 \nn \\ & \quad
 - \frac{s_{123}^2}{2s_{12}s_{13}} z_1 \bigg[ \frac{1+z_1^2}{(1-z_2)(1-z_3)} - \eps\bigg(1+2 \frac{1-z_2}{1-z_3}\bigg)-\eps^2 \bigg]\bigg\} + (2 \lra 3)
\nn \\
P_{ggq} &= C_F^2 \bigg\{\frac{s_{123}^2}{2s_{13}s_{23}} z_3 \bigg[\frac{1+z_3^2}{z_1z_2}-\eps \frac{z_1^2+z_2^2}{z_1z_2} - \eps(1+\eps)\bigg] +(1-\eps)\bigg[\eps-(1-\eps)\frac{s_{23}}{s_{13}}\bigg]
+ \frac{s_{123}}{s_{13}} \bigg[\frac{z_3(1-z_1)+(1-z_2)^3}{z_1z_2}
\nn \\ & \quad
 - \eps(z_1^2+z_1z_2+z_2^2) \frac{1-z_2}{z_1z_2} + \eps^2(1+z_3) \bigg] \bigg\}
+ C_F C_A \bigg\{(1-\eps)\bigg(
 \frac{[z_1 (s_{12}+2s_{23}) - z_2 (s_{12}+2s_{13})]^2}{4(z_1 + z_2)^2 s_{12}^2}
+\frac14-\frac{\eps}{2}\bigg) 
\nn \\ & \quad
+ \frac{s_{123}^2}{2s_{12}s_{13}}
\bigg[\frac{2z_3+(1-\eps)(1-z_3)^2}{z_2} 
+ \frac{2(1-z_2)+(1-\eps)z_2^2}{1-z_3}\bigg] - \frac{s_{123}^2}{4s_{13}s_{23}} z_3 \bigg[\frac{2z_3+(1-\eps)(1-z_3)^2}{z_1z_2} + \eps(1-\eps)\bigg] 
\nn \\ & \quad
+ \frac{s_{123}}{2s_{12}} \bigg[(1-\eps)\frac{z_1(2-2z_1+z_1^2)-z_2(6-6z_2+z_2^2)}{z_2(1-z_3)}+2\eps \frac{z_3(z_1-2z_2)-z_2}{z_2(1-z_3)}\bigg]
+ \frac{s_{123}}{2s_{13}} \bigg[(1-\eps)\frac{(1-z_2)^3+z_3^2-z_2}{z_2(1-z_3)} 
\nn \\ & \quad
- \eps\bigg(\frac{2(1-z_2)(z_2-z_3)}{z_2(1-z_3)} - z_1 + z_2 \bigg) 
- \frac{z_3(1-z_1)+(1-z_2)^3}{z_1z_2} + \eps(1-z_2)\bigg(\frac{z_1^2+z_2^2}{z_1z_2}-\eps\bigg)\bigg]\bigg\} + (1 \lra 2)
\,.\end{align}
The real-virtual contributions have the same two-body phase-space in \eq{jet_ps} and can be written as a correction to the splitting function $P_{qg}^\zero$ (we use the explicit form given in \cite{Sborlini:2013jba})
\begin{align}
 P_{qg}^{(1)} &= 
 \Big(\frac{\mu^2 e^{\ga_E}}{s}\Big)^{\eps}
  \frac{2g^2}{(4\pi)^2}
 \frac{\pi\, \Ga(1-\eps)}{\eps \tan(\pi \eps) \Ga(1-2\eps)}\,
 C_F \bigg\{\Big[\frac{1+z^2}{1-z}-\eps (1-z)\Big]
 \Big[ (C_F - C_A) \Big(1-\frac{\eps^2}{1-2\eps}\Big) 
 \\ & \quad 
 + (C_A-2C_F) {}_2F_1\Big(1,-\eps;1-\eps;-\frac{1-z}{z}\Big)
 - C_A\, {}_2F_1\Big(1,-\eps;1-\eps;-\frac{z}{1-z}\Big) + C_F \Big]
 + (C_F - C_A)\frac{z(1+z)}{1-z} \frac{\eps^2}{1-2\eps} \bigg\}
\,.\nn\end{align}
\end{widetext}

\subsection{Calculational Technique}

The real-virtual corrections to the (fragmenting) jet function only involve a two particle final state. Their calculation proceeds along the same lines as in \sec{LO} and is straightforward to carry out. 
The double real emission contributions are more challenging and have been calculated in two ways.
In the direct phase space integration, we start by performing the integration over the invariants $s_{ij}$
in $4-2\eps$ dimensions using the analytic results in the appendix of Ref.~\cite{Kosower:2003np}.
We carry out the integrals over the momentum fractions by first extracting the singular behavior in the soft/collinear limits, expanding in $\epsilon$ using the plus distribution expansion in \eq{plus_exp} and integrating the regularized expressions. Hypergeometric functions are expanded in $\eps$ with the aid of the \HypExp and \HPL packages~\cite{Maitre:2005uu,Huber:2005yg}.
Some additional details are given in \secs{jet_nnlo}{fragjet_nnlo} along with the presentation of the results.

Alternatively, we use the reverse-unitarity approach to phase space integrals 
\cite{Anastasiou:2002yz, Anastasiou:2002qz, Anastasiou:2003yy, Anastasiou:2003ds}
to perform a reduction to master integrals for the jet function and the fragmenting jet function integrals separately.
We use both \FIRE~\cite{Smirnov:2008iw} and \Reduze~\cite{Studerus:2009ye,vonManteuffel:2012np} for this purpose.
For the jet function, the resulting master integrals are obtained by performing the phase space integration for 
arbitrary $\eps$. For the fragmenting jet function, we use a combination of direct integration and differential equations
\cite{Kotikov:1990kg,Kotikov:1991hm,Kotikov:1991pm,Remiddi:1997ny,Caffo:1998yd,Caffo:1998du,Gehrmann:1999as}
to obtain the master integrals. Additional details are provided in \apps{jet_mi}{fragjet_mi}.

\subsection{Bare Jet Function Calculation}
\label{sec:jet_nnlo}

In the jet function the total invariant mass $s$ is fixed, the other phase-space variables are integrated over and the contributions from the various channels are summed,
\begin{align}
 J_{q,\text{bare}}(s) &= 
 \int\! \dPhiColl{2}(s')\, \siColl{2}(s')\, \de(s-s')
 \\ & \quad 
 +  \sum_{ij} S_{ijq} \int\! \dPhiColl{3}\, \siColl{3,ijq}\, \de(s-s_{123})
 + \ord{\al_s^3}\,,
\nn \end{align}
where $ij$ runs over $\{gg,\bar u u, \bar d d, \dots\}$ and $S_{ijq}$ is an identical particle factor.

We start with $P_{\bar q' q' q}$ which only has a collinear divergence described by $(1-z_3)^{-1-2\eps}$. After expanding this in $\eps$ using \eq{plus_exp} the remaining integrals are regular, 
\begin{widetext}
\begin{align} \label{eq:jet_QQq}
 & \int\! \dPhiColl{3}\, \siColl{3,\bar q'q'q}\, \de(s-s_{123})
 \nn \\ & \quad
 = \int\! \df s_{12}\, \df s_{13}\, \df s_{23}\, \df z_1\, \df z_2\, \df z_3\,  \de(s_{123}-s_{12}-s_{13}-s_{23})\, \de(1-z_1-z_2-z_3)\,
 \frac{4\Theta(-\De)(-\De)^{-\frac12-\eps}}{(4\pi)^{5-2\eps} \Ga(1-2\eps)}
 \Big(\frac{\mu^2 e^{\ga_E}}{4\pi}\Big)^{2\eps}\, \frac{4g^4}{s_{123}^2} P_{\bar q' q' q}
\nn \\ & \quad
 = \frac{\al_s^2 C_F T_F}{(4\pi)^2}   \frac{(\mu^2 e^{\ga_E})^{2\eps}}{\Ga(1-2\eps)} \frac{4}{s_{123}^{1+2 \eps}}
 \int_0^1\! \df z_1 \int_0^{1-z_1}\!\!\df z_3\, \frac{1}{(1-z_3)^{1+2\eps}} 
\nn \\ & \qquad \times
 \frac{(1\!+\!z_3^2) [z_1^2 \!+\! (1\!-\!z_1\!-\!z_3)^2]\!-\!2 \eps [(1\!-\!z_3\!+\!z_3^2) (1\!-\!z_3)^2\!-\!(1\!-\!6 z_3\!+\!z_3^2) z_1 (1\!-\! z_1 \!-\! z_3)] \!+\!\eps^2 (1\!-\!z_3)^4}{\eps (2 \eps-1) z_1^{\eps} z_3^{\eps} (1-z_1-z_3)^{\eps} (1-z_3)^{3-2\eps}}
\nn\\ & \quad
 = \frac{\al_s^2 C_F T_F}{(4\pi)^2}   \frac{(\mu^2 e^{\ga_E})^{2\eps}}{\Ga(1-2\eps)} \frac{4}{s_{123}^{1+2 \eps}}
 \int_0^1\! \df z_3 \int_0^{1-z_3}\!\!\df z_1\, 
 \Big[-\frac{1}{2\eps} \de(1\!-\!z_3) + \cL_0(1\!-\!z_3) +\ord{\eps} \Big]
\bigg\{\!-\!\frac1\eps\,\frac{(1\!+\!z_3^2) [z_1^2 \!+\! (1\!-\!z_1\!-\!z_3)^2]}{(1-z_3)^3}
\nn \\ & \qquad \times
\Big[1 + 2\eps + \eps \ln \frac{(1\!-\!z_3)^2}{z_1 z_3(1\!-\!z_1\!-\!z_3)} \Big]
+ \frac{2 [(1\!-\!z_3\!+\!z_3^2) (1\!-\!z_3)^2\!-\!(1\!-\!6 z_3\!+\!z_3^2) z_1 (1\!-\! z_1 \!-\! z_3)]}{(1-z_3)^3} + \ord{\eps}
\bigg\}
\nn\\ & \quad
 = \frac{\al_s^2 C_F T_F}{(4\pi)^2} \frac{\mu^{4\eps}}{s_{123}^{1+2 \eps}}
\bigg[\frac{8}{3 \eps^2}+\frac{76}{9 \eps}+\frac{746}{27}-\frac{20}{9} \pi ^2+\eps \Big(\frac{7081}{81}-\frac{190}{27} \pi ^2-\frac{256}{9} \zeta_3 \Big) + \ord{\eps^2} \bigg]
\,.\end{align}
On the second to last line we suppressed the $\ord{\eps}$ terms for brevity, though they are of course necessary to obtain the final expression.
For $q \to \bar q q q$ we get a contribution equal to \eq{jet_QQq}, as well as an additional interference contribution described by $P_{\bar qqq}^\text{(id)}$.
This interference contribution has neither collinear nor soft divergences in $z_i$, so we may directly expand in $\eps$ 
\begin{align} \label{eq:jet_qqq}
 & \frac12 \int\! \dPhiColl{3}\, \siColl{3,\bar qqq\text{(id)}}\, \de(s-s_{123})
 \nn \\ & \quad
 = \frac12 \int\! \df s_{12}\, \df s_{13}\, \df s_{23}\, \df z_1\, \df z_2\, \df z_3\,  \de(s_{123}-s_{12}-s_{13}-s_{23})\, \de(1-z_1-z_2-z_3)\,
 \frac{4\Theta(-\De)(-\De)^{-\frac12-\eps}}{(4\pi)^{5-2\eps} \Ga(1-2\eps)}
 \Big(\frac{\mu^2 e^{\ga_E}}{4\pi}\Big)^{2\eps}\, \frac{4g^4}{s_{123}^2} P_{\bar qqq}^\text{(id)}
\nn \\ & \quad
 = \frac{\al_s^2 C_F (C_F - \frac{1}{2}C_A)}{(4\pi)^2}   \frac{(\mu^2 e^{\ga_E})^{2\eps}}{\Ga(1-2\eps)} \frac{4}{s_{123}^{1+2 \eps}}
 \int_0^1\! \df z_1 \int_0^{1-z_1}\!\!\df z_2\, \frac{z_1^{-\eps} z_2^{-\eps} (1-z_1-z_2)^{-\eps}}{\eps(1-2\eps) (1-z_2) (z_1+z_2)^2}
\nn \\ & \qquad \times
\Big[(1-2\eps)z_1^{-\eps} (1\!-\!z_2)^{\eps}(z_1\!+\!z_2)^{1+\eps} [1\!+\!z_1^2\!-\!\eps (1\!-\!z_2)(2\!+\!z_1\!-\!z_2)\!-\!\eps^2 (1\!-\!z_2) (z_1\!+\!z_2)] \, _2F_1\Big(\!-\!\eps,-\!\eps;1\!-\!\eps;\frac{z_2 (1\!-\!z_1\!-\!z_2)}{(1\!-\!z_2) (z_1\!+\!z_2)}\Big)
\nn \\ & \qquad \quad
-(z_1 + z_2) (z_1^2+1)
+\eps (3 z_1^3+6 z_1^2z_2+4 z_1z_2^2+z_2^3-z_1^2-4 z_1z_2+ z_2^2+5 z_1+z_2)
\nn \\ &\qquad \quad
-\eps^2 (2 z_1^3+6 z_1^2 z_2 +6 z_1 z_2^2 +2 z_2^3 -4 z_1 z_2 +4 z_1)
- \eps^3 (z_1+z_2)^2 (1- z_2) \Big]
\nn\\ & \quad
 = \frac{\al_s^2 C_F (C_F - \frac{1}{2}C_A)}{(4\pi)^2} \frac{\mu^{4\eps}}{s_{123}^{1+2 \eps}}
\big[13 - 2 \pi^2 + 8 \zeta_3+ \, \eps \left( \frac{175}{2} - 4 \pi^2 -84 \zeta_3 + \frac{11}{15} \pi^4\right)+ \ord{\eps^2} \big]
\,.\end{align}
The symmetry factor $S_{\bar q qq}=1/2$ for identical quarks cancels against the permutation $(2 \lra 3)$ inside $P_{\bar qqq}^\text{(id)}$. We will separate the calculation of $P_{ggq}$ by color structure. 
The $C_F^2$ color structure has two soft divergences $z_1^{-1-\eps} z_2^{-1-\eps}$, yielding
\begin{align} \label{eq:jet_cf2}
 & \frac12 \int\! \dPhiColl{3}\, \siColl{3,ggq,C_F^2}\, \de(s-s_{123})
 \nn \\ & \quad
= \frac 12 \int\! \df s_{12}\, \df s_{13}\, \df s_{23}\, \df z_1\, \df z_2\, \df z_3\,  \de(s_{123}-s_{12}-s_{13}-s_{23})\, \de(1-z_1-z_2-z_3)\,
 \frac{4\Theta(-\De)(-\De)^{-\frac12-\eps}}{(4\pi)^{5-2\eps} \Ga(1-2\eps)}
 \Big(\frac{\mu^2 e^{\ga_E}}{4\pi}\Big)^{2\eps}\, \frac{4g^4}{s_{123}^2} P_{ggq}^{C_F^2}
\nn \\ & \quad
 = \frac{\al_s^2 C_F^2}{(4\pi)^2}   \frac{(\mu^2 e^{\ga_E})^{2\eps}}{\Ga(1-2\eps)} \frac{4}{s_{123}^{1+2 \eps}}
 \int_0^1\! \df z_1 \int_0^{1-z_1}\!\!\df z_2\, \frac{1}{z_1^{1+\eps}z_2^{1+\eps}} 
 \times \frac{(1-z_1-z_2)^{-\eps}}{\eps(1-2\eps)(1-z_2)^2}
\Big[(2\eps\!-\!1)(1\!-\!z_1)^{\eps} (1\!-\!z_2)^{2\!+\!\eps}(1\!-\!z_1\!-\!z_2)^{-\eps} 
\nn \\ & \qquad \quad \times
[(1\!-\!\eps) (z_1\!+\!z_2)^2 \!+\! 2(1\!-\!z_1\!-\!z_2)\!+\! \eps (1\!-\!\eps) z_1 z_2] \, _2F_1\Big(\!-\!\eps,-\!\eps;1\!-\!\eps;\frac{z_1 z_2}{(1\!-\!z_1)(1\!-\!z_2)}\Big)
\nn \\ & \qquad \quad 
\!-\!z_2^4\!+\!4 z_2^3\!-\!z_1^2\!-\!2 z_1 z_2\!-\!7 z_2^2\!+\!2 z_1\!+\!6 z_2\!-\!2
\!+\!\eps (3 z_2^4\!+\!z_1 z_2^3\!-\!z_1^2 z_2\!-\!z_1 z_2^2\!-\!10 z_2^3\!+\!3 z_1^2\!+\!4 z_1 z_2\!+\!15 z_2^2\!-\!4 z_1\!-\!12 z_2\!+\!4)
\nn \\ & \qquad \quad
\!+\!\eps^2 (-\!z_1^2 z_2^2\!-\!2 z_1 z_2^3\!-\!2 z_2^4\!+\!2 z_1^2 z_2 \!+\!2 z_1 z_2^2 \!+\!4 z_2^3 \!-\!2 z_1^2\!-\!2 z_2^2)
\!+\!\eps^3 z_1 z_2 (1\!-\!z_2) (2\!-\!z_1\!-\!z_2) \Big]
\nn\\ & \quad
 = \frac{\al_s^2 C_F^2}{(4\pi)^2} \frac{\mu^{4\eps}}{s_{123}^{1+2 \eps}}
\bigg[-\frac{16}{\eps^3}-\frac{24}{\eps^2}+\frac{1}{\eps} \Big(-75+\frac{40}{3} \pi^2 \Big)-\frac{417}{2} + 20 \pi ^2 +\frac{464}{3}\zeta_3 
\nn \\& \qquad \qquad \qquad \qquad
+ \, \eps \left( - \frac{2275}{4} + \frac{125}{2} \pi^2 + 256 \zeta_3 - \frac{122}{45} \pi^4 \right)+ \ord{\eps^2} \bigg]
\,,\end{align}
where the symmetry factor of 1/2 cancels against the permutation $(1 \lra 2)$. The calculation of the $C_F C_A$ color structure is more complicated and is split up into parts (a) - (e) which have different singular structures,
\begin{align} \label{eq:jet_cfca}
 & \frac12 \int\! \dPhiColl{3}\, \siColl{3,ggq,C_F C_A}\, \de(s-s_{123})
 \nn \\ & 
= \int\! \df s_{12}\, \df s_{13}\, \df s_{23}\, \df z_1\, \df z_2\, \df z_3\,  \de(s_{123}-s_{12}-s_{13}-s_{23})\, \de(1-z_1-z_2-z_3)\,
 \frac{4\Theta(-\De)(-\De)^{-\frac12-\eps}}{(4\pi)^{5-2\eps} \Ga(1-2\eps)}
 \nn \\ & \quad \times 
 \Big(\frac{\mu^2 e^{\ga_E}}{4\pi}\Big)^{2\eps} \frac{4g^4}{s_{123}^2}\,  
 C_F C_A \bigg\{(1\!-\!\eps)\bigg(
 \frac{[z_1 (s_{12}\!+\!2s_{23}) \!-\! z_2 (s_{12}\!+\!2s_{13})]^2}{4(z_1 \!+\! z_2)^2 s_{12}^2}
\!+\! \frac14 \!-\! \frac{\eps}{2}\bigg) 
 & &(a) \nn \\ & \qquad
- \frac{s_{123}^2}{4s_{13}s_{23}} z_3 \bigg[\frac{2z_3+(1-\eps)(1-z_3)^2}{z_1z_2} + \eps(1-\eps)\bigg] + \frac{s_{123}}{2s_{13}} \bigg[- \frac{z_3(1-z_1)+(1-z_2)^3}{z_1z_2} + \eps(1-z_2)\bigg(\frac{z_1^2+z_2^2}{z_1z_2}-\eps\bigg)\bigg]
 & &(b) \nn \\ & \qquad
+ \frac{s_{123}}{2s_{12}} \bigg[(1-\eps)\frac{z_1(2-2z_1+z_1^2)-z_2(6-6z_2+z_2^2)}{z_2(1-z_3)}+2\eps \frac{z_3(z_1-2z_2)-z_2}{z_2(1-z_3)}\bigg] 
 + \frac{s_{123}}{2s_{13}} \bigg[(1-\eps)\frac{(1-z_2)^3+z_3^2-z_2}{z_2(1-z_3)} 
 & &(c) \nn \\ & \qquad
- \eps\bigg(\frac{2(1\!-\!z_2)(z_2\!-\!z_3)}{z_2(1\!-\!z_3)} \!-\! z_1 \!+\! z_2 \bigg) \bigg]_{(c)}
+ \bigg[\frac{s_{123}^2}{2s_{12}s_{13}}
 \frac{2(1\!-\!z_2)\!+\!(1\!-\!\eps)z_2^2}{1\!-\!z_3}\bigg]_{(d)}
 + \bigg[\frac{s_{123}^2}{2s_{12}s_{13}}
\frac{2z_3\!+\!(1\!-\!\eps)(1\!-\!z_3)^2}{z_2} \bigg]_{(e)} \bigg\}
\nn \\ & 
 = \frac{\al_s^2 C_F C_A}{(4\pi)^2}   \frac{(\mu^2 e^{\ga_E})^{2\eps}}{\Ga(1-2\eps)} \frac{4}{s_{123}^{1+2 \eps}}
 \int_0^1\! \df z_1\,\df z_2\, \df z_3\, \de(1-z_1-z_2-z_3)
\nn \\ & \quad  \times \bigg\{
 \frac{1}{(1-z_3)^{1+2\eps}} \times
 \frac{(\eps-1) z_1^{-\eps} z_3^{-\eps} (1-z_1-z_3)^{-\eps}}{2 \eps (1-2 \eps)(1-z_3)^{3-2 \eps}} 
\big[4 z_1 z_3 (1-z_1-z_3) 
 & &(a) \nn \\ & \qquad +
\eps [-(1-z_3)^2 (1+z_3^2)+2 z_1(1+z_3)^2 (1- z_1 -z_3)]+ \eps^2 (1-z_3)^4 \big]
\nn \\ & \quad +
 \frac{1}{z_1^{1+\eps} z_2^{1+\eps}} \times \frac{(1\!-\!z_1\!-\!z_2)^{-\eps}}{2\eps(1\!-\!z_2)}
\Big[(1\!-\!z_1)^\eps (1\!-\!z_2)^{1+\eps} (1\!-\!z_1\!-\!z_2)^{-\eps} 
[(1\!-\!\eps) (z_1\!+\!z_2)^2 \!+\! 2(1\!-\!z_1\!-\!z_2)\!+\! \eps (1\!-\!\eps) z_1 z_2]
 & &(b) \nn \\ & \qquad  \times
 {}_2F_1\Big(\!-\!\eps,-\!\eps;1\!-\!\eps;\frac{z_1 z_2}{(1\!-\!z_1)(1\!-\!z_2)} \Big)
 \!-\!z_2^3\!+\!z_1^2\!+\!z_1 z_2\!+\!3 z_2^2\!-\!2z_1\!-\!4 z_2\!+\!2 
 \!-\!\eps (1\!-\!z_2) (z_1^2\!+\!z_2^2) + \eps^2 z_1 z_2 (1\!-\!z_2)
\Big]
\nn \\ & \quad +
\frac{(1\!-\!z_2\!-\!z_3)^{-\eps}}{z_2^{1+\eps}(1\!-\!z_3)} \times \frac{z_3^{-\eps}}{2\eps(1\!-\!z_2)(1\!-\!z_3)} \Big[
\!-\! 2 z_2^4 \!-\! 4 z_2^3 z_3 \!-\! 3 z_2^2 z_3^2 \!-\! z_2 z_3^3 \!+\! 10 z_2^3 \!+\! 8 z_2^2 z_3 \!+\! 4 z_2 z_3^2 \!+\! 2 z_3^3 \!-\! 17 z_2^2 \!-\! 7 z_2 z_3 \!-\! 2 z_3^2
 & &(c) \nn \\ & \qquad 
\!+\! 12 z_2 \!+\! 2 z_3 \!-\! 2 \!+\! \eps (2 z_2^4 \!+\!  4 z_2^3 z_3 \!+\! 5 z_3^2 z_2^2 \!+\!2 z_2 z_3^3 \!-\!10 z_2^3 \!-\! 16 z_2^2 z_3 \!-\! 11 z_2 z_3^2 \!-\! 2 z_3^3 \!+\! 15 z_2^2 \!+\! 18 z_2 z_3 \!+\! 6 z_3^2 \!-\! 9 z_2 \!-\! 6 z_3+ 2)\Big]
\nn\\ & \quad 
- \frac{(1-z_1-z_3)^{-\eps}}{z_1^{1+2 \eps}(1-z_3)^{1-\eps}} \times \frac{1}{\eps}
 z_3^{-\eps} (z_1+z_3)^{\eps} \Big[2 (z_1+z_3)+(1-\eps) (1-z_1-z_3)^2\Big] \, _2F_1\Big(-\eps,-\eps;1-\eps;\frac{z_3 (1-z_1-z_3)}{(1-z_3) (z_1+z_3)}\Big)
 & &(d) \nn\\ & \quad 
+ \frac{(z_1+z_2)^\eps}{z_1^{1+2\eps} z_2^{1+\eps}} \times
 \frac{(1-z_2)^{\eps} (1\!-\!z_1\!-\!z_2)^{-\eps}}{\eps}  \Big[2(z_1+z_2-1)-(1-\eps)(z_1+z_2)^2\Big] \, _2F_1\Big(-\eps,-\eps;1-\eps;\frac{z_2 (1\!-\!z_1\!-\!z_2)}{(1\!-\!z_2) (z_1\!+\!z_2)}\Big)
 & &(e) \nn\\ & 
 = \frac{\al_s^2 C_F C_A}{(4\pi)^2} \frac{\mu^{4\eps}}{s_{123}^{1+2 \eps}}
\bigg\{\bigg[\frac{2}{3 \eps^2}+\frac{22}{9 \eps}+\frac{242}{27} -\frac{5}{9}\pi^2+\eps \Big(\frac{2401}{81}-\frac{55}{27}\pi ^2-\frac{64}{9}\zeta_3 \Big)+ \ord{\eps^2} \bigg]
 & &(a) \nn\\ & \quad +
\bigg[\frac{8}{\eps^3}+\frac{12}{\eps^2}+\frac{1}{\eps}\Big(\frac{77}{2}-\frac{20 }{3} \pi^2\Big)+\frac{435}{4} -10 \pi^2-\frac{232}{3} \zeta_3 - 37.8\, \eps + \ord{\eps^2}\bigg]
 & &(b) \nn\\ & \quad +
\bigg[
-\frac{4}{\eps^3}-\frac{14}{\eps^2}+\frac{1}{\eps}\Big(-\!\frac{245}{6}\!+\!2\pi^2\Big)-\frac{4295}{36} \!+\!\frac{35}{3} \pi^2 \!+\! \frac{128}{3}\zeta_3 +\eps \Big(-\!\frac{75851}{216} \!+\! \frac{1225}{36}\pi^2 \!+\!\frac{448}{3} \zeta_3 \!-\! \frac{31}{90}\pi^4 \Big) + \ord{\eps^2} \bigg]
 & &(c) \nn\\ &\quad
+ \bigg[
 -\frac{2}{\eps^3}-\frac{3}{\eps^2}+\frac{1}{\eps}(\pi ^2-13)-41+\frac{3}{2}\pi ^2-\frac{20 }{3}\zeta_3 - 134.8\, \eps + \ord{\eps^2} \bigg]
 & &(d) \nn \\ &\quad
 +\bigg[-\frac{6}{\eps^3}-\frac{9}{\eps^2}+\frac{1}{\eps}(-29+5 \pi^2)-75+\frac{15}{2}\pi^2 + 64 \zeta_3 + 118.6\, \eps + \ord{\eps^2} \bigg]
 & &(e) \nn\\ & 
 = \frac{\al_s^2 C_F C_A}{(4\pi)^2} \frac{\mu^{4\eps}}{s_{123}^{1+2 \eps}}
\bigg[-\frac{4}{\eps^3} - \frac{40}{3 \eps^2}+ \frac{1}{\eps}\Big(-\frac{377}{9}+\frac{8}{3}\pi^2\Big) - \frac{3175}{27} + \frac{91}{9} \pi^2 + \frac{68}{3} \zeta_3 
\nn \\ & \qquad \qquad \qquad \qquad
+ \, \eps \left( - \frac{51337}{162} + \frac{1741}{54} \pi^2 + \frac{902}{9} \zeta_3 - \frac{41}{90} \pi^4 \right)+ \ord{\eps^2} \bigg] 
\,.\end{align}
\end{widetext}
In the last line we have displayed the analytic result obtained from the reduction to master integrals which agrees with the partially numerical result $(a)+\dotsb+(e)$ within integration errors.
We have kept the labels (a) - (e) throughout to allow one to keep track of the various terms contributing to $P_{ggq}^{C_F C_A}$. The first factor for each term shows the singular structure, which we expand using \eq{plus_exp}. For terms (c) through (e) it is convenient to first perform a change of variables:
\begin{align}
(c)\quad &r = (1-z_3)^2\,, \quad w = \frac{z_2}{1-z_3}
\,, \\ &
 \int_0^1\! \df z_3 \int_0^{1-z_3}\! \df z_2\, \frac{(1\!-\!z_2\!-\!z_3)^{-\eps}}{z_2^{1+\eps}(1\!-\!z_3)} 
\, \nn \\ & \quad
= \int_0^1\! \df r \int_0^1\! \df w\, \frac{(1-w)^{-\eps}}{2r^{1+\eps} w^{1+\eps}}
\,, \nn \\ 
(d)\quad &r = (1-z_3)^2\,, \quad v = \frac{z_1}{1-z_3}
\,, \nn \\ & 
 \int_0^1\! \df z_3 \int_0^{1-z_3}\! \df z_1\,  \frac{(1-z_1-z_3)^{-\eps}}{z_1^{1+2 \eps}(1-z_3)^{1-\eps}} 
\, \nn \\ & \quad
 = \int_0^1\! \df r \int_0^1\! \df v \, \frac{(1-v)^{-\eps}}{2r^{1+\eps} v^{1+2\eps}}
\,,\nn \\
(e)\quad &a = z_1+z_2\,, \quad b = \frac{z_1-z_2}{z_1+z_2}
\,, \nn \\ &
 \int_0^1\! \df z_1 \int_0^{1-z_1}\! \df z_2\, \frac{(z_1+z_2)^\eps}{z_1^{1+2\eps} z_2^{1+\eps}} 
\, \nn \\ & \quad
 = \int_0^1\! \df a \int_{-1}^1\! \df b\, \frac{2^{1+3\eps}}{a^{1+2\eps}(1-b)^{1+\eps}(1+b)^{1+2\eps}}
\,.\nn \end{align}
This eliminates the overlap of divergences which would prevent an expansion in plus distributions. For example, the divergences in term (e) at $(a,b) = (0,1)$ and $(0,-1)$ would otherwise coincide at $(z_1,z_2)=(0,0)$.

The calculation of the real-virtual corrections follows similar (but simpler) steps:
\begin{widetext}
\begin{align}\label{eq:jet_rv}
 & \int\! \dPhiColl{2}(s)\, \siColl{2}{}^{\one}(s')\, \de(s-s')
 \nn \\ & \quad
= \int_0^1\!\df z\, \frac{[z(1-z)s]^{-\eps}}{(4\pi)^{2-\eps} \Ga(1-\eps)}\, \Big(\frac{\mu^2 e^{\ga_E}}{4\pi}\Big)^\eps\,\frac{2g^2}{s}\, P_{qg}^\one
\nn \\ & \quad
= \frac{\al_s^2 C_F}{(4\pi)^2}\, \frac{(\mu^2 e^{\ga_E})^{2\eps}}{\Ga(1\!-\!2\eps)} \frac{4}{s^{1+2\eps}}
\int_0^1\!\df z\, \frac{1}{(1\!-\!z)^{1+2\eps}} \times z^{-\eps} (1\!-\!z)^\eps
 \frac{\pi}{\eps \tan(\pi \eps)}\,
 \bigg\{[1\!+\!z^2\!-\!\eps (1\!-\!z)^2]
 \Big[ (C_F \!-\! C_A) \Big(1\!-\!\frac{\eps^2}{1\!-\!2\eps}\Big) 
 \nn \\ & \qquad \quad
 \!+\! (C_A\!-\!2C_F) {}_2F_1\Big(1,-\eps;1\!-\!\eps;\!-\!\frac{1\!-\!z}{z}\Big)
 \!-\! C_A\, {}_2F_1\Big(1,-\eps;1\!-\!\eps;\!-\!\frac{z}{1\!-\!z}\Big) \!+\! C_F \Big]
 \!+\! (C_F \!-\! C_A) \frac{\eps^2}{1\!-\!2\eps} z(1\!+\!z) \bigg\}
\nn \\ & \quad
 = \frac{\al_s^2 C_F}{(4\pi)^2} \frac{\mu^{4\eps}}{s^{1+2 \eps}}\bigg\{ 
 C_F \bigg[ \frac{1}{\eps}\Big(10 \!-\! \frac{8}{3} \pi^2\Big) \!+\! 40 \!-\! 80 \zeta_3 
 + \, \eps \left( 122 - \frac{25}{3} \pi^2 - \frac{8}{15} \pi^4 \right)\bigg]
 \nn \\ & \qquad \qquad \qquad \qquad
 + C_A \bigg[ \frac{4}{\eps^3}\!+\!\frac{6}{\eps^2}\!+\!\frac{1}{\eps}(16\!-\!2 \pi ^2)\!+\!40 \!-\! 5 \pi ^2 \!+\! \frac{64}{3} \zeta_3 \!
 +  \eps \left( 94 - \frac{40}{3} \pi^2 - 28 \zeta_3 + \frac{91}{90} \pi^4 \right) \bigg] + \ord{\eps^2} \bigg\}
\,.\end{align}
Adding up Eqs.~\eqref{eq:jet_QQq}, \eqref{eq:jet_qqq}, \eqref{eq:jet_cf2}, \eqref{eq:jet_cfca} and \eqref{eq:jet_rv}, we obtain the two-loop contribution to the bare jet function,
\begin{align} \label{eq:J_bare}
  J_{q,\text{bare}}(s)  &= 
  \de(s) + 
  Z_\al \frac{\al_sC_F}{4\pi} \frac{\mu^{2\eps}}{s^{1+\eps}} 
  \bigg[- \frac{4}{\eps} - 3 +\eps(-7+\pi^2)+\eps^2\Big(-14+\frac{3}{4}\pi^2+\frac{28}{3} \zeta_3\Big) + \eps^3\Big(-28+\frac{7}{4}\pi ^2 + 7 \zeta_3+\frac{1}{24}\pi^4 \Big)\bigg]
  \nn \\ & \quad
  + Z_\al^2 \frac{\al_s^2 C_F}{(4\pi)^2} \frac{\mu^{4\eps}}{s^{1+2 \eps}} \bigg\{ C_F \bigg[
  -\frac{16}{\eps^3}-\frac{24}{\eps^2}+\frac{1}{\eps}\Big(-65+\frac{32}{3}\pi ^2\Big)-\frac{311}{2} +18 \pi ^2 + \frac{248}{3}  \zeta_3 
  + \eps \Big(-\frac{1437}{4} + \frac{301}{6} \pi^2 
  \nn \\ & \quad
  - \frac{113}{45} \pi^4 + 172 \zeta_3\Big) \bigg] 
  + C_A \bigg[-\frac{22}{3 \eps^2}+\frac{1}{\eps}\Big(-\frac{233}{9}+\frac{2}{3}\pi ^2\Big)-\frac{4541}{54}+\frac{55}{9}\pi^2+40 \zeta_3+ \eps \Big(-\frac{86393}{324} + \frac{1129}{54} \pi^2 
  \nn \\ & \quad 
  + \frac{17}{90} \pi^4 + \frac{1028}{9} \zeta_3\Big) \bigg]
  + T_F n_f \bigg[\frac{8}{3 \eps^2}+\frac{76}{9 \eps}+\frac{746}{27}-\frac{20}{9} \pi ^2+\eps \Big(\frac{7081}{81}-\frac{190}{27} \pi ^2-\frac{256}{9} \zeta_3 \Big) \bigg] + \ord{\eps^2} \bigg\} + \ord{\al_s^3}
\,.\end{align}
Here we also included all the relevant orders in $\eps$ of the one-loop jet function beyond the finite terms in \eq{J_q_1}, and have taken the renormalization of the coupling constant into account
\begin{align}
  Z_\al = 1 - \frac{\al_s}{4\pi} \frac{\beta_0}{\eps} + \ord{\al_s^2}
  = 1 - \frac{\al_s}{4\pi} \frac{1}{\eps}\Big(\frac{11}{3} C_A - \frac{4}{3} T_F n_f\Big) + \ord{\al_s^2}
\,.\end{align}

\subsection{Renormalized Jet Function}
\label{sec:jet_nnlo_ren}

The final result for the jet function can be obtained from \eq{J_bare} by expanding $s$ in plus distributions using \eq{plus_exp} and renormalizing
\begin{align} \label{eq:Jrenormalization}
  J_q(s,\mu) &= \int_0^\infty\! \df s'\, Z_{J_q}(s',\mu) J_{q,\text{bare}}(s-s')
  \,, \\ 
  Z_{J_q}(s',\mu) &= \de(s') + \frac{\al_s C_F}{4\pi} \Big[\frac{4}{\eps} \frac{1}{\mu^2} \cL_0\Big(\frac{s'}{\mu^2}\Big) - \de(s')\Big(\frac{4}{\eps^2}+\frac{3}{\eps} \Big) \Big] +  
\nn \\ & \quad
  \frac{\al_s^2C_F}{(4 \pi)^2}  \bigg\{ 
  C_F \bigg[
	\frac{1}{\mu^2} \cL_1 \Big( \frac{s'}{\mu^2} \Big) 
		\frac{16}{\eps^2}
	+\frac{1}{\mu^2} \cL_0 \Big( \frac{s'}{\mu^2} \Big) \Big(
		-\frac{16}{\eps^3} -\frac{12}{\eps^2}
		\Big)
	\nn \\ & \quad	\qquad \qquad \qquad 
  	+ \de(s') \bigg( 
  		\frac{8}{\eps^4}  + \frac{12}{\eps^3}  
		+ \frac{1}{\eps^2} \Big(\frac{9}{2}- \frac{4}{3}\pi^2 \Big) 
		+ \frac{1}{\eps} \Big(-\frac{3}{4}+\pi^2 - 12 \zeta_3 \Big)  \bigg)
  \bigg]
  \nn  \\ & 	\qquad \qquad 
  +C_A \bigg[ 
  	\frac{1}{\mu^2} \cL_0 \Big( \frac{s'}{\mu^2} \Big) \bigg(
		- \frac{22}{3} \frac{1}{\eps^2}
		+ \frac{1}{\eps} \Big( \frac{134}{9} - \frac{2}{3} \pi^2 \Big)
		\bigg)  
	\nn \\ & \quad \qquad \qquad \qquad 
  	+ \de(s') \bigg(
  		\frac{11}{\eps^3}  
		+ \frac{1}{\eps^2} \Big(-\frac{35}{18}+ \frac{\pi^2}{3}\Big) 
		+ \frac{1}{\eps} \Big(-\frac{1769}{108}-\frac{11}{18} \pi^2 + 20 \zeta_3 \Big)  \bigg) 	
\bigg]
    \nn  \\ & \qquad \qquad 
  +T_F n_f \bigg[ 
	\frac{1}{\mu^2} \cL_0 \Big( \frac{s'}{\mu^2} \Big) \bigg(
		\frac{8}{3} \frac{1}{\eps^2}
		- \frac{40}{9} \frac{1}{\eps} 
		\bigg)
     + \de(s') \bigg( 
  		- \frac{4}{\eps^3}  
		+ \frac{2}{9} \frac{1}{\eps^2}
		+ \frac{1}{\eps} \Big(\frac{121}{27}+\frac{2}{9} \pi^2 \Big)  \bigg) 
   \bigg]
   \bigg\}
  + \ord{\al_s^3}
\,.\nn\end{align}
To obtain the contribution induced by the one-loop renormalization, it is easiest to use \eq{plus_conv}. The remaining $\ord{\eps^0}$ terms give the renormalized jet function
\begin{align} \label{eq:J_q_2}
J_q^\two(s,\mu) &= \frac{\al_s^2C_F}{(4\pi)^2}
\bigg\{
8 C_F \frac{1}{\mu^2} \cL_3\Big(\frac{s}{\mu^2}\Big) + 
 \frac{1}{\mu^2} \cL_2\Big(\frac{s}{\mu^2}\Big) \bigg[- 18 C_F -\frac{22}{3} C_A +\frac{8}{3} T_F n_f\bigg] +
 \frac{1}{\mu^2} \cL_1\Big(\frac{s}{\mu^2}\Big) \bigg[C_F \Big(37 - \frac{20}{3} \pi^2\Big)
 \nn \\ & \quad +
  C_A  \Big(\frac{367}{9} + \frac{4}{3} \pi^2\Big)  
    - \frac{116}{9} T_F n_f \bigg] + 
  \frac{1}{\mu^2} \cL_0\Big(\frac{s}{\mu^2}\Big) \bigg[
  C_F \Big(- \frac{45}{2} + 7 \pi^2 - 8 \zeta_3\Big) +
  C_A \Big(- \frac{3155}{54} + \frac{22}{9} \pi^2 +
       40 \zeta_3\Big) 
 \nn \\ & \quad +       
 T_F n_f \Big(\frac{494}{27}-\frac{8}{9}\pi^2\Big) \bigg] +
  \de(s) \bigg[ C_F \Big(\frac{205}{8}-\frac{67}{6}\pi^2 + \frac{14}{15}\pi^4 - 18 \zeta_3\Big)
  + C_A \Big(\frac{53129}{648}-\frac{208}{27}\pi^2-\frac{17}{180}\pi^4 -\frac{206}{9}\zeta_3\Big)
 \nn \\ & \quad 
  +T_F n_f\Big(-\frac{4057}{162}+\frac{68}{27}\pi^2+\frac{16}{9}\zeta_3 \Big)\bigg]
\bigg\}
\,.\end{align}
This reproduces the well-known result of Ref.~\cite{Becher:2006qw}. 

From $Z_{J_q}$ we can calculate the anomalous dimension,
\begin{align}
  \ga_{J_q}(s,\mu) &= \int\! \df s'\, Z_{J_q}^{-1}(s-s',\mu)\, \mu\frac{\df}{\df\mu}\, Z_{J_q}(s',\mu)
  \nn \\ &
  = \frac{\al_s C_F}{4\pi} \bigg\{\bigg[-8 + \frac{\al_s}{4\pi}  \bigg(C_A\Big(-\frac{536}{9}+\frac{8}{3}\pi^2\Big)+ \frac{160}{9}\, T_F n_f\bigg) \bigg] \frac{1}{\mu^2} \cL_0\Big(\frac{s}{\mu^2}\Big)
  \nn \\ & \qquad
  + \bigg[6+ \frac{\al_s}{4\pi} \bigg(
  C_F\Big(3 - 4\pi^2 + 48 \zeta_3 \Big)
  + C_A \Big(\frac{1769}{27}+\frac{22}{9} \pi^2  -80 \zeta_3\Big)
  +T_F n_f \Big(- \frac{484}{27} -\frac{8}{9} \pi ^2 \Big)
  \bigg)\bigg] \de(s) \bigg\}
\,.\end{align}
Here we used $\mu\, \df \al_s/\df \mu = -2 \al_s [\eps + \al_s \beta_0/(4\pi) + \ord{\al_s^2} ]$,
the derivative of plus distributions in \eq{plus_deriv} and the convolution identities in appendix B of Ref.~\cite{Ligeti:2008ac}. This expression for $\ga_J$ is in agreement with Ref.~\cite{Neubert:2004dd}.

\subsection{Bare Fragmenting Jet Function}
\label{sec:fragjet_nnlo}

In the Fragmenting Jet Function we measure both the invariant mass of the jet and the momentum fraction of one of the partons,
\begin{align}
 \cG_{q,\text{bare}}^k(s) &= 
 \int\! \dPhiColl{2}(s',z')\, \siColl{2}(s',z')\, \de(s-s')
 [\de_{k,q} \de(z-z') + \de_{k,g} \de(z+z'-1)]
 \nn \\ & \quad
 +  \sum_{ij} S_{ijq} \int\! \dPhiColl{3}\, \siColl{3,ijq}\, \de(s-s_{123})
  [\de_{k,i} \de(z-z_1) + \de_{k,j} \de(z-z_2) 
  + \de_{k,q} \de(z-z_3)] + \ord{\al_s^3}\,,
\end{align}
where $ij$ runs over $\{gg,\bar u u, \bar d d, \dots\}$.
At variant with the jet function case, the momentum fraction z is not integrated over.
Below we give the results in a form regular for both $z\to0$ and $z\to1$, which were obtained using the reverse-unitarity approach to phase space integrals. The regularity at $z = 0$ is not necessary since it never enters in \eq{matching}, but it allows an additional check with the corresponding contribution to the bare jet function. 
For the direct integration approach, we performed the calculation for $0<z<1$, which simplifies the calculation compared to the jet function by \eg removing the complication of overlapping singularities in $P_{ggq}^{C_F C_A}$. We then turned the result into plus distributions and fixed the coefficient of $\de(1-z)$ using the quark-number sum rule in \eq{Jsumrules}.

Starting with $P_{\bar q' q' q}$, we first consider the case where the momentum fraction of $q'$ or $\bar q'$ is measured,
\begin{align}
 & \int\! \dPhiColl{3}\, \siColl{3,\bar q'q'q}\, \de(s-s_{123}) \de(z-z_1) = \int\! \dPhiColl{3}\, \siColl{3,\bar q'q'q}\, \de(s-s_{123}) \de(z-z_2)
 \\ & \quad
 = \frac{\al_s^2 C_F T_F}{(4\pi)^2} \frac{\mu^{4\eps}}{s_{123}^{1+2 \eps}}
 \Bigg\{
\frac{1}{\eps^2}\frac{8}{3}\delta \left(z\right)
+\frac{1}{\eps}\left[-\frac{4}{3}\delta \left(z\right)-\frac{16}{3}\cL_0 \left(z\right)-8\left(1+z\right)\text{H}_{0}+\frac{4}{3}\left(4z^2+3z-3\right)\right]
\nonumber \\ & \qquad
-\frac{4}{9}\left(3+\pi^2\right)\delta \left(z\right)+\frac{8}{3}\cL_0 \left(z\right)+\frac{32}{3}\cL_1 \left(z\right)+24\left(1+z\right)\text{H}_{0,0}-16\left(1+z\right)\text{H}_{2}-\frac{8}{3}\frac{1-z}{z}\left(4z^2+7z+4\right)\text{H}_{1}
\nonumber \\ & \qquad
+4\left(3+z\right)\text{H}_{0}-\frac{8}{3}\left(z^2-8z+8-\pi^2\left(1+z\right)\right)
+\eps\left[\frac{2}{9}\delta \left(z\right)\left(-32\zeta_3+\pi^2\right)+\frac{8}{9}\left(3+\pi^2\right)\cL_0 \left(z\right)-\frac{16}{3}\cL_1 \left(z\right)
\nonumber \right. \\ & \qquad  \left.
-\frac{32}{3}\cL_2 \left(z\right)
+16\left(1+z\right)\zeta_3-56\left(1+z\right)\text{H}_{0,0,0}-32\left(1+z\right)\text{H}_{2,1}+16\left(1+z\right)\text{H}_{2,0}-\frac{16}{3}\frac{1-z}{z}\left(4z^2+7z+4\right)\text{H}_{1,1}
\nonumber \right. \\ & \qquad  \left.
+\frac{8}{3}\frac{1-z}{z}\left(4+4z^2+7z\right)\text{H}_{1,0}-\frac{4}{3}\left(8z^2+15z+21\right)\text{H}_{0,0}+48\left(1+z\right)\text{H}_{3}+\frac{8}{3}\frac{1}{z}\left(3z^2+9z+8\right)\text{H}_{2}
\nonumber \right. \\ & \qquad  \left.
+\frac{16}{3}\frac{1\!-\!z}{z}\left(z^2-7z+1\right)\text{H}_{1}-\frac{4}{3}\left(10z-22+\pi^2(1\!+\!z)\right)\text{H}_{0}-\frac{2}{9}\left(12z^2-126z+126+\pi^2(20z^2\!+\!21z\!+\!3)\right)\right]

\!+\!\ord{\eps^2} \Bigg\}
\,.\nn\end{align}
The harmonic polylogarithms $\text{H}$ are defined in \app{hpl}, and we have suppressed their argument $z$ for brevity.
Measuring the momentum fraction of $q$ leads to
\begin{align}
 & \int\! \dPhiColl{3}\, \siColl{3,\bar q'q'q}\, \de(s-s_{123}) \de(z-z_3)
 \\ &\quad 
 = \frac{\al_s^2 C_F T_F}{(4\pi)^2} \frac{\mu^{4\eps}}{s^{1+2 \eps}}
  \Bigg\{
\frac{1}{\eps^2}\frac{8}{3}\delta \left(1-z\right)
+\frac{1}{\eps}\left[\frac{40}{9}\delta \left(1-z\right)-\frac{16}{3}\cL_0 \left(1-z\right)+\frac{8}{3}\left(1+z\right)\right]
-\frac{4}{27}\left(-56+9\pi^2\right)\delta \left(1-z\right)
\nonumber \\ & \qquad
-\frac{80}{9}\cL_0 \left(1-z\right)+\frac{32}{3}\cL_1 \left(1-z\right)+\frac{16}{3}\left(1+z\right)\text{H}_{1}+\frac{8}{3}\frac{1+z^2}{1-z}\text{H}_{0}+\frac{40}{9}\left(1+z\right)
\nonumber \\ & \qquad
+\eps\left[-\frac{4}{81}\delta \left(1-z\right)\left(-328+252\zeta_3+45\pi^2\right)+\frac{8}{27}\left(-56+9\pi^2\right)\cL_0 \left(1-z\right)+\frac{160}{9}\cL_1 \left(1-z\right)-\frac{32}{3}\cL_2 \left(1-z\right)
\nonumber \right. \\ & \qquad \left. 
+\frac{32}{3}\left(1\!+\!z\right)\text{H}_{1,1}
+\frac{8}{9}\frac{1\!+\!z^2}{1\!-\!z} \left(
6 \text{H}_{1,0}-3\text{H}_{0,0}+6\text{H}_{2}+5\text{H}_{0}
\right)
+\frac{80}{9}\left(1\!+\!z\right)\text{H}_{1}
-\frac{4}{27}\left(-74z-38+9\pi^2(1\!+\!z)\right)\right]

\!+\!\ord{\eps^2} \Bigg\}
\,.\nn\end{align}
Moving on to $P_{\bar qqq}^\text{(id)}$, the measurement of the momentum fraction of $\bar q$ yields
\begin{align}
 & \frac12 \int\! \dPhiColl{3}\, \siColl{3,\bar qqq\text{(id)}}\, \de(s-s_{123}) \de(z-z_1)
 \\ &\quad 
 = \frac{\al_s^2 C_F (C_F - \frac{1}{2}C_A)}{(4\pi)^2} \frac{\mu^{4\eps}}{s^{1+2 \eps}}
   \Bigg\{
\frac{1+z^2}{1+z} \left(
8\text{H}_{0,0}-16\text{H}_{-1,0}-\frac{4}{3}\pi^2
\right)
+16\left(1-z\right) + 8 \left(1+z \right) \text{H}_{0} 
\nonumber \\ & \qquad
+\eps\left[
\frac{1+z^2}{1+z} \left(
-40\text{H}_{0,0,0}+48\text{H}_{-1,0,0}-32\text{H}_{-1,2}+32\text{H}_{-2,0}+16\text{H}_{3}+\frac{16}{3}\pi^2\text{H}_{-1}-24\zeta_3
\right)
-32\left(1+z\right)\text{H}_{0,0}
\nonumber \right. \\ & \qquad \left.
+16\left(1+z\right)\text{H}_{-1,0}+16\left(1+z\right)\text{H}_{2}+32\left(1-z\right)\text{H}_{1}+4\left(z-7\right)\text{H}_{0}-\frac{4}{3}\pi^2\left(1+z\right)+4\left(1-z\right)\right]

 +\ord{\eps^2} \Bigg\}
\,.\nn\end{align}
When the momentum fraction of a quark is measured, we find
\begin{align}
 & \frac12 \int\! \dPhiColl{3}\, \siColl{3,\bar qqq\text{(id)}}\, \de(s-s_{123}) [\de(z-z_2) + \de(z-z_3)]
 \\ &\quad 
 = \frac{\al_s^2 C_F (C_F - \frac{1}{2}C_A)}{(4\pi)^2} \frac{\mu^{4\eps}}{s^{1+2 \eps}}
 \Bigg\{
\frac{4}{3}\frac{1+z^2}{1-z}\left(6\text{H}_{1,0}+\pi^2\right)-4\frac{2-5z^2}{1-z}\text{H}_{0}+4\left(8z-7\right)
\nonumber \\ & \qquad
+\eps\left[
\frac{1+z^2}{1-z} \left(
24\zeta_3-24\text{H}_{1,0,0}-16\text{H}_{2,0}+16\text{H}_{1,2}-\frac{8}{3} \pi^2 \text{H}_{1}-\frac{8}{3} \pi^2 \text{H}_{0}
\right)
-8\left(1-z\right)\text{H}_{1,0} +8\left(8z-7\right)\text{H}_{1}
\nonumber \right. \\ & \qquad \left.
+ \frac{2-5z^2}{1-z} \left(
12 \text{H}_{0,0} - 8 \text{H}_{2}
\right)
+4\frac{1}{1-z}\left(10z^2-13z+9\right)\text{H}_{0}
-\frac{4}{3}\pi^2\frac{1}{1-z}\left(6z^2-2z-1\right)+4\left(17z-15\right)
\right]

+\ord{\eps^2} \Bigg\}
\,.\nn\end{align}
Next we consider the $C_F^2$ color structure in $q^* \to ggq$.
Measuring the momentum fraction of one of the gluons,
\begin{align}
 & \frac12 \int\! \dPhiColl{3}\, \siColl{3,ggq,C_F^2}\, \de(s-s_{123}) [\de(z-z_1) + \de(z-z_2)]
 \\ &\quad 
 = \frac{\al_s^2 C_F^2}{(4\pi)^2} \frac{\mu^{4\eps}}{s^{1+2 \eps}}
 \Bigg\{
-\frac{1}{\eps^3}32\delta \left(z\right)
+\frac{1}{\eps^2}\left[-24\delta \left(z\right)+32\cL_0 \left(z\right)+16\left(z-2\right)\right]
+\frac{1}{\eps}\left[8\left(2\pi^2-7\right)\delta \left(z\right)+24\cL_0 \left(z\right)
\nonumber \right. \\ & \qquad \left.
-32\cL_1 \left(z\right)+32\frac{1}{z}\left(z^2-2z+2\right)\text{H}_{1}-12\left(z-2\right)\text{H}_{0}-2\left(16+z\right)\right]
+\frac{4}{3}\delta \left(z\right)\left(112\zeta_3+9\pi^2-84\right)
\nonumber \\ & \qquad
-8\left(2\pi^2-7\right)\cL_0 \left(z\right)
-24\cL_1 \left(z\right)+16\cL_2 \left(z\right)+8\frac{1}{z}\left(z^2-2z+2\right)\left(8\text{H}_{1,1}-3\text{H}_{1,0}\right)+4\left(z-2\right)\text{H}_{0,0}
\nonumber \\ & \qquad
-8\frac{1}{z}\left(3z^2-6z+8\right)\text{H}_{2}-4\frac{1}{z}\left(z^2+16z-12\right)\text{H}_{1}+2\left(16+7z\right)\text{H}_{0}-\frac{28}{3}\pi^2\left(z-2\right)+6\left(z-9\right)
\nonumber \\ & \qquad
+\eps\left[-\frac{4}{3}\delta \left(z\right)\left(168-84\zeta_3+\pi^4-21\pi^2\right)-\frac{4}{3}\cL_0 \left(z\right)\left(112\zeta_3+9\pi^2-84\right)+8\left(2\pi^2-7\right)\cL_1 \left(z\right)
\nonumber \right. \\ & \qquad \left.
+12\cL_2 \left(z\right)
-\frac{16}{3}\cL_3 \left(z\right)-\frac{248}{3}\left(z-2\right)\zeta_3+128\frac{1}{z}\left(z^2-2z+2\right)\text{H}_{1,1,1}-48\frac{1}{z}\left(z^2-2z+2\right)\text{H}_{1,1,0}
\nonumber \right. \\ & \qquad \left.
+8\frac{1}{z}\left(z^2-2z+2\right)\text{H}_{1,0,0}+12\left(z-2\right)\text{H}_{0,0,0}-16\frac{1}{z}\left(3z^2-6z+8\right)\text{H}_{2,1}+8\frac{1}{z}\left(3z^2-6z+8\right)\text{H}_{2,0}
\nonumber \right. \\ & \qquad \left.
-48\frac{1}{z}\left(z^2-2z+2\right)\text{H}_{1,2}-8\frac{1}{z}\left(z^2+16z-12\right)\text{H}_{1,1}-4\frac{1}{z}\left(z^2-16z+12\right)\text{H}_{1,0}-2\left(16+19z\right)\text{H}_{0,0}
\nonumber \right. \\ & \qquad \left.
+8\frac{1}{z}\left(8+z^2-2z\right)\text{H}_{3}+4\frac{1}{z}\left(-12+7z^2+16z\right)\text{H}_{2}+\left(4\frac{1}{z}\left(3z^2-27z+28\right)-\frac{56}{3}\pi^2\frac{1}{z}\left(z^2-2z+2\right)\right)\text{H}_{1}
\nonumber \right. \\ & \qquad \left.
+\left(6\left(5+z\right)+\frac{26}{3}\pi^2\left(z-2\right)\right)\text{H}_{0}+2\left(-52+3z\right)+\pi^2\left(16-3z\right)\right]

+\ord{\eps^2} \Bigg\}
\,.\nn\end{align}
When instead the momentum fraction of the quark is measured, we find
\begin{align}
 &\frac12 \int\! \dPhiColl{3}\, \siColl{3,ggq,C_F^2}\, \de(s-s_{123}) \de(z-z_3)
 \\ &\quad 
 = \frac{\al_s^2 C_F^2}{(4\pi)^2} \frac{\mu^{4\eps}}{s^{1+2 \eps}}
 \Bigg\{
-\frac{1}{\eps^3}16\delta \left(1-z\right)
+\frac{1}{\eps^2}\left[32\cL_0 \left(1-z\right)-16\left(1+z\right)\right]
+\frac{1}{\eps}\Big[
8\pi^2\delta \left(1-z\right)-64\cL_1 \left(1-z\right)-32\left(1+z\right)\text{H}_{1}
\nonumber \\ & \qquad
-4\frac{5+3z^2}{1-z}\text{H}_{0}-8\left(1-z\right)
\Big]
+\frac{224}{3}\delta \left(1-z\right)\zeta_3-16\pi^2\cL_0 \left(1-z\right)+64\cL_2 \left(1-z\right)-64\left(1+z\right)\text{H}_{1,1}-32\frac{1+z^2}{1-z}\text{H}_{1,0}
\nonumber \\ & \qquad
+4\frac{5-z^2}{1-z}\text{H}_{0,0}-8\frac{5+3z^2}{1-z}\text{H}_{2}-16\left(1-z\right)\text{H}_{1}+8\left(1-z\right)\text{H}_{0}+\frac{28}{3}\pi^2\left(1+z\right)+4\left(1-z\right)
+\eps\left[
-\frac{2}{3}\pi^4\delta \left(1-z\right)
\nonumber \right. \\ & \qquad \left.
-\frac{448}{3} \zeta_3 \cL_0 \left(1-z\right)+32\pi^2\cL_1 \left(1-z\right)-\frac{128}{3}\cL_3 \left(1-z\right)+\frac{8}{3}\frac{37-25z^2}{1-z}\zeta_3-128\left(1+z\right)\text{H}_{1,1,1}-4\frac{1-13z^2}{1-z}\text{H}_{0,0,0}
\nonumber \right. \\ & \qquad \left.
+32  \frac{1+z^2}{1-z}\left(
\text{H}_{1,0,0}-2\text{H}_{1,2}-2\text{H}_{1,1,0}
\right)
+\frac{5+3z^2}{1-z}\left(
-16\text{H}_{2,1}+8\text{H}_{2,0}
\right)
+ \left(1-z \right) \left(
-32\text{H}_{1,1}+16\text{H}_{1,0}+16\text{H}_{2}
\right)
\nonumber \right. \\ & \qquad \left.
+8\frac{5\!-\!z^2}{1\!-\!z}\text{H}_{3}
+\left(16\pi^2\left(1\!+\!z\right)+8\left(1\!-\!z\right)\right)\text{H}_{1}+\left(\frac{2}{3}\pi^2\frac{15+17z^2}{1-z}+4\left(1\!+\!3z\right)\right)\text{H}_{0}
+4\pi^2\left(1\!-\!z\right)+28\left(1\!-\!z\right)
\right]

+\ord{\eps^2} \Bigg\}
\,.\nn\end{align}
For the $C_F C_A$ color structure of $q^* \to ggq$, when the momentum fraction of one of the gluons is measured, we find
\begin{align}
 & \frac12 \int\! \dPhiColl{3}\, \siColl{3,ggq,C_F C_A}\, \de(s-s_{123}) [\de(z-z_1) + \de(z-z_2)]
 \\ &\quad 
 = \frac{\al_s^2 C_F C_A}{(4\pi)^2} \frac{\mu^{4\eps}}{s^{1+2 \eps}} 
 \Bigg\{
-\frac{1}{\eps^3}8\delta \left(z\right)
+\frac{1}{\eps^2}\left[-\frac{62}{3}\delta \left(z\right)+8\cL_0 \left(z\right)+4\left(z-2\right)\right]
+\frac{1}{\eps}\left[\frac{2}{3}\left(-3+4\pi^2\right)\delta \left(z\right)+\frac{124}{3}\cL_0 \left(z\right)
\nonumber \right. \\ & \qquad \left.
+8\frac{1}{z}\left(z^2-2z+2\right)\text{H}_{1}+8\left(4+z\right)\text{H}_{0}-\frac{8}{3}\left(2z^2+3z+12\right)\right]
+\frac{1}{9}\delta \left(z\right)\left(48\zeta_3+31\pi^2\right)+4\left(1-\pi^2\right)\cL_0 \left(z\right)
\nonumber  \\ & \qquad 
-\frac{248}{3}\cL_1 \left(z\right)-16\cL_2 \left(z\right)+ 8 \frac{z^2-2z+2}{z} \left(
2 \text{H}_{1,1} - \text{H}_{1,0}
\right)
 + 8 \frac{z^2+2z+2}{z}\text{H}_{-1,0}
-40\left(2+z\right)\text{H}_{0,0}+16\left(4+z\right)\text{H}_{2}
\nonumber  \\ & \qquad 
+\frac{8}{3}\frac{1}{z}\left(31-4z^3-6z^2-24z\right)\text{H}_{1}-4\left(4+3z\right)\text{H}_{0}+\frac{4}{3}\left(4+6z^2-13z\right)-\frac{2}{3}\pi^2\left(6+7z\right)
\nonumber  \\ & \qquad 
+\eps\left[
\frac{1}{45}\delta \left(z\right)\left(-60+2480\zeta_3-9\pi^4+15\pi^2\right)-\frac{2}{9}\cL_0 \left(z\right)\left(-48\zeta_3+31\pi^2\right)+\frac{8}{3}\left(-3+2\pi^2\right)\cL_1 \left(z\right)+\frac{248}{3}\cL_2 \left(z\right)
\nonumber \right. \\ & \qquad \left.
+\frac{64}{3}\cL_3 \left(z\right)-\frac{16}{3}\left(5+2z\right)\zeta_3
+ \frac{z^2-2z+2}{z} \left(
32\text{H}_{1,1,1}-24 \text{H}_{1,1,0}+8\text{H}_{1,0,0}-16\text{H}_{1,2}
\right)
\nonumber \right. \\ & \qquad \left.
+ \frac{z^2+2z+2}{z} \left(
-24 \text{H}_{-1,0,0}+16\text{H}_{-1,2}-16\text{H}_{-2,0}-\frac{8}{3} \pi^2 \text{H}_{-1}
\right)
+8\left(22+15z\right)\text{H}_{0,0,0}+32\left(4+z\right)\text{H}_{2,1}
\nonumber \right. \\ & \qquad \left.
+ \frac{8}{3} \frac{1}{z}\left(4z^3+6z^2+24z-31\right) \left(
 \text{H}_{1,0}-2\text{H}_{1,1}
\right)
+\frac{4}{3}\left(8z^2+45z+84\right)\text{H}_{0,0}-8z\text{H}_{-1,0}-16\frac{1}{z}\left(5z^2+10z+4\right)\text{H}_{3}
\nonumber \right. \\ & \qquad \left.
-\frac{8}{3}\frac{1}{z}\left(9z^2+12z+62\right)\text{H}_{2}
-8\frac{1}{z}\left(3z^2+6z+2\right)\text{H}_{2,0}+\left(\frac{4}{3}\left(13z-16\right)+4\pi^2z\right)\text{H}_{0}
\nonumber \right. \\ & \qquad \left.
-\left(4\pi^2\frac{1}{z}\left(z^2-2z+2\right)+\frac{8}{3}\frac{\left(1-z\right)\left(2z-3\right)\left(1+3z\right)}{z}\right)\text{H}_{1}
+\frac{4}{9}\pi^2\left(10z^2+21z+72\right)+\frac{8}{3}\left(1-z\right)
\right]

+\ord{\eps^2} \Bigg\}
\,.\nn\end{align}
For the momentum fraction of the quark we find
\begin{align}
 &\frac12 \int\! \dPhiColl{3}\, \siColl{3,ggq,C_F C_A}\, \de(s-s_{123}) \de(z-z_3)
  \\ &\quad 
 = \frac{\al_s^2 C_F C_A}{(4\pi)^2} \frac{\mu^{4\eps}}{s^{1+2 \eps}} 
 \Bigg\{
-\frac{1}{\eps^3}4\delta \left(1-z\right)
+\frac{1}{\eps^2}\left[-\frac{22}{3}\delta \left(1-z\right)+8\cL_0 \left(1-z\right)-4\left(1+z\right)\right]
\nonumber \\ & \qquad
+\frac{1}{\eps}\left[\frac{2}{9}\left(12\pi^2-67\right)\delta \left(1-z\right)+\frac{44}{3}\cL_0 \left(1-z\right)-16\cL_1 \left(1-z\right)-8\left(1+z\right)\text{H}_{1}-\frac{2}{3}\left(17+5z\right)\right]
\nonumber \\ & \qquad
+\frac{1}{27}\delta \left(1-z\right)\left(1044\zeta_3+99\pi^2-808\right)-\frac{88}{3}\cL_1 \left(1-z\right)-\frac{4}{9}\left(12\pi^2-67\right)\cL_0 \left(1-z\right)+16\cL_2 \left(1-z\right)
\nonumber \\ & \qquad
-16\left(1+z\right)\text{H}_{1,1}
- 4 \frac{1+z^2}{1-z} \left( 
\text{H}_{1,0}+\text{H}_{0,0}
\right)
-\frac{4}{3}\left(17+5z\right)\text{H}_{1}-\frac{2}{3}\frac{11+2z^2}{1-z}\text{H}_{0}-\frac{2}{3}\pi^2\frac{5z^2-3}{1-z}-\frac{2}{9}\left(94+49z\right)
\nonumber \\ & \qquad
+\eps\left[
\frac{1}{810}\delta \left(1-z\right)\left(27720\zeta_3+153\pi^4+6030\pi^2-48560\right)-\frac{2}{27}\cL_0 \left(1-z\right)\left(1044\zeta_3+99\pi^2-808\right)
\nonumber \right. \\ & \qquad \left.
+\frac{8}{9}\left(12\pi^2-67\right)\cL_1 \left(1-z\right)+\frac{88}{3}\cL_2 \left(1-z\right)-\frac{32}{3}\cL_3 \left(1-z\right)-\frac{8}{3}\frac{13z^2-16}{1-z}\zeta_3-32\left(1+z\right)\text{H}_{1,1,1}
\nonumber \right. \\ & \qquad \left.
+ \frac{1+z^2}{1-z} \left(
-8\text{H}_{1,1,0}
+4\text{H}_{0,0,0}
-4\text{H}_{1,0,0}
-8\text{H}_{3}
-8\text{H}_{1,2}
\right)
-\frac{8}{3}\left(17+5z\right)\text{H}_{1,1}-\frac{8}{3}\frac{1}{1-z}\left(4+4z^2+3z\right)\text{H}_{1,0}
\nonumber \right. \\ & \qquad \left.
-\frac{2}{3}\frac{1}{1-z}\left(10z^2+12z-17\right)\text{H}_{0,0}-\frac{4}{3}\frac{2z^2+11}{1-z}\text{H}_{2}+\left(-\frac{4}{9}\left(94+49z\right)+\frac{16}{3}\pi^2\left(1+z\right)\right)\text{H}_{1}
\nonumber \right. \\ & \qquad \left.
-\left(\frac{2}{9}\frac{1}{1-z}\left(4z^2+9z+49\right)-\frac{4}{3}\pi^2\frac{1+z^2}{1-z}\right)\text{H}_{0}-\frac{1}{3}\pi^2\frac{1}{1-z}\left(7z^2+20z-21\right)-\frac{2}{27}\left(629+233z\right)
\right]

+\ord{\eps^2} \Bigg\}
\,.\nn\end{align}
For the real-virtual corrections, we do not need to perform any integrals, but simply expand in $\eps$. We first consider the case where the momentum fraction $z$ of the quark is measured,
\begin{align}
& \int\! \dPhiColl{2}(s',z')\, \siColl{2}{}^\one(s',z')\, \de(s-s')
 \de(z-z') 
\nn \\ & \quad
 = \frac{\al_s^2C_F}{(4\pi)^2} \frac{\mu^{4\eps}}{s^{1+2 \eps}}\bigg\{ 
 C_F \bigg[
 \frac{z^2\!+\!1}{1\!-\!z} \bigg(\frac{8}{\eps}\ln z \!+\! 8 \text{Li}_2(z) \!-\!12 \ln^2 z \!-\! \frac{4}{3} \pi ^2\bigg)
\!-\!8(1\!-\!z) \ln z \!-\!4 \bigg] 
\!+\! C_A \bigg[\frac{4}{\eps^3}\delta (1\!-\!z) 
\!-\!\frac{4}{\eps^2} (1\!+\!z^2) \cL_0(1\!-\!z) 
\nn \\ & \qquad
\!+\! \frac{1}{\eps} \Big(8(1\!+\!z^2) \cL_1(1\!-\!z) \!-\! 2 \pi ^2 \delta (1\!-\!z)\!+\!4(1\!-\!z) \Big)
\!-\!8(1\!+\!z^2) \cL_2(1\!-\!z)
\!+\! 2\pi^2 (1\!+\!z^2) \cL_0(1\!-\!z)
\!-\!\frac{32}{3} \zeta_3 \delta (1\!-\!z)
\nn \\ & \qquad
\!+\! \frac{z^2\!+\!1}{1\!-\!z} \left( \!-\!8 \text{Li}_2(z)\!-\!8  \ln (1\!-\!z) \ln z \!+\!4 \ln ^2 z \!+\!\frac{4}{3} \pi ^2 \right)
\!-\!8 (1\!-\!z) \ln(1\!-\!z)\!+\!4
\bigg]
+ \ord{\eps} \bigg\}
\,.\end{align}
The expression for when the momentum fraction of the gluon is measured instead can be obtained by $z\to 1-z$.

\subsection{Renormalization and Matching of Fragmenting Jet Function}
\label{sec:fragjet_nnlo_ren}

As stated in \eq{renormalization}, the renormalization of the fragmenting jet function does not depend on the momentum fraction $z$ and is identical to that of the jet function in \eqref{eq:Jrenormalization}. The $1/\eps$ poles that remain after renormalization are IR divergences, which cancel in the matching in \eq{matching},
\begin{align} \label{eq:matching_G_2}
  \cG_q^{i\two}(s,z,\mu) &= \cJ_{qi}^\two(s,z,\mu) +   
    \sum_j \int_z^1\! \frac{\df z'}{z'}\, \cJ_{qj}^\one\Big(s,\frac{z}{z'},\mu \Big) D_j^{i\one}(z',\mu) +
    \de(s) D_q^{i\two}(z,\mu)
\,.\end{align}
We have worked out the second and third term on the right-hand side of \eq{matching_G_2} in \app{IR} and verified that their poles agree with those of the renormalized fragmenting jet function. The finite terms of $\cG_q^{i\two}$ minus the finite contribution from the convolution of $\cJ_{qj}^\one$ with $ D_j^{i,\one}$ give the two-loop matching coefficients $\cJ_{qi}^\two$, which are given below.

Starting with $\cJ_{qq}^\two$, we separate its contributions by color structure,
\begin{equation}
 \cJ_{qq}^\two (s, z, \mu)= \frac{\alpha_s^2 C_F}{(4 \pi)^2}\, \frac{1}{\mu^2} \Big( 
 C_F\, \cGbar_{qq, C_F}^\two + C_A\, \cGbar_{qq, C_A}^\two + T_F\, \cGbar_{qq'}^\two + n_f T_F\, \cGbar_{qq, T_F}^\two
\Big)
\,,\end{equation}
where for later convenience we keep the secondary-quark contribution $\cGbar_{qq'}^\two$ separate. The ingredients are given by
\begin{align}
\cGbar_{qq, C_F}^\two &=
\delta \left(\frac{s}{\mu^2}\right)\left[
\frac{7 \pi^4}{30} \delta \left(1-z\right)
+ 32 \, \zeta_3 \, \cL_0 \left(1-z\right)
-  \frac{20 \pi^2}{3}  \cL_1 \left(1-z\right)
+8 \cL_3 \left(1-z\right)
+24\left(1+z\right)\text{H}_{1,1,1}
\nonumber \right. \\ & \quad \left.
+16\frac{1+z^2}{1-z}\text{H}_{1,1,0}
+4\frac{1+z^2}{1-z}\text{H}_{1,0,0}
-10 \frac{3+5 z^2}{1-z}\text{H}_{0,0,0}
+8\frac{2+z^2}{1-z}\text{H}_{2,1}
-12 \frac{1+z^2}{1-z}\text{H}_{2,0}
+8\frac{1+z^2}{1-z}\text{H}_{1,2}
\nonumber \right. \\ & \quad \left.
-12  (1-z)\text{H}_{1,0}
-\frac{2}{1-z}\left(12-14z-7z^2\right)\text{H}_{0,0}
+4\frac{1+5 z^2}{1-z}\text{H}_{3}
+12\frac{1}{1-z}\left(1-z-z^2\right)\text{H}_{2}
\nonumber \right. \\ & \quad \left.
+ \frac{2}{3 (1-z) } \left( \pi^2 \left( -3 + 7 z^2 \right) + 33 - 69 z + 36 z^2  \right)\text{H}_{1}
- \frac{2}{3 (1-z) } \left( \pi^2 \left( 3 + 5 z^2 \right) + 21 - 42 z + 39 z^2  \right)\text{H}_{0}
\nonumber \right. \\ & \quad \left.
-\frac{2}{3}\pi^2\frac{5-7z-z^2}{1-z}
- 4 (16+10 z^2) \zeta_3 \frac{1}{1-z}
+34(1-z)
\right]
\nonumber \\ & \quad 
+\cL_0 \left(\frac{s}{\mu^2}\right)\left[
32 \, \zeta_3 \, \delta \left(1-z\right)
+24 \, \cL_2 \left(1-z\right)
-\frac{20}{3}\pi^2\cL_0 \left(1-z\right)
-\frac{4(3+z^2)}{1-z}\text{H}_{2}
-24\left(1+z\right)\text{H}_{1,1}
\nonumber \right. \\ & \quad \left.
+4 \pi^2\left(1+z\right)
-24\left(1-z\right)
-8\frac{1+z^2}{1-z}\text{H}_{1,0}
-16\frac{1+2 z^2}{1-z}\text{H}_{0,0}
-\frac{4}{1-z}\left(4-5z-2z^2\right)\text{H}_{0}
\right]
\nonumber \\ & \quad
+\cL_1 \left(\frac{s}{\mu^2}\right)\left[
48\cL_1 \left(1-z\right)-\frac{20}{3}\pi^2\delta \left(1-z\right)+24\left(1+z\right)\text{H}_{1}
+4(1+z)\text{H}_{0}
\right]
\nonumber \\ & \quad
+\cL_2 \left( \frac{s}{\mu^2}\right)\left[24\cL_0 \left(1-z\right)-12\left(1+z\right) \right]
+8 \, \cL_3 \left(\frac{s}{\mu^2}\right)\delta \left(1-z\right)
,\end{align}
\begin{align}
\cGbar_{qq, C_A}^\two &=
\delta \left(\frac{s}{\mu^2}\right)\left[
\frac{2}{27}\left(378\zeta_3 +33\pi^2-404\right)\cL_0 \left(1-z\right)+\frac{1}{162}\left(-18\pi^4-1980\zeta_3 -603\pi^2+4856\right)\delta \left(1-z\right)
\nonumber \right. \\ & \quad \left.
-\frac{22}{3}\cL_2 \left(1-z\right)-\frac{4}{9}\left(3\pi^2-67\right)\cL_1 \left(1-z\right)+\frac{22}{3}\left(1+z\right)\text{H}_{1,1}-4\left(1+z\right)\text{H}_{2}+\frac{2}{27}\left(85+319z\right)-\frac{2}{9}\pi^2\left(4+z\right)
\nonumber \right. \\ & \quad \left.
+2 \, \zeta_3 \frac{-5+9z^2}{1-z}-4\frac{1\!+\!z^2}{1\!-\!z}\text{H}_{1,1,0}+10\frac{1\!+\!z^2}{1\!-\!z}\text{H}_{0,0,0}-4\frac{1\!+\!z^2}{1\!-\!z}\text{H}_{2,0}+4\frac{1\!+\!z^2}{1\!-\!z}\text{H}_{1,2}-\frac{1}{3}\frac{1}{1\!-\!z}\left(-35+12z+z^2\right)\text{H}_{0,0}
\nonumber \right. \\ & \quad \left.
+\left(\frac{22}{9}\left(-1+14z\right)-\frac{2}{3}\pi^2\frac{3+z^2}{1-z}\right)\text{H}_{1}+\left(\frac{2}{9}\frac{1}{1-z}\left(107-129z+80z^2\right)-\frac{2}{3}\pi^2\frac{1+z^2}{1-z}\right)\text{H}_{0}
\right]
\nonumber \\ & \quad
+\cL_0 \left(\frac{s}{\mu^2}\right)\left[
\frac{2}{27}\left(378\zeta_3 +33\pi^2-404\right)\delta \left(1-z\right)-\frac{44}{3}\cL_1 \left(1-z\right)-\frac{4}{9}\left(3\pi^2-67\right)\cL_0 \left(1-z\right)-\frac{22}{3}\left(1+z\right)\text{H}_{1}
\nonumber \right. \\ & \quad \left.
+4\left(1+z\right)\text{H}_{0}-\frac{4}{9}\left(-10+77z\right)+\frac{2}{3}\pi^2\left(1+z\right)+4\frac{1+z^2}{1-z}\text{H}_{0,0}
\right]
\nonumber \\ & \quad
+\cL_1 \left(\frac{s}{\mu^2}\right)\left[
-\frac{44}{3}\cL_0 \left(1-z\right)-\frac{4}{9}\left(3\pi^2-67\right)\delta \left(1-z\right)+\frac{22}{3}\left(1+z\right)
\right]
-\frac{22}{3}\cL_2 \left(\frac{s}{\mu^2}\right)\delta \left(1-z\right)
,\end{align}
\begin{align}
\cGbar_{qq, T_F}^\two &=
\delta \left(\frac{s}{\mu^2}\right)\left[
\frac{2}{81}\left(180\zeta_3 +45\pi^2-328\right)\delta \left(1-z\right)+\frac{8}{3}\cL_2 \left(1-z\right)-\frac{80}{9}\cL_1 \left(1-z\right)-\frac{8}{27}\left(3\pi^2-28\right)\cL_0 \left(1-z\right)
\nonumber \right. \\ & \quad \left.
-\frac{8}{3}\left(1\!+\!z\right)\text{H}_{1,1}-\frac{16}{9}\left(1+4z\right)\text{H}_{1}-\frac{4}{27}\left(19+37z\right)+\frac{4}{9}\pi^2\left(1\!+\!z\right)-\frac{4}{3}\frac{1\!+\!z^2}{1\!-\!z}\text{H}_{0,0}-\frac{4}{9}\frac{1}{1\!-\!z}\left(11-12z+11z^2\right)\text{H}_{0}
\right]
\nonumber \\ & \quad 
+\cL_0 \left(\frac{s}{\mu^2}\right)\left[
\frac{16}{3}\cL_1 \left(1-z\right)-\frac{80}{9}\cL_0 \left(1-z\right)-\frac{8}{27}\left(3\pi^2-28\right)\delta \left(1-z\right)+\frac{8}{3}\left(1+z\right)\text{H}_{1}+\frac{16}{9}\left(1+4z\right)
\right]
\nonumber \\ & \quad 
+\cL_1 \left(\frac{s}{\mu^2}\right)\left[\frac{16}{3}\cL_0 \left(1-z\right)-\frac{80}{9}\delta \left(1-z\right)-\frac{8}{3}\left(1+z\right)\right]
+\frac{8}{3}\cL_2 \left(\frac{s}{\mu^2}\right)\delta \left(1-z\right)
,\end{align}
\begin{align} \label{eq:Gqqprime}
\cGbar_{qq'}^\two &=
\delta \left(\frac{s}{\mu^2}\right) \left[
\frac{-2 \pi^2 \left(2+3 z\right)}{3}
-\frac{2 \left(1-z\right) \left(107+239 z+287 z^2\right)}{27 z}
+\left(\frac{4 \pi^2 \left(1+z\right)}{3}
+\frac{-4 \left(7+90 z+81 z^2+31 z^3\right)}{9 z}\right) \text{H}_{0}
\nonumber \right. \\ & \quad \left.
+\frac{4 \left(1-z\right) \left(7+67 z+25 z^2\right) }{9 z} \text{H}_{1}
-\frac{4 \left(4-3 z-12 z^2-4 z^3\right) }{3 z} \text{H}_{2}
-16 \left(1+z\right) \text{H}_{3}
+\frac{2 \left(8-3 z-15 z^2\right) }{3 z} \text{H}_{0,0}
\nonumber \right. \\ & \quad \left.
+\frac{4 \left(1-z\right) \left(4+7 z+4 z^2\right)}{3 z}  \text{H}_{1,1}
+8 \left(1+z\right) \text{H}_{2,1}
+20 \left(1+z\right) \text{H}_{0,0,0}
+8 \left(1+z\right) \zeta_3 
\right]
\nonumber  \\ & \quad
+\cL_0 \left(\frac{s}{\mu^2}\right) \left[
\frac{4 \pi^2 \left(1+z\right)}{3}
-\frac{4 \left(1-z\right) \left(7+67 z+25 z^2\right)}{9 z}
+\frac{4 \left(4-3 z-12 z^2-4 z^3\right) \text{H}_{0}}{3 z}
-8 \left(1+z\right) \text{H}_{2}
\nonumber \right. \\ & \quad \left.
-\frac{4 \left(1-z\right) \left(4+7 z+4 z^2\right) \text{H}_{1}}{3 z}
+16 \left(1+z\right) \text{H}_{0,0}
\right]
+\cL_1 \left(\frac{s}{\mu^2}\right) \left[\frac{4 \left(1-z\right) \left(4+7 z+4 z^2\right)}{3 z}+8 \left(1+z\right) \text{H}_{0}\right]
.\end{align}

The matching coefficient $\cJ_{q\bar q}$ that describes the contribution from the anti-quark fragmentation function (of the same flavor as the quark) has the following form
\begin{equation}
 \cJ_{q\bar{q}}^\two (s, z, \mu)= \frac{\alpha_s^2 C_F}{(4 \pi)^2}\, \frac{1}{\mu^2} \left[ 
 \left(C_F-C_A/2 \right)  \cGbar_{q\bar{q}, (id)}^\two + T_F \cGbar_{qq'}^\two
 \right]
.\end{equation}
The secondary quark contribution $\cGbar_{qq'}^\two$ was given above in \eq{Gqqprime}. The interference contribution is given by
\begin{align}
\cGbar_{q\bar{q}, (id)}^\two &=
\delta \left(\frac{s}{\mu^2}\right)\left[
16\left(1+z\right)\text{H}_{0,0}-8\left(1+z\right)\text{H}_{-1,0}-8\left(1+z\right)\text{H}_{2}-16\left(1-z\right)\text{H}_{1}-2\left(-7+z\right)\text{H}_{0}+\frac{2}{3}\pi^2\left(1+z\right)
\nonumber \right. \\ & \quad \left.
+ \frac{1+z^2}{1+z} \left(
12\zeta_3 
+20\text{H}_{0,0,0}
-24\text{H}_{-1,0,0}
+16\text{H}_{-1,2}
-16\text{H}_{-2,0}
-8\text{H}_{3}
-\frac{8}{3} \pi^2 \text{H}_{-1}
\right)
-2\left(1-z\right)
\right]
\nonumber \\ & \quad
+\cL_0 \left(\frac{s}{\mu^2}\right)\left[
8\left(1+z\right)\text{H}_{0}+16\left(1-z\right)+8\frac{1+z^2}{1+z}\text{H}_{0,0}-16\frac{1+z^2}{1+z}\text{H}_{-1,0}-\frac{4}{3}\pi^2\frac{1+z^2}{1+z}
\right]
.\end{align}

The contribution involving fragmentation from an (anti)quark of a different flavor $q' \neq q$ is given by
\begin{equation}
\cJ_{qq'}^\two (s, z, \mu) =  \cJ_{q \bar{q}'}^\two (s, z, \mu)= \frac{\alpha_s^2 C_F T_F}{(4 \pi)^2}\, \frac{1}{\mu^2}\, \cGbar_{qq'}^\two
\,.\end{equation}

For the matching $\cJ_{qg}$ onto gluon fragmentation functions, we have
\begin{equation}
 \cJ_{qg}^\two (s, z, \mu)= \frac{\alpha_s^2 C_F}{(4 \pi)^2}\, \frac{1}{\mu^2} \Big( 
 C_F\, \cGbar_{qg, C_F}^\two + C_A\, \cGbar_{qg, C_A}^\two 
 \Big)
\,,\end{equation}
with
\begin{align}
 \cGbar_{qg, C_F}^\two &=
\delta \left(\frac{s}{\mu^2}\right) \left[
-24+19 z
+\frac{\pi^2 \left(6-16 z+z^2\right)}{3 z}
+\left(\frac{-4 \pi^2 \left(3-4 z+2 z^2\right)}{3 z}+\frac{28+3 z-13 z^2}{z}\right) \text{H}_{0}
\nonumber \right. \\ & \quad \left.
+\left(\frac{4 \pi^2 \left(2-2 z+z^2\right)}{z}-\frac{4 \left(7-7 z+3 z^2\right)}{z}\right) \text{H}_{1}
+\frac{4 \left(3-2 z^2\right)}{z} \text{H}_{2}
-\left(16-13 z\right) \text{H}_{0,0}
-\frac{2 \left(12-16 z+z^2\right)}{z} \text{H}_{1,1}
\nonumber \right. \\ & \quad \left.
-\frac{16}{z} \text{H}_{3}
+\frac{4 \left(2-2 z+z^2\right)}{z} \text{H}_{1,2}
-\frac{8 \left(2-2 z+z^2\right)}{z} \text{H}_{2,0}
+\frac{4 \left(12-10 z+5 z^2\right)}{z} \text{H}_{2,1}
+10 \left(2-z\right) \text{H}_{0,0,0}
\nonumber \right. \\ & \quad \left.
+\frac{16 \left(2-2 z+z^2\right) \text{H}_{1,0,0}}{z}
-\frac{4 \left(2-2 z+z^2\right) \text{H}_{1,1,0}}{z}
-\frac{20 \left(2-2 z+z^2\right) \text{H}_{1,1,1}}{z}
+\frac{4 \left(4-2 z+z^2\right) \zeta_3 }{z}
\right]
\nonumber \\ & \quad
 +\cL_0 \left(\frac{s}{\mu^2}\right) \left[
 \frac{-4 \pi^2 \left(5-6 z+3 z^2\right)}{3 z}
 +\frac{2 \left(14-15 z+6 z^2\right)}{z}
 -\frac{12}{z} \text{H}_{0}
 +\frac{2 \left(12-16 z+3 z^2\right)}{z} \text{H}_{1}
  +8 \left(2-z\right) \text{H}_{0,0}
 \nonumber \right. \\ & \quad \left.
 -\frac{4 \left(8-6 z+3 z^2\right)}{z} \text{H}_{2}
 +\frac{8 \left(2-2 z+z^2\right)}{z} \text{H}_{1,0}
 +\frac{16 \left(2-2 z+z^2\right)}{z} \text{H}_{1,1}
 \right]
+ \cL_1 \left(\frac{s}{\mu^2}\right)
\left[
\frac{-2 \left(12-16 z+3 z^2\right)}{z}
 \nonumber \right. \\ & \quad \left.
+\frac{4 \left(4-2 z+z^2\right)}{z} \text{H}_{0}
-\frac{16 \left(2-2 z+z^2\right)}{z} \text{H}_{1}
\right]
+\cL_2 \left(\frac{s}{\mu^2}\right) \frac{12 \left(2-2 z+z^2\right) }{z}
,\end{align}
\begin{align}
 \cGbar_{qg, C_A}^\two &=
\delta \left(\frac{s}{\mu^2}\right) \left[
\frac{2 \pi^2 \left(-3-6 z+z^2\right)}{3 z}
+\frac{2 \left(1169-723 z-204 z^2-269 z^3\right)}{27 z}
+\frac{4 \pi^2 \left(2+2 z+z^2\right)}{3 z} \text{H}_{-1}
\nonumber \right. \\ & \quad \left.
+\left(\frac{-8 \pi^2 \left(1+z+z^2\right)}{3 z}
+\frac{2 \left(152+405 z+171 z^2+62 z^3\right)}{9 z}\right) \text{H}_{0}
-\frac{2 \left(26+57 z-48 z^2-26 z^3\right)}{9 z} \text{H}_{1}
\nonumber \right. \\ & \quad \left.
+\frac{4 \left(40-6 z-3 z^2-4 z^3\right)}{3 z} \text{H}_{2}
+4 \, z \, \text{H}_{-1,0}
-\frac{2 \left(80+48 z+15 z^2\right)}{3 z} \text{H}_{0,0}
-12 \, z \, \text{H}_{1,0}
-\frac{4 \left(31-24 z-4 z^3\right)}{3 z} \text{H}_{1,1}
\nonumber \right. \\ & \quad \left.
-\frac{8 \left(2+2 z+z^2\right)}{z} \text{H}_{-1,2}
+\frac{8 \left(2+2 z+z^2\right)}{z} \text{H}_{-2,0}
+\frac{8 \left(6+4 z+5 z^2\right)}{z} \text{H}_{3}
+\frac{8 \left(2-2 z+z^2\right)}{z} \text{H}_{1,2}
\nonumber \right. \\ & \quad \left.
-\frac{8 \left(2-2 z+z^2\right)}{z} \text{H}_{2,0}
-\frac{4 \left(6+2 z+5 z^2\right)}{z} \text{H}_{2,1}
+\frac{12 \left(2+2 z+z^2\right)}{z} \text{H}_{-1,0,0}
-\frac{20 \left(4+2 z+3 z^2\right)}{z} \text{H}_{0,0,0}
\nonumber \right. \\ & \quad \left.
-\frac{12 \left(2-2 z+z^2\right)}{z} \text{H}_{1,0,0}
+\frac{16 \left(2-2 z+z^2\right)}{z} \text{H}_{1,1,0}
-\frac{4 \left(2-2 z+z^2\right)}{z} \text{H}_{1,1,1}
-\frac{8 \left(8-z+5 z^2\right) \zeta_3 }{z}
\right]
\nonumber  \\ & \quad
+\cL_0 \left(\frac{s}{\mu^2}\right) \left[
\frac{-4 \pi^2 \left(2+z\right)}{3}
+\frac{4 \left(13+33 z-24 z^2-13 z^3\right)}{9 z}
-\frac{8 \left(20-3 z-3 z^2-2 z^3\right)}{3 z} \text{H}_{0}
\nonumber \right. \\ & \quad \left.
+\frac{4 \left(1-z\right) \left(31+7 z+4 z^2\right)}{3 z} \text{H}_{1}
+\frac{16 \left(1+z+z^2\right)}{z} \text{H}_{2}
+\frac{8 \left(2+2 z+z^2\right)}{z} \text{H}_{-1,0}
-\frac{8 \left(6+4 z+5 z^2\right)}{z} \text{H}_{0,0}
\nonumber \right. \\ & \quad \left.
-\frac{16 \left(2-2 z+z^2\right)}{z} \text{H}_{1,0}
+\frac{8 \left(2-2 z+z^2\right)}{z} \text{H}_{1,1}
\right] 
\nonumber \\ & \quad
+\cL_1 \left(\frac{s}{\mu^2}\right) \left[
\frac{-4 \left(1-z\right) \left(31+7 z+4 z^2\right)}{3 z}
-\frac{16 \left(1+z+z^2\right)}{z} \text{H}_{0}
-\frac{8 \left(2-2 z+z^2\right)}{z} \text{H}_{1}
\right]
.\end{align}

We have verified that these results satisfy the quark number and momentum sum rules of \eq{Jsumrules}, providing an important cross check. In terms of the ingredients above, these sum rules read: 
\begin{align}
\int\! \df z \Big[ \cJ_{qq}^\two-  \cJ_{q\bar{q}}^\two \Big]  (s, z, \mu) &= J_q^\two(s,\mu) 
\,,\nn \\
\int\! \df z \, z \left[ \cJ_{qq}^\two + \cJ_{q\bar{q}}^\two
	+ (n_f-1) \Big(\cJ_{qq'}^\two + \cJ_{q \bar{q}'}^\two \Big) + \cJ_{qg}^\two\right](s, z, \mu) &= J_q^\two(s,\mu) 
\,.\end{align}

\end{widetext}

\section{Discussion and Conclusion}
\label{sec:discussion}

In this paper, we have pointed out that beam and jet functions in SCET can be calculated by integrating the well-known QCD splitting functions over the appropriate collinear phase space.
To demonstrate the utility of this approach, we have first shown that it reduces the computation of NLO beam and jet functions to expansions in $\eps$, using the (fragmenting) quark jet function and the (TMD) quark beam function as examples.
At NNLO, we have calculated the quark fragmenting jet function for the first time. This result is checked by verifying the cancellation of IR poles in the matching onto fragmentation functions and by using sum rules that relate it to the known quark jet function.

More general beam and jet functions than what we have considered in this paper have found applications in phenomenology.
Beam functions differential in both the transverse virtuality and the transverse momentum~\cite{Mantry:2009qz,Jain:2011iu,Larkoski:2014tva} entered in a calculation of the Higgs $p_T$ spectrum~\cite{Mantry:2009qz} and a recent event shape study in deep inelastic scattering~\cite{Kang:2013nha}. 
The generalization of the fragmenting jet function where both momentum fractions in the double real contribution are measured would for example enter in the description of jet charge at NNLO~\cite{Krohn:2012fg,Waalewijn:2012sv}. 
In addition, one could study jet functions and beam functions in the presence of a jet algorithm, see \eg Refs.~\cite{Ellis:2010rwa,Becher:2013xia,Stewart:2013faa}.
Splitting functions in dense QCD matter were calculated in Ref.~\cite{Fickinger:2013xwa}, and so one could envision including medium effects on jet functions in this way.
The approach discussed here is certainly advantageous at NLO, but it will depend on the details of the measurement whether that remains true at NNLO.

\begin{acknowledgments}
WW thanks G.~Ovanesyan for discussions. 
MR is grateful to A.~von~Manteuffel for providing him with a preliminary version of \Reduze and to ETH Z\"urich and CERN for their hospitality.
We thank A.~Gehrmann-De Ridder and C.~Lee for comments on the manuscript, and E.~Mereghetti for discussions concerning the results of~\cite{Bauer:2013bza}.
MR is supported by ERC Advanced Grant no.~320651, \textit{HEPGAME}.
WW is supported by a Marie Curie International Incoming Fellowship within the 7th European Community Framework Program (PIIF-GA-2012-328913).
\end{acknowledgments}

\appendix

\section{Plus Distributions}
\label{app:plus}

The plus distributions are defined as
\begin{align} \label{eq:plus_def}
 \cL_n(a) &\equiv \Big[\frac{\ln^n a}{a}\Big]_+ 
 \nn \\ &
 = \lim_{b\to0} \Big[\frac{\theta(a-b) \ln^n a}{a} + \de(a-b) \frac{\ln^{n+1} b}{n+1}\Big]
\end{align}
and satisfy the boundary condition
\begin{align}
 \int_0^1\! \df a\, \cL_n(a) = 0
\,.\end{align}
We will use the following plus distribution expansion
\begin{align} \label{eq:plus_exp}
  \frac{1}{a^{1+\eps}} &= -\frac{1}{\eps} \de(a) + \cL_0(a) -\eps \cL_1(a) 
  \nn \\ & \quad 
  + \frac{\eps^2}{2} \cL_2(a) - \frac{\eps^3}{6} \cL_3(a) + \ord{\eps^4}
\,, \end{align}
as well as
\begin{align} \label{eq:plus_conv}
  \int_0^a\! \df b\, \cL_0(b)\, (a-b)^{-1-\eps}
  &= \int_0^1\! \df b \, \frac{(a-b)^{-1-\eps}-a^{-1-\eps}}{b}
  \nn \\ & \quad
  + \int_1^a\! \df b \, \frac{(a-b)^{-1-\eps}}{b}
  \nn \\ &
  = \Big[-\psi(-\eps)-\ga_E -\frac{\df}{\df \eps}\Big] \frac{1}{a^{1+\eps}}
\,.\end{align}
The transverse momentum plus distributions can be converted to impact-parameter space using the Fourier transforms
\begin{align} \label{eq:plus_fourier}
  4\pi \int\! \frac{\df^2 \vec k_\perp^{\,2}}{(2\pi)^2}\, e^{\img \vec b_\perp \cdot \vec k_\perp}\, \de^2(\vec k_\perp^{\,2}) &= 1
  \,, \nn \\ 
  4\pi \int\! \frac{\df^2 \vec k_\perp^{\,2}}{(2\pi)^2}\, e^{\img \vec b_\perp \cdot \vec k_\perp}\,  \cL_0(\vec k_\perp^{\,2}) &= - L_\perp
  \,, \nn \\ 
  4\pi \int\! \frac{\df^2 \vec k_\perp^{\,2}}{(2\pi)^2}\, e^{\img \vec b_\perp \cdot \vec k_\perp}\,  \cL_1(\vec k_\perp^{\,2}) &= \frac{1}{2} L_\perp^2
\,,\end{align}
where $L_\perp = \ln (\vec b_\perp^{\,2} e^{2\ga_E}/4)$.
The following derivative of plus distributions is useful for calculating the anomalous dimensions
\begin{align} \label{eq:plus_deriv}
  \mu \frac{\df}{\df \mu} \Big[\frac{1}{\mu^2} \cL_n\Big(\frac{s}{\mu^2}\Big) \Big]
  = \begin{cases}
  -\frac{2n}{\mu^2} \cL_{n-1}\big(\frac{s}{\mu^2}\big) & n >0\,, \\
  -2 \de(s) & n=0\,.
  \end{cases}
\end{align}

\section{Harmonic Polylogarithms}
\label{app:hpl}

Harmonic polylogarithms \cite{Remiddi:1999ew} reduce to logarithms for weight one:
\begin{align}
\text{H}_1(z) &= - \ln \left(1-z\right)
\,, \nn \\
\text{H}_0 (z) &= \ln z
\,, \nn \\
\text{H}_{-1} (z) &= \ln \left( 1+z \right)
\,.\end{align}
Writing higher weights as vectors $b_1, \dotsc, b_w \equiv \vec{b}$, with $b_i \in \{1,0,-1\}$,
the harmonic polylogarithms of weight $w$ are defined through
\begin{equation}
\text{H}_{\vec{0}_w}(z) = \frac{1}{w!} \ln^w z
\,,\end{equation}
and recursively for $\vec{c} = (a, \vec{b} ) \neq \vec{0}_w$ by
\begin{equation}
\text{H}_{a, \vec{b}}\, (z) = \int_0^z \text{d} t f_a( t) \text{H}_{\vec{b}} \left( t \right)
\end{equation}
with the integration kernels
\begin{equation}
	f_1 (t) = \frac{1}{1-t}, \quad
	f_0 (t) = \frac{1}{t}, \quad
	f_{-1} (t) = \frac{1}{1+t} \, .
\end{equation}
Weight vectors with zeros to the left of $1$ or $-1$ are abbreviated:
\begin{equation}
	\text{H}_{\dotsc, \vec{0}_w, \pm 1, \dotsc}(z) \equiv \text{H}_{\dotsc, \pm (1+w), \dotsc}(z)
\,,\end{equation}
and we omit the argument $z$ if there is no potential for confusion.

\section{Quark Jet Function at NNLO}
\label{app:jet_mi}

\begin{widetext}
The integral reduction for the jet function calculation in \sec{jet_nnlo} can be performed for the kinematics $q \to p_1, p_2, p_3, p_4$, keeping $s_{123}$ fixed. 
After integral reduction the expansion to first order in $s_{123}/s_{1234}$ is performed.
We have
\begin{align}
	 \int\! \dPhiColl{3}\, \siColl{3,\bar q'q'q}\, \de(s-s_{123})
	&= \frac{ \alpha_s^2 C_F T_F}{(4 \pi)^2}\, \frac{ \mu^{4 \eps}}{s^{1+2\eps}}\,
\frac{2(1-\eps)^2(2-4\eps+\eps^2)}{\eps^2(3-2\eps)} I_1

\,, \nn \\
	\frac12 \int\! \dPhiColl{3}\, \siColl{3,\bar qqq\text{(id)}}\, \de(s-s_{123})
	&= \frac{ \alpha_s^2 C_F (C_F- C_A/2)}{(4 \pi)^2}\, \frac{ \mu^{4 \eps}}{s^{1+2\eps}}\, I_{\bar{q} q q}^\text{(id)}
\,, \nn \\
	\frac12 \int\! \dPhiColl{3}\, \siColl{3,qgg}\, \de(s-s_{123})
	&= \frac{ \alpha_s^2 C_F}{(4 \pi)^2}\, \frac{ \mu^{4 \eps}}{s^{1+2\eps}} \left(C_F I_{qgg}^{C_F^2} + C_A I_{qgg}^{C_F C_A} \right)
\,,\end{align}
which have the following decomposition into master integrals
\begin{align}
I_{\bar{q} q q}^\text{(id)} &=
\frac{ 4 - 38 \eps + 136 \eps^2 - 207 \eps^3 + 120 \eps^4 - 6 \eps^5 - 9 \eps^6 + 2 \eps^7 }{2 \eps^3 (1-\eps)(1-2\eps)} I_1
+ \frac{1-6\eps+6\eps^2}{\eps(1-\eps)} I_2
- \frac{1-6\eps+6\eps^2}{\eps(1-\eps)} I_3
\nn \\ & \quad
- \frac{1-9\eps+9\eps^2}{(1-2\eps)(1-\eps)} I_4
+ I_6

\,, \nn \\
I_{qgg}^{C_F^2} &= 
\frac{8-112\eps+607\eps^2-1541\eps^3+1772\eps^4-647\eps^5-134\eps^6+8\eps^7}{2 \eps^3 (1-4\eps)(1-2 \eps)} I_1
+ \frac{1-\eps}{1-4\eps} I_4
+ \frac{1-\eps}{1-4\eps} I_5
+ I_7

\,, \nn \\
I_{qgg}^{C_F C_A} &= 
\frac{96 - 1148 \eps + 5476 \eps^2 - 13443 \eps^3 + 18033 \eps^4 - 12497 \eps^5 + 3524 \eps^6 - 132 \eps^7 + 16 \eps^8}{ 4 \eps^3 (1 - 4 \eps) (1 - 2 \eps) (-3 + 2 \eps) } I_1
\nn \\ & \quad
+ \frac{ 1-6\eps+4\eps^2-8\eps^3}{2(1-4\eps)(1-2\eps)} I_4
- \frac{1-\eps}{2(1-4\eps)} I_5
-\frac{1}{2} I_7

\,.\end{align}

The integrals are defined as
\begin{equation}
	I_k = \mathcal{N}_I\, \Gamma \left( 1-2\eps \right) 2^{8-4\eps} \pi^{4-\eps} s^{-1+2 \eps} \int\! \dPhiColl{3}\, \de (s_{123} - s)\, b_k
\end{equation}
where we have $\mathcal{N}_I = 4 e^{2 \eps \ga_E}/\Gamma \left( 1-2\eps \right)$ and
\begin{equation}
\vec{b} = \left\{
1,
\frac{1}{1-z_1},
\frac{1}{1-z_1} \frac{1}{1-z_2},
\frac{s_{123}}{s_{23} (1-z_3) }, 
\frac{s_{123}}{s_{13} (1-z_1) z_3},
\frac{s_{123}^2}{s_{12} s_{23} (1-z_1) (1-z_3)},
\frac{s_{123}^2}{s_{12} s_{13} z_2 z_3}
\right\}
.\end{equation}
They evaluate to
\begin{align}
I_1/\mathcal{N}_I &= 
\frac {\Gamma \left ( 1 - \eps \right)^3} {(1 - 2 \eps) \Gamma \left ( 3 - 3 \eps \right) }

\,,\\
I_2/\mathcal{N}_I &=
\frac{\Gamma \left( 1-\eps \right)^3}{(1-2 \eps)^2 \Gamma \left( 2 - 3 \eps \right)}
\,,\\
I_3/\mathcal{N}_I &=
\frac{ \Gamma \left( 1 - \eps \right)^3 }{ (1-2\eps )^2 \Gamma \left( 2-3 \eps \right) }
{}_3 F_2\left(1, 1-2 \eps, 1-\eps, 2-3 \eps, 2-2\eps; 1 \right)

\,,\\
I_4/\mathcal{N}_I &=
- \frac{ \Gamma \left( 1-\eps \right)^3 }{ \eps \left( 1-2\eps \right) \Gamma \left( 2-3 \eps \right) }
{}_3F_2 \left( 1, 1-2\eps, 1-\eps, 2-3\eps, 2-2\eps ; 1 \right)

\,,\\
I_5/\mathcal{N}_I &=
- \frac{ \Gamma \left( 1-\eps \right)^3 }{ \eps^3 \Gamma \left( 1-3\eps \right)}
{}_3F_2 \left( 1, -2\eps, -\eps, 1-3\eps, 1-2\eps ; -1 \right)

\,,\\
I_6/\mathcal{N}_I &=
- \frac{ \Gamma \left( 1-2\eps \right) \Gamma \left( 1-\eps \right)^2 }{ \eps^3 \Gamma \left( 1-4 \eps \right) }
{}_4F_3 \left( 1-\eps, -2\eps, -2\eps, -2\eps, 1-2\eps, 1-2\eps, -4\eps; 1 \right)

\,,\\
I_7/\mathcal{N}_I &=
-\frac{ 6 }{\eps^3 } \frac{ \Gamma \left( 1-\eps \right)^3 }{ \Gamma \left( 1-3 \eps \right) }
+ \frac{1}{\eps^3 } 
	\frac{ \Gamma \left( 1-2 \eps \right)^2 \Gamma \left( 1-\eps \right) \Gamma \left( 1+\eps \right) }{ \Gamma \left( 1-4\eps \right) } 
	{}_3F_2 \left( -2\eps, -2\eps, -2\eps, 1-2\eps, -4\eps ; 1 \right)
\nonumber \\ & \qquad
-\frac{2}{\eps \left(1-\eps \right) \left( 1+\eps \right) }
	\frac{ \Gamma \left( 1-\eps \right)^3 }{ \Gamma \left( 1-3 \eps \right) }
	{}_4F_3 \left( 1, 1-\eps, 1-\eps, 1-\eps, 1-3\eps, 2-\eps, 2+\eps ; 1 \right)

\,.\end{align}
\end{widetext}

\newpage \phantom{a} \newpage \phantom{a} \newpage

\section{Quark Fragmenting Jet Function at NNLO}
\label{app:fragjet_mi}

Similar to the jet function case, the integral reduction for the fragmenting jet function in \sec{fragjet_nnlo} can be performed for the kinematics 
$q \to p_1, p_2, p_3, p_4$, keeping $s_{123}$ and $s_{14}$ fixed. 
After the integral reduction, the expansion to first order in $s_{123}/s_{1234}$ is performed.
(The integrated splitting functions in terms of master integrals are given in an auxiliary file.)
We denote
\begin{align}
	F[b] &= \mathcal{N} \int \! \text{d} \Phi_{1\to4}\, \delta( s - s_{123} )\, \delta \big( s_{14} - z (s_{1234}-s_{123}) \big)\, b
	\nn \\ &
	= \sum_{j} \left( \frac{s_{123}}{s_{1234}} \right)^j F[b]^{(j)}
\end{align}
where $\mathcal{N}$ is a normalization factor (not equal to $\mathcal{N}_I$).
At variance with the jet function, for some integrals subleading coefficients in  $s_{123}/s_{1234}$ remain after reduction and expansion.

The integrals are obtained either by direct integration or by solving the differential equation in $z$ (in some cases order by order in $\eps$).
The integrals show up multiplied by various powers of $z$ and $1-z$. We pull out factors of either $z$ or $1-z$ from the integrals such that 
$z^{-1-a\eps}$ or $(1-z)^{-1-a\eps}$ can be expanded in distributions using \eq{plus_exp}.
The expansion of factors $z^{-1-a\eps}$ is not necessary for the calculation at hand, but it allows us to check that the integral over $z$ reproduces the corresponding contribution to the jet function.

For all the integrals there is an integer $m$ such that $(1-z)^{m+2\eps} F \xrightarrow{z \to 1} k$ at arbitrary $\eps$.
The constant can be determined by direct integration in $4-2\eps$ dimensions (starting for example from the explicit parametrization of the $1 \to 4$ phase space in Ref.~\cite{GehrmannDeRidder:2003bm}) in all cases.
The analogous statement is not true for $z \to 0$. In almost all cases, the leading behavior at $z=0$ was determined from the explicit phase space parametrization.
The exception to this is $F\left[ 1/(s_{12} s_{13} s_{124}) \right]$ for which the form with explicit factors of $z$ was determined from the 
form with explicit factors of $1-z$, using the corresponding contribution to the jet function as additional input.

The normalization of the integrals is the same as that for the jet function, such that \eg $\int \df z\, F \left[ 1 \right]^\one = I_1$.
For the integrals which have been determined only to finite order in $\eps$, we don't display all the required orders below, since the full expressions 
are contained in a file accompanying this paper.

\begin{widetext}
\begin{align}
	F [1]^\one / \mathcal{N}_I &= z^{-\eps} (1-z)^{1-2\eps} \frac{1}{(1-2 \eps)^2} \frac{ \Gamma (1-\eps)^2}{\Gamma (1-2\eps)}
	\,, \\
	F [1]^\two  / \mathcal{N}_I &= - z^{-\eps} (1-z)^{1-2\eps}  \frac{ \Gamma (1-\eps)^2}{\Gamma (2-2\eps)}
	\,, \\
	F [1]^{(3)}  / \mathcal{N}_I &= - z^{-\eps} (1-z)^{1-2\eps}  \, \eps \, \frac{ \Gamma (1-\eps)^2}{\Gamma (2-2\eps)}
	\,, \\
	F \Big[ \frac{1}{s_{12}} \Big]^{(0)} /\mathcal{N}_I & = -z^{-\eps} (1-z)^{1-2\eps} \frac{1}{\eps} \frac{ \Gamma (1-\eps)^2}{\Gamma(2-2\eps)}
		{}_2F_1 \left( 1, 1-\eps, 2-2\eps; 1-z \right)

	\,, \\
	F \Big[ \frac{1}{s_{12}} \Big]^{(1)} /\mathcal{N}_I &= z^{-\eps}(1-z)^{1-2\eps} \frac{1}{\eps} 
		\frac{ \Gamma(1-\eps)^2}{\Gamma(1-2\eps)}
		{}_2F_1 \left( 1, 1-\eps, 2-2\eps; 1-z \right)

	\,, \\
	F \Big[ \frac{1}{s_{12}} \Big]^{(2)} /\mathcal{N}_I &= z^{-\eps}(1-z)^{1-2\eps} 
		\frac{ \Gamma(1-\eps)^2}{\Gamma(1-2\eps)}
		{}_2F_1 \left( 1, 1-\eps, 2-2\eps; 1-z \right) 

	\,, \\
	F \Big[ \frac{1}{s_{12}} \Big]^{(3)} /\mathcal{N}_I &= z^{-\eps}(1-z)^{1-2\eps} \frac{1+2\eps}{3}
		\frac{ \Gamma(1-\eps)^2}{\Gamma(1-2\eps)}
		{}_2F_1 \left( 1, 1-\eps, 2-2\eps; 1-z \right) 

	\,, \\
	F \Big[ \frac{1}{s_{124}} \Big]^{(4)} / \mathcal{N}_I &= \frac{(1+4\eps)(3-9\eps-8\eps^2+8\eps^3)}{12\eps(1-2\eps)(3-2\eps)} \frac{\Gamma(1-\eps)^2}{\Gamma(1-2\eps)}
		\nonumber \\ & \quad
		\times
		\left[
			z^{-\eps}(1-z)^{-1-2\eps}(1+z)
			- 2 z^{1-\eps}(1-z)^{-1-2\eps} {}_2F_1 \left( 1, -\eps, -2\eps; 1-z \right)
		\right] 

	\,, \\
	F \Big[ \frac{s_{12}}{s_{124}} \Big]^{(2)} / \mathcal{N}_I &= z^{-\eps}(1-z)^{1-2\eps}  \frac{1}{2 (1-2\eps)^2} \frac{ \Gamma ( 1-\eps)^2 }{\Gamma ( 1-2\eps) }

	\,, \\
	F \Big[ \frac{s_{12}}{s_{124}} \Big]^{(3)}/\mathcal{N}_I &= - z^{-\eps}(1-z)^{1-2\eps} \frac{2-7\eps+4\eps^2}{2(1-2\eps)^2(3-2\eps)}
		 \frac{ \Gamma(1-\eps)^2}{\Gamma (1-2\eps)} 

	\,, \\
	F \Big[ \frac{s_{12}}{s_{124}} \Big]^{(4)}/\mathcal{N}_I &= - z^{-\eps}(1-z)^{1-2\eps} \frac{1+\eps-12\eps^2+8\eps^3}{4(1-2\eps)^2(3-2\eps)}
		 \frac{ \Gamma(1-\eps)^2}{\Gamma (1-2\eps)} 

	\,, \\
	F \Big[ \frac{s_{13}}{s_{124}} \Big]^{(2)}/\mathcal{N}_I &= \frac{1}{2 \eps (1-2 \eps)^2}
		\frac{\Gamma(1-\eps)^2}{\Gamma(1-2\eps)} \left[
			-z^{-\eps}(1-z)^{-1-2\eps} \left( (1+z)^2 -\eps(1+6z+z^2) \right)
			\nonumber \right. \\ & \qquad \qquad \left.
			+2 (1-2\eps) z^{1-\eps}(1-z)^{-1-2\eps}(1+z) 
				{}_2F_1 \left(1,-\eps,-2\eps;1-z\right)
		\right]

	\,, \\
	F \Big[ \frac{s_{13}}{s_{124}} \Big]^{(3)} / \mathcal{N}_I &= \frac{1}{2 \eps (1-2\eps)^2(3-2\eps)} \frac{\Gamma(1-\eps)^2}{\Gamma(1-2\eps)} \left[
			z^{-\eps}(1-z)^{-1-2\eps} 
				\nonumber \right. \\ & \qquad  \left.
				(2(1+z)^2 - 3 \eps (3+z)(1+3 z) + \eps^2 (11+58 z+11 z^2) - 4 \eps^3 (1+6 z + z^2))
			\nonumber \right. \\ & \qquad  \left.
			-4 (1-2 \eps) (1-4\eps+2\eps^2) z^{1-\eps}(1-z)^{-1-2\eps} (1+z) {}_2F_1 (1, -\eps, -2\eps;1-z)
		\right] 

	\,, \\
	F \Big[ \frac{s_{13}}{s_{124}} \Big]^{(4)} / \mathcal{N}_I &= \frac{1}{4 \eps (1-2\eps)^2(3-2\eps)} \frac{\Gamma(1-\eps)^2}{\Gamma(1-2\eps)} \left[
			z^{-\eps}(1-z)^{-1-2\eps} 
			\nonumber \right. \\ & \qquad  \left.
			((1+z)^2 -4\eps z -\eps^2(13+46 z+13 z^2)
			 +20 \eps^3 (1+6 z + z^2) 
			- 8 \eps^4 (1+6 z + z^2))
			\nonumber \right. \\ & \qquad  \left.
			-2 (1-2 \eps) (1+\eps-16\eps^2+8\eps^3) z^{1-\eps}(1-z)^{-1-2\eps} (1+z) {}_2F_1 (1, -\eps, -2\eps;1-z)
		\right]

	\,, \\
	F \Big[ \frac{1}{s_{13}s_{124}} \Big]^{(0)}/\mathcal{N}_I &= -\frac{2}{\eps^2} \frac{z^{-2\eps}}{1+z} \Gamma(1-\eps) \Gamma(1+\eps)
		+ \frac{2}{\eps^2} \frac{z^{-\eps}(1-z)^{1-2\eps}}{1+z} \frac{ \Gamma(1-\eps)^2}{\Gamma(1-2\eps)} {}_2F_1(1,1-\eps,1+\eps;z)

	\,, \\
	F \Big[ \frac{1}{s_{23} s_{24} s_{124}} \Big]^{(0)} / \mathcal{N}_I &= \frac{1}{\eps^2} z^{-1-2\eps}(1-z)^{-1-2\eps} \frac{\Gamma(1-\eps)^2}{\Gamma(1-2\eps)} {}_2F_1(-2\eps, -\eps, 1-2\eps; 1-z)

	\,, \\
	F \Big[ \frac{1}{s_{14} s_{34} s_{234}} \Big]^{(0)} / \mathcal{N}_I &= \frac{1}{\eps^2} \frac{\Gamma(1-\eps)^2}{\Gamma(1-2\eps)} z^{-\eps} (1-z)^{-1-2\eps}
		- \frac{1}{\eps} \frac{\Gamma(1-\eps)^2}{\Gamma(2-2\eps)} z^{-\eps}(1-z)^{-2\eps} {}_2F_1 (1,1-\eps,2-2\eps;1-z)

	\,, \\
	F \Big[ \frac{1}{s_{12} s_{34} s_{134}} \Big]^{(0)} / \mathcal{N}_I &=  -\frac{1}{\eps^2} \frac{z^{-2-\eps} (1-z)^{-2\eps}(1-3 z)}{1+z} \frac{\Gamma(1-\eps)^2}{\Gamma(1-2\eps)}
		+ \frac{1}{\eps^2} \frac{z^{-1-2\eps}(1-z)}{1+z} \Gamma(1-\eps) \Gamma(1+\eps)
		\nonumber \\ & \quad
		+ \frac{1}{\eps^2} \frac{z^{-2-\eps} (1-z)^{2-2\eps}}{1+z} \frac{\Gamma(1-\eps)^2}{\Gamma(1-2\eps)} {}_2F_1(1,-\eps,\eps;z)

	\,, \\
	F \Big[ \frac{1}{s_{12} s_{24} s_{234} } \Big]^{(0)} / \mathcal{N}_I &= \frac{1}{\eps^2} z^{-\eps} (1-z)^{-1-2\eps} \frac{\Gamma(1-\eps)^2}{\Gamma(1-2\eps)} {}_2F_1(1,1-\eps,1-2\eps;1-z)

	\,, \\
	F \Big[ \frac{1}{s_{12} s_{13} s_{24}} \Big]^{(-1)}  / \mathcal{N}_I &=  \frac{2}{\eps^2}z^{-1-\eps}(1-z)^{-2\eps} \frac{ \Gamma(1-\eps)^2}{\Gamma(1-2\eps)}
		+ \frac{1}{\eps^3} z^{-1-2\eps}\frac{ \Gamma(1-\eps)^3 \Gamma(1+\eps)^3}{\Gamma(1-2\eps)\Gamma(2\eps)}
		\nn \\ & \quad
		+ \frac{2}{\eps} z^{-1-2\eps} \Gamma(1-\eps)\Gamma(1+\eps) \log z
		- \frac{2}{\eps^2} z^{-1-\eps} \frac{\Gamma(1-\eps)^2}{\Gamma(1-2\eps)} {}_3F_2( \eps, \eps, 2\eps, 1+\eps, 1+\eps;z)

	\,, \\
	F \Big[ \frac{1}{s_{12} s_{23} s_{134}} \Big]^{(-1)} / \mathcal{N}_I &=
		(1-z)^{-2 \eps} \left( \frac{1}{\eps^2} \frac{1}{z} - \frac{2}{\eps} \frac{\log z}{z} + \ord{\eps^0} \right)

		\nn \\ &= 
		z^{-1-2\eps} \left( \frac{1}{\eps^2} - \frac{2}{\eps} \log(1-z) + \ord{\eps^0} \right)

	, \\
	F \Big[ \frac{1}{s_{12} s_{23} s_{34}} \Big]^{(-1)} / \mathcal{N}_I &=
	(1-z)^{-1-2\eps} \left( \frac{3}{\eps^2} - \frac{2}{\eps}  \log z + \ord{\eps^0} \right)

		\\ &= 
		z^{-2 \eps} \left( - \frac{1}{\eps^2} \frac{1}{1-z} + \frac{2}{\eps} \frac{\log (1-z)}{1-z} + \ord{\eps^0} \right) 
			+ z^{-\eps} \left( \frac{4}{\eps^2} \frac{1}{1-z} - \frac{8}{\eps} \frac{\log (1-z) }{1-z} + \ord{\eps^0} \right)

	\nn , \\
	F \Big[ \frac{1}{s_{12} s_{13} s_{124}} \Big]^{(-1)} / \mathcal{N}_I &=
	(1-z)^{1-2\eps} \left( \frac{2}{\eps} \frac{\log z }{z (1-z)} + \ord{\eps^0} \right)

		\nn \\ &= 
		2 z^{-1-2\eps} \Gamma(1-\eps) \Gamma(\eps) \log(z) 
		+ z^{-1-2\eps} \left( 4 \text{H}_{2} + 4 \text{H}_{-1,0} - \frac{\pi^2}{3} + \ord{\eps^1} \right)

.\end{align}
\end{widetext}

\section{Matching Corrections for the NNLO Quark Fragmenting Jet Function}
\label{app:IR}

In addition to the two-loop matching coefficient $\cJ_{qi}^\two(s,z,\mu)$ that we want to extract, the RHS of \eq{matching_G_2} also involves the two-loop fragmentation function and a cross term between the one-loop matching coefficient and the one-loop fragmentation function, which we work out in this appendix. The IR divergences provide an important cross check on our results and the finite terms enter in the determination of the matching coefficients.

In pure dimensional regularization all radiative corrections to the fragmentation function are scaleless and vanish,
\begin{align}
D_{i,\text{bare}}^j(z) = \de_{i,j} \de(1-z)
\,.\end{align}
Beyond the tree-level contribution, the renormalized fragmentation function thus only consists of $1/\eps_\text{IR}$ poles, which must exactly cancel the $1/\eps_\text{UV}$ poles in the bare fragmentation function, 
\begin{align}
  D_i^j(z,\mu) &= \int_z^1\! \frac{\df z'}{z'}\, Z^D_{ik}\Big(\frac{z}{z'},\mu\Big) D_{k,\text{bare}}^j(z') = Z_{ij}^D(z,\mu)
\,.\end{align}
We can thus obtain the renormalized fragmentation function from its known renormalization,
\begin{align} \label{eq:D_ren}
  \mu \frac{\df}{\df \mu} D_i^j(z,\mu) &= \sum_k \int_z^1\! \frac{\df z'}{z'}\, \ga^D_{ik}(z',\mu) D_{k,\text{bare}}^j\Big(\frac{z}{z'}\Big)
\,, \nn\\
  \ga^D_{ik}(z',\mu) & = \sum_\ell \int_{z'}^1\! \frac{\df z''}{z''}\, (Z^D)^{-1}_{i\ell}\Big(\frac{z'}{z''},\mu\Big)\, \mu \frac{\df}{\df\mu} Z^D_{\ell k}(z'',\mu)
\nn \\ 
& = \frac{\al_s}{\pi} p^\zero_{ki}(z) + \frac{\al_s^2}{2\pi^2}\, p^\one_{ki}(z) + \ord{\al_s^3}
\,. \end{align}
The splitting functions $p_{ki}$ that enter here are collected in \eqs{split0}{split1}, and are not the same as in the main text (though they are obviously related). To avoid confusion we denote them with a lower case $p$. Note that the convention for the indices $ki$ in the subscript is also different from the main text.
From \eq{D_ren} we can obtain the renormalization $Z$ factor and thus the fragmentation function,
\begin{align} \label{eq:D_pert}
  D_i^j(z,\mu) &= \de_{i,j}\de(1-z) - \frac{\al_s}{2\pi}\, \frac{1}{\eps}\, p_{ji}^\zero(z)
  \nn \\ & \quad  
  + \Big(\frac{\al_s}{2\pi}\Big)^2 \bigg[-\frac{1}{2\eps}\, p_{ji}^\one(z) 
  +  \frac{\beta_0}{4\eps^2}\, p_{ji}^\zero(z)
  \nn \\ & \quad  
  + \frac{1}{2\eps^2} \int_z^1\! \frac{\df z'}{z'}\, p_{jk}^\zero(z') p_{ki}^\zero\Big(\frac{z}{z'}\Big)\bigg]
\,.\end{align}

The LO splitting functions are given by~\cite{Altarelli:1977zs}
\begin{align} \label{eq:split0}
  p_{qq}^\zero(z) &= C_F \Big[(1+z^2)\, \cL_0(1-z) + \frac{3}{2}\, \de(1-z)\Big]
  \,, \nn \\
  p_{gq}^\zero(z) &= C_F \,\frac{1+(1-z)^2}{z}
  \,, \nn \\
  p_{gg}^\zero(z)
  &= 2C_A \Big[z\, \cL_0(1-z) + \frac{1-z}{z} +  z(1-z) \Big]
  \nn \\ & \quad
  + \frac{1}{2} \beta_0\, \delta(1-z)
  \,,\nn\\
  p_{qg}^\zero(z) &= T_F \big[z^2+(1-z)^2\big]
  \,,
\end{align}
where $z\to 1$ (but not $z\to 0$) is regulated.
The convolutions of splitting functions that enter in \eq{D_pert} are given by 
\begin{widetext}
\begin{align}
  \int_z^1\! \frac{\df z'}{z'} p_{qq}^\zero(z') p_{qq}^\zero\Big(\frac{z}{z'}\Big) &=
  C_F^2 \bigg[ 4 (1+z^2) \cL_1(1\!-\!z) +6 \cL_0(1\!-\!z)+\Big(\frac{9}{4}-\frac{2\pi^2}{3}\Big)\de(1\!-\!z)+\Big(3z+3-\frac{4}{1\!-\!z}\Big)\ln z-z-5 \bigg]
  \,,\nn \\
  \int_z^1\! \frac{\df z'}{z'} p_{qg}^\zero(z') p_{gq}^\zero\Big(\frac{z}{z'}\Big) &=
  C_F T_F \bigg[2 (1+z) \ln z -\frac{4}{3}z^2 - z +1 + \frac{4}{3 z}\bigg]
  \,,\nn \\
  \int_z^1\! \frac{\df z'}{z'} p_{gq}^\zero(z') p_{qq}^\zero\Big(\frac{z}{z'}\Big) &=
  C_F^2 \bigg[2 \bar p_{gq}^\zero(z) \ln (1-z)+(2-z) \ln z-\frac{z}{2}+2\bigg]
  \,,\nn \\
  \int_z^1\! \frac{\df z'}{z'} p_{gg}^\zero(z') p_{gq}^\zero\Big(\frac{z}{z'}\Big) &=
2 C_F C_A \bigg[\bar p_{gq}^\zero(z) \ln (1-z)-\Big(2z + 2 + \frac{2}{z} \Big) \ln z + \frac{2}{3}z^2+\frac{1}{2}z + 4 -\frac{31}{6 z} \bigg]  + \frac{\beta_0}{2}\, p_{gq}^\zero(z)
\,,\end{align}
where $\bar p_{ij}^\zero(z)$ denote the splitting functions in \eq{split0} without the overall color factor.
The NLO splitting functions for $0<z<1$ are\footnote{The published version of Ref.~\cite{Furmanski:1980cm} contains some misprints~\cite{Ellis:1991qj}.}~\cite{Curci:1980uw,Furmanski:1980cm} (given in electronic form \eg accompanying~\cite{Mitov:2006wy})
\begin{align}\label{eq:split1}
p_{qq}^\one(z) & = P_{qq}^{V\one}(z) + P_{qq}^{S\one}(z)
\,, \qquad
p_{q\bar q}^\one(z) = P_{q\bar q}^{V\one}(z) + P_{qq}^{S\one}(z)
\,, \qquad
p_{qq'}^\one(z) = P_{q\bar q'}^\one(z) = P_{qq}^{S\one}(z)
\,,  \\
p_{qq}^{V\one}(z) &= 
  C_F^2 \bigg[\Big(2 \ln z \ln (1-z) + \frac32 \ln z - 2 \ln z^2\Big) \bar p^\zero_{qq}(z) 
  - \Big(\frac72 + \frac32 z\Big) \ln z + \frac12 (1 + z) \ln^2 z - 
      5 (1 - z)\bigg] 
  \nn \\ & \quad      
      + C_F C_A \bigg[\Big(\frac12 \ln^2 z + \frac{11}{6} \ln z + \frac{67}{18} - \frac{\pi^2}{6}\Big) \bar p^\zero_{qq}(z)
       + (1 + z) \ln z + \frac{20}{3} (1 - z)\bigg] 
  \nn \\ & \quad       
       +    C_F T_F n_f \bigg[-\Big(\frac23 \ln z + \frac{10}{9}\Big) \bar p^\zero_{qq}(z) - \frac{4}{3} (1 - z)\bigg]
\,,\nn \\
p_{q\bar q}^{V\one}(z) &= 
  C_F \Big(C_F - \frac{C_A}{2}\Big) \big[2 \bar p^\zero_{qq}(-z) S_2(z) + 2 (1 + z) \ln z + 4 (1 - z)\big]
\,,\nn \\
p_{qq}^{S\one}(z) &= 
 C_F T_F \bigg[(1+z) \ln ^2 z +\Big(-\frac{8}{3}z^2-9 z-5\Big) \ln z +  \frac{56}{9} z^2 + 4 z - 8 -\frac{20}{9 z}\bigg]
\,,\nn \\
p_{gq}^\one(z) &= 
  C_F^2 \bigg[-\frac12 + \frac92 z + \Big(-8 + \frac12 z\Big) \ln z + 
      2 z \ln (1-z) + \Big(1 - \frac12 z\Big) \ln^2 z
    +   \Big(\ln^2 (1-z) + 
         4 \ln z \ln (1-z) 
    \nn \\ & \quad               
         + 8 \text{Li}_2(1\!-\!z) - \frac{4 \pi^2}{3}\Big) \bar p^\zero_{gq}(z)\bigg]
    + C_F C_A \bigg[ \frac{62}{9} - \frac{35}{18} z - \frac{44}{9} z^2   
    + \Big(2 \!+\! 12 z \!+\! \frac83 z^2\Big) \ln z 
    -  2 z \ln (1\!-\!z) - (4 \!+\! z) \ln^2 z
    \nn \\ & \quad           
    + \bar p^\zero_{gq}(-z) S_2(z) + \Big(-2 \ln z \ln (1-z) - 3 \ln z - \frac32 \ln^2 z - 
         \ln^2 (1-z) - 8 \text{Li}_2(1-z) + \frac{7 \pi^2}{6} + \frac{17}{18}\Big) \bar p^\zero_{gq}(z)\bigg]
\,,\nn\end{align}
where 
\begin{align}
S_2(z) = -2 \text{Li}_2(-z) + \frac12 \ln^2 z - 2 \ln z \ln(1 + z) - \frac{\pi^2}{6}
\,.\end{align}
The endpoint contribution for $z=1$ in \eq{split1} can be fixed by the sum rules for the fragmentation functions.
The convolutions of matching coefficients with the LO splitting functions that enter in \eq{matching_G_2} are given by
\begin{align}
 \int_z^1\! \frac{\df z'}{z'}\, \cJ_{qi}^\one\Big(s,\frac{z}{z'},\mu \Big) p_{ji}^\zero(z')
 &=
 \frac{\al_s C_F}{2\pi}\,\frac{1}{\mu^2} \bigg\{
  \bigg[2\cL_1\Big(\frac{s}{\mu^2}\Big) - \eps\, \cL_2\Big(\frac{s}{\mu^2}\Big)\bigg] \de_{i,q}\, p_{ji}^\zero(z)
  +  \bigg[\cL_0\Big(\frac{s}{\mu^2}\Big) - \eps\, \cL_1\Big(\frac{s}{\mu^2}\Big)\bigg]
   \\ & \quad \times
    \bigg[\frac{1}{C_F}\int_z^1\! \frac{\df z'}{z'} p_{ji}^\zero(z') p_{iq}^\zero\Big(\frac{z}{z'}\Big) - \frac32\,   \de_{i,q}\, p_{ji}^\zero(z) \bigg]
  + \bigg[\de\Big(\frac{s}{\mu^2}\Big) - \eps\, \cL_0\Big(\frac{s}{\mu^2}\Big)\bigg] J_{qij}(z) + \ord{\eps^2} \bigg\}
\,,\nn\end{align}
where 
\begin{align}
J_{qqq}(z)
 &= C_F \bigg[6 \cL_2(1-z) + 3 \cL_1(1-z) - \pi^2 \cL_0(1-z) + \Big(-\frac{\pi^2}{4} + 4\zeta_3\Big) \de(1-z) 
  \\ & \quad  
 - (1+z) \Big(\text{Li}_2(z) +3 \ln^2(1-z) + \frac{3}{2} \ln (1-z) \Big) 
 - \frac{3 z^2+1}{2 (1-z)} \ln^2 z
 + \frac{z^2+6 z-1}{2 (1-z)} \ln z 
  + \frac12 (1-z) 
 +\frac{2 \pi^2}{3}(1+z) 
 \nn \\ & \quad
 + \eps \bigg\{ 
  -\frac83 \cL_3(1-z)- \frac32 \cL_2(1-z)+ \frac{4\pi^2}{3} \cL_1(1-z)+\Big(\frac{\pi^2}{4} - \frac{16 \zeta_3}{3}\Big) \cL_0(1-z) 
  - \left( \frac{\pi^4}{90} + \zeta_3 \right) \de(1-z)
  \nn \\ & \quad
  +\frac{7 z^2+5}{1-z}\,\text{Li}_3(z) - \frac{3 z^2+1}{1-z}\,\text{Li}_3(1-z)
  -\frac{2 z^2+2}{1-z}\, \text{Li}_3\Big(\frac{z}{z-1}\Big)
  +(z+1) \text{Li}_2(z) \ln (1-z)
  - \frac{5 z^2+3}{1-z}\, \text{Li}_2(z) \ln z  
  \nn \\ & \quad
 -\frac{3 z^2-5}{3(1- z)} \ln ^3(1-z) 
 -\frac{5 z^2+3}{2 (1-z)} \ln^2(1-z) \ln z
 +\frac{3 z^2+1}{6(1-z)}\ln ^3z
  +(-5 z+4) \text{Li}_2(z) + \Big(\frac74 z- \frac14\Big) \ln ^2(1-z)
  \nn \\ & \quad
  +\frac{5 z^2-16 z+5}{2(1- z)} \ln(1-z) \ln z
  -\frac{z^2+6 z-1}{4 (1-z)}\,\ln ^2z
  +\Big(\frac{7 z^2-3}{6 (1-z)}\,\pi ^2+\frac12 z - \frac12\Big) \ln (1-z)
  \nn \\ & \quad  
  +\Big(-\frac{7 z^2+5}{12 (1-z)}\pi ^2+\frac92 z+ \frac12 \Big) \ln z+ \Big(\frac{13}{24}z-\frac{5}{8}\Big)\pi ^2
  -\frac{29 z^2+7}{3 (1-z)}\, \zeta_3
  -4z +4  
 \bigg\}\bigg]
\,, \nn \\
J_{qgq}(z) &=
T_F \bigg[-2(1+z) \text{Li}_2(z)+(1+z) \ln^2 z +\Big(-\frac{4}{3}z^2-z+1+\frac{4}{3 z}\Big) \ln (1-z)
\nn \\ & \quad
+\Big(\frac{4}{3}z^2+5 z+4+\frac{4}{3 z}\Big) \ln z
-\frac{31}{9}z^2 +\Big(\frac{2}{3}+\frac{\pi ^2}{3}\Big) z +\frac{13}{9 z}+\frac{4}{3} + \frac{\pi ^2}{3} 
 \nn \\ & \quad
 + \eps \bigg\{ (z+1) \Big[2\text{Li}_3(1-z) - 2\text{Li}_3(z) - \frac13 \ln^3 z+ 2\text{Li}_2(z) \big( \ln (1-z) + \ln z\big) + \ln^2(1-z) \ln z 
  \nn \\ & \quad 
 + \frac{\pi^2}{6} \big(\ln z - 2\ln (1-z)\big)+ 2\zeta_3 \Big]
+ \Big(\frac{8}{3}z^2+6 z+3\Big) \text{Li}_2(z)  + \Big(\frac{2}{3}z^2+\frac{1}{2}z-\frac{1}{2}-\frac{2}{3 z}\Big) \ln^2(1-z) 
\nn \\ & \quad
+\Big(\frac{4}{3}z^2+z-1-\frac{4}{3 z}\Big) \ln(1-z) \ln z
-\Big(\frac{2}{3}z^2+\frac{5}{2}z+2+\frac{2}{3 z}\Big) \ln ^2 z +  \Big(\frac{31}{9} z^2-\frac{2}{3} z-\frac{4}{3}-\frac{13}{9 z}\Big) \ln (1-z)
  \nn \\ & \quad
-\Big(\frac{31}{9}z^2+\frac{32}{3}z+\frac{19}{3}+\frac{13}{9 z}\Big) \ln z
+ \Big(-\frac{5}{9}z^2-\frac{13}{12}z-\frac{5}{12}+\frac{1}{9 z}\Big) \pi ^2
 +\frac{269}{54}z^2 +\frac{47}{18}z-\frac{107}{18}-\frac{89}{54 z}
 \bigg\}\bigg]
 \,, \nn \\
J_{qg\bar q}(z) &= J_{qgq'}(z) = J_{qg\bar q'}(z) = J_{qgq}(z)
\,, \nn \\
J_{qqg}(z)
 &=C_F \bigg[\Big(3 z-6+\frac{4}{z}\Big) \text{Li}_2(z)+\bar p_{gq}^\zero(z) \ln (1-z) \big(\ln(1-z) + 2\ln z\big)+\Big(-\frac{1}{2}\,z + 1\Big) \ln ^2 z
 \nn \\ & \quad
 +\Big(-2 z+5-\frac{3}{z}\Big) \ln (1-z)+(z+5) \ln z - \frac{2 \pi ^2}{3}\, z +\frac{7-\pi ^2}{z} -7 +\frac{4 \pi ^2}{3}
 \nn \\ & \quad
 + \eps \bigg\{
 -\Big(3 z-6+\frac{4}{z}\Big) \Big[\text{Li}_3(1-z) - \text{Li}_3(z)  +\text{Li}_2(z) \big(\ln (1-z) + \ln z)\Big]
 + \Big(3 z+\frac{3}{z} \Big) \text{Li}_2(z) 
\nn \\ & \quad   
 +\Big(-\frac{1}{3}z+\frac{2}{3}-\frac{2}{3 z}\Big) \ln ^3(1-z)
 +\Big(-\frac{5}{2}z+5-\frac{4}{z}\Big) \ln^2(1-z) \ln z
 +\Big(-z+2-\frac{2}{z}\Big) \ln(1-z) \ln ^2 z
 \nn \\ & \quad  
 +\Big(\frac{1}{6}z-\frac{1}{3}\Big) \ln ^3 z
 +\Big(z-\frac{5}{2} +\frac{3}{2 z} \Big) \ln ^2(1-z)
 +\Big(2 z-5+\frac{3}{z}\Big) \ln(1-z) \ln z
 -\Big(\frac{1}{2}z+\frac{5}{2}\Big) \ln ^2 z
\nn \\ & \quad  
+\Big[ \Big(\frac{2}{3}z-\frac{4}{3}+\frac{1}{z}\Big)\pi ^2+7-\frac{7}{z}\Big] \ln (1-z)
+\Big[\Big(-\frac{1}{12}z+\frac{1}{6}\Big)\pi ^2 -4 z+5\Big] \ln (z)
+\Big(-\frac{2}{3}z+\frac{5}{12}-\frac{3}{4 z}\Big) \pi ^2
\nn \\ & \quad  
+\Big(-\frac{11}{3}z+\frac{22}{3}-\frac{16}{3 z}\Big) \zeta_3
  +6 z-20+\frac{14}{z}
  \bigg\}\bigg] 
\,, \nn \\
J_{qgg}(z)
 &= 2C_A \bigg[\Big(6-\frac{2}{z}\Big) \text{Li}_2(z)
 + \bar p_{gq}^\zero \ln (1-z) \big(\ln(1-z) - \ln z\big) -\Big(z+1+\frac{1}{z}\Big) \ln^2 z
 \nn \\ & \quad
 + \Big(\frac{2}{3}z^2 + \frac{3}{2}z +4 - \frac{31}{6 z}\Big) \ln (1-z)
-\Big(\frac{2}{3}z^2+\frac{7}{2}z+\frac{11}{3z}+3\Big) \ln z  +\frac{31}{18}z^2  -\Big(\frac{5}{6}+\frac{\pi ^2}{6}\Big) z-\frac{2}{9 z}-\frac{2}{3}-\frac{2 \pi ^2}{3}
 \nn \\ & \quad
 + \eps \bigg\{
 \Big(6 z+\frac{8}{z}\Big) \text{Li}_3(z)
 + \Big(2 z-4+\frac{4}{z}\Big) \text{Li}_3(1-z)
 +\Big(2 z+2+\frac{2}{z}\Big) \Big( \text{Li}_3\Big(\frac{z-1}{z}\Big) -  \text{Li}_2(z) \ln z \Big)
 \nn \\ & \quad 
 +\Big(-6+\frac{2}{z}\Big)  \text{Li}_2(z) \ln (1-z)
+\Big(-\frac{1}{2}z+1-\frac{1}{z}\Big) \ln ^3(1-z) 
 +\Big(-3+\frac{1}{z}\Big) \ln^2(1-z) \ln z
 \nn \\ & \quad  
+\Big(\frac{3}{2}z+\frac{2}{z}\Big) \ln(1-z) \ln ^2 z  
 + \Big(-\frac{4}{3} z^2-2 z-7+\frac{3}{2 z}\Big) \text{Li}_2(z) 
+\Big(-\frac{1}{3}z^2-\frac{5}{4}z-2+\frac{31}{12 z}\Big) \ln ^2(1-z) 
 \nn \\ & \quad
+\Big(-\frac{2}{3}z^2+\frac{1}{2}z-4+\frac{31}{6 z}\Big) \ln(1-z)\ln z
+\Big(\frac{1}{3}z^2+\frac{7}{4}z+\frac{3}{2}+\frac{11}{6 z}\Big) \ln ^2 z
+\Big[ \Big(\frac{1}{4}z+\frac{1}{2}+\frac{1}{6 z}\Big) \pi ^2
\nn \\ & \quad
-\frac{31}{18}z^2+\frac{5}{6}z+\frac{2}{3}+\frac{2}{9 z}\Big] \ln (1-z)
+\Big[ \Big(-\frac{2}{3}z-\frac{1}{6}-\frac{5}{6 z}\Big)\pi ^2+\frac{31}{18}z^2+\frac{35}{6}z+\frac{29}{3}+\frac{67}{18 z}\Big] \ln z
 \nn \\ & \quad
+ \Big(\frac{5}{18}z^2+\frac{13}{24}z+\frac{3}{2}-\frac{49}{72 z}\Big) \pi ^2
+\Big(-7 z+2-\frac{10}{z}\Big) \zeta_3
-\frac{269}{108} z^2 -\frac{83}{36}z-\frac{217}{36} +\frac{1169}{108 z}
 \bigg\}
 \bigg]
 \nn \\ & \quad
 + \frac{\beta_0}{2} \bigg[\bar p_{gq}^\zero(z) \big(\ln (1-z) + \ln z\big) + z
 + \eps \bigg\{
 \bar p_{gq}^\zero(z) \Big[-\frac{1}{2} (\ln (1-z) + \ln z)^2 + \frac{\pi ^2}{12}\Big]-z \big(\ln (1-z)+\ln z \big)
 \bigg\} \bigg]
\,,\nn\end{align}
\end{widetext}
The $\ord{\eps}$ terms in $J_{qij}$ are needed, because they give a finite correction to the matching coefficients.
We have also used the \MT package~\cite{Hoeschele:2013gga} to perform the necessary convolutions.

\bibliography{splitNNLO}

\begin{thebibliography}{85}%
\makeatletter
\providecommand \@ifxundefined [1]{%
 \@ifx{#1\undefined}
}%
\providecommand \@ifnum [1]{%
 \ifnum #1\expandafter \@firstoftwo
 \else \expandafter \@secondoftwo
 \fi
}%
\providecommand \@ifx [1]{%
 \ifx #1\expandafter \@firstoftwo
 \else \expandafter \@secondoftwo
 \fi
}%
\providecommand \natexlab [1]{#1}%
\providecommand \enquote  [1]{``#1''}%
\providecommand \bibnamefont  [1]{#1}%
\providecommand \bibfnamefont [1]{#1}%
\providecommand \citenamefont [1]{#1}%
\providecommand \href@noop [0]{\@secondoftwo}%
\providecommand \href [0]{\begingroup \@sanitize@url \@href}%
\providecommand \@href[1]{\@@startlink{#1}\@@href}%
\providecommand \@@href[1]{\endgroup#1\@@endlink}%
\providecommand \@sanitize@url [0]{\catcode `\\12\catcode `\$12\catcode
  `\&12\catcode `\#12\catcode `\^12\catcode `\_12\catcode `\%12\relax}%
\providecommand \@@startlink[1]{}%
\providecommand \@@endlink[0]{}%
\providecommand \url  [0]{\begingroup\@sanitize@url \@url }%
\providecommand \@url [1]{\endgroup\@href {#1}{\urlprefix }}%
\providecommand \urlprefix  [0]{URL }%
\providecommand \Eprint [0]{\href }%
\providecommand \doibase [0]{http://dx.doi.org/}%
\providecommand \selectlanguage [0]{\@gobble}%
\providecommand \bibinfo  [0]{\@secondoftwo}%
\providecommand \bibfield  [0]{\@secondoftwo}%
\providecommand \translation [1]{[#1]}%
\providecommand \BibitemOpen [0]{}%
\providecommand \bibitemStop [0]{}%
\providecommand \bibitemNoStop [0]{.\EOS\space}%
\providecommand \EOS [0]{\spacefactor3000\relax}%
\providecommand \BibitemShut  [1]{\csname bibitem#1\endcsname}%
\let\auto@bib@innerbib\@empty
\bibitem [{\citenamefont {Bauer}\ \emph {et~al.}(2000)\citenamefont {Bauer},
  \citenamefont {Fleming},\ and\ \citenamefont {Luke}}]{Bauer:2000ew}%
  \BibitemOpen
  \bibfield  {author} {\bibinfo {author} {\bibfnamefont {C.~W.}\ \bibnamefont
  {Bauer}}, \bibinfo {author} {\bibfnamefont {S.}~\bibnamefont {Fleming}}, \
  and\ \bibinfo {author} {\bibfnamefont {M.~E.}\ \bibnamefont {Luke}},\ }\href
  {\doibase 10.1103/PhysRevD.63.014006} {\bibfield  {journal} {\bibinfo
  {journal} {Phys. Rev.}\ }\textbf {\bibinfo {volume} {D63}},\ \bibinfo {pages}
  {014006} (\bibinfo {year} {2000})},\ \Eprint
  {http://arxiv.org/abs/hep-ph/0005275} {hep-ph/0005275} \BibitemShut {NoStop}%
\bibitem [{\citenamefont {Bauer}\ \emph {et~al.}(2001)\citenamefont {Bauer},
  \citenamefont {Fleming}, \citenamefont {Pirjol},\ and\ \citenamefont
  {Stewart}}]{Bauer:2000yr}%
  \BibitemOpen
  \bibfield  {author} {\bibinfo {author} {\bibfnamefont {C.~W.}\ \bibnamefont
  {Bauer}}, \bibinfo {author} {\bibfnamefont {S.}~\bibnamefont {Fleming}},
  \bibinfo {author} {\bibfnamefont {D.}~\bibnamefont {Pirjol}}, \ and\ \bibinfo
  {author} {\bibfnamefont {I.~W.}\ \bibnamefont {Stewart}},\ }\href {\doibase
  10.1103/PhysRevD.63.114020} {\bibfield  {journal} {\bibinfo  {journal} {Phys.
  Rev.}\ }\textbf {\bibinfo {volume} {D63}},\ \bibinfo {pages} {114020}
  (\bibinfo {year} {2001})},\ \Eprint {http://arxiv.org/abs/hep-ph/0011336}
  {hep-ph/0011336} \BibitemShut {NoStop}%
\bibitem [{\citenamefont {Bauer}\ and\ \citenamefont
  {Stewart}(2001)}]{Bauer:2001ct}%
  \BibitemOpen
  \bibfield  {author} {\bibinfo {author} {\bibfnamefont {C.~W.}\ \bibnamefont
  {Bauer}}\ and\ \bibinfo {author} {\bibfnamefont {I.~W.}\ \bibnamefont
  {Stewart}},\ }\href {\doibase 10.1016/S0370-2693(01)00902-9} {\bibfield
  {journal} {\bibinfo  {journal} {Phys. Lett.}\ }\textbf {\bibinfo {volume}
  {B516}},\ \bibinfo {pages} {134} (\bibinfo {year} {2001})},\ \Eprint
  {http://arxiv.org/abs/hep-ph/0107001} {hep-ph/0107001} \BibitemShut {NoStop}%
\bibitem [{\citenamefont {Bauer}\ \emph {et~al.}(2002)\citenamefont {Bauer},
  \citenamefont {Pirjol},\ and\ \citenamefont {Stewart}}]{Bauer:2001yt}%
  \BibitemOpen
  \bibfield  {author} {\bibinfo {author} {\bibfnamefont {C.~W.}\ \bibnamefont
  {Bauer}}, \bibinfo {author} {\bibfnamefont {D.}~\bibnamefont {Pirjol}}, \
  and\ \bibinfo {author} {\bibfnamefont {I.~W.}\ \bibnamefont {Stewart}},\
  }\href {\doibase 10.1103/PhysRevD.65.054022} {\bibfield  {journal} {\bibinfo
  {journal} {Phys. Rev.}\ }\textbf {\bibinfo {volume} {D65}},\ \bibinfo {pages}
  {054022} (\bibinfo {year} {2002})},\ \Eprint
  {http://arxiv.org/abs/hep-ph/0109045} {hep-ph/0109045} \BibitemShut {NoStop}%
\bibitem [{\citenamefont {Abbate}\ \emph {et~al.}(2011)\citenamefont {Abbate},
  \citenamefont {Fickinger}, \citenamefont {Hoang}, \citenamefont {Mateu},\
  and\ \citenamefont {Stewart}}]{Abbate:2010xh}%
  \BibitemOpen
  \bibfield  {author} {\bibinfo {author} {\bibfnamefont {R.}~\bibnamefont
  {Abbate}}, \bibinfo {author} {\bibfnamefont {M.}~\bibnamefont {Fickinger}},
  \bibinfo {author} {\bibfnamefont {A.~H.}\ \bibnamefont {Hoang}}, \bibinfo
  {author} {\bibfnamefont {V.}~\bibnamefont {Mateu}}, \ and\ \bibinfo {author}
  {\bibfnamefont {I.~W.}\ \bibnamefont {Stewart}},\ }\href {\doibase
  10.1103/PhysRevD.83.074021} {\bibfield  {journal} {\bibinfo  {journal} {Phys.
  Rev.}\ }\textbf {\bibinfo {volume} {D83}},\ \bibinfo {pages} {074021}
  (\bibinfo {year} {2011})},\ \Eprint {http://arxiv.org/abs/1006.3080}
  {arXiv:1006.3080} \BibitemShut {NoStop}%
\bibitem [{\citenamefont {Becher}\ and\ \citenamefont
  {Bell}(2012{\natexlab{a}})}]{Becher:2012qc}%
  \BibitemOpen
  \bibfield  {author} {\bibinfo {author} {\bibfnamefont {T.}~\bibnamefont
  {Becher}}\ and\ \bibinfo {author} {\bibfnamefont {G.}~\bibnamefont {Bell}},\
  }\href {\doibase 10.1007/JHEP11(2012)126} {\bibfield  {journal} {\bibinfo
  {journal} {JHEP}\ }\textbf {\bibinfo {volume} {1211}},\ \bibinfo {pages}
  {126} (\bibinfo {year} {2012}{\natexlab{a}})},\ \Eprint
  {http://arxiv.org/abs/1210.0580} {arXiv:1210.0580} \BibitemShut {NoStop}%
\bibitem [{\citenamefont {Stewart}\ \emph
  {et~al.}(2014{\natexlab{a}})\citenamefont {Stewart}, \citenamefont
  {Tackmann},\ and\ \citenamefont {Waalewijn}}]{Stewart:2014nna}%
  \BibitemOpen
  \bibfield  {author} {\bibinfo {author} {\bibfnamefont {I.~W.}\ \bibnamefont
  {Stewart}}, \bibinfo {author} {\bibfnamefont {F.~J.}\ \bibnamefont
  {Tackmann}}, \ and\ \bibinfo {author} {\bibfnamefont {W.~J.}\ \bibnamefont
  {Waalewijn}},\ }\href@noop {} {\  (\bibinfo {year} {2014}{\natexlab{a}})},\
  \Eprint {http://arxiv.org/abs/1405.6722} {arXiv:1405.6722} \BibitemShut
  {NoStop}%
\bibitem [{\citenamefont {Stewart}\ \emph
  {et~al.}(2010{\natexlab{a}})\citenamefont {Stewart}, \citenamefont
  {Tackmann},\ and\ \citenamefont {Waalewijn}}]{Stewart:2010tn}%
  \BibitemOpen
  \bibfield  {author} {\bibinfo {author} {\bibfnamefont {I.~W.}\ \bibnamefont
  {Stewart}}, \bibinfo {author} {\bibfnamefont {F.~J.}\ \bibnamefont
  {Tackmann}}, \ and\ \bibinfo {author} {\bibfnamefont {W.~J.}\ \bibnamefont
  {Waalewijn}},\ }\href {\doibase 10.1103/PhysRevLett.105.092002} {\bibfield
  {journal} {\bibinfo  {journal} {Phys. Rev. Lett.}\ }\textbf {\bibinfo
  {volume} {105}},\ \bibinfo {pages} {092002} (\bibinfo {year}
  {2010}{\natexlab{a}})},\ \Eprint {http://arxiv.org/abs/1004.2489}
  {arXiv:1004.2489} \BibitemShut {NoStop}%
\bibitem [{\citenamefont {Berends}\ and\ \citenamefont
  {Giele}(1988)}]{Berends:1987me}%
  \BibitemOpen
  \bibfield  {author} {\bibinfo {author} {\bibfnamefont {F.~A.}\ \bibnamefont
  {Berends}}\ and\ \bibinfo {author} {\bibfnamefont {W.}~\bibnamefont
  {Giele}},\ }\href {\doibase 10.1016/0550-3213(88)90442-7} {\bibfield
  {journal} {\bibinfo  {journal} {Nucl. Phys.}\ }\textbf {\bibinfo {volume}
  {B306}},\ \bibinfo {pages} {759} (\bibinfo {year} {1988})}\BibitemShut
  {NoStop}%
\bibitem [{\citenamefont {Mangano}\ and\ \citenamefont
  {Parke}(1991)}]{Mangano:1990by}%
  \BibitemOpen
  \bibfield  {author} {\bibinfo {author} {\bibfnamefont {M.~L.}\ \bibnamefont
  {Mangano}}\ and\ \bibinfo {author} {\bibfnamefont {S.~J.}\ \bibnamefont
  {Parke}},\ }\href {\doibase 10.1016/0370-1573(91)90091-Y} {\bibfield
  {journal} {\bibinfo  {journal} {Phys. Rept.}\ }\textbf {\bibinfo {volume}
  {200}},\ \bibinfo {pages} {301} (\bibinfo {year} {1991})},\ \Eprint
  {http://arxiv.org/abs/hep-th/0509223} {hep-th/0509223} \BibitemShut {NoStop}%
\bibitem [{\citenamefont {Kosower}(1999)}]{Kosower:1999xi}%
  \BibitemOpen
  \bibfield  {author} {\bibinfo {author} {\bibfnamefont {D.~A.}\ \bibnamefont
  {Kosower}},\ }\href {\doibase 10.1016/S0550-3213(99)00251-5} {\bibfield
  {journal} {\bibinfo  {journal} {Nucl. Phys.}\ }\textbf {\bibinfo {volume}
  {B552}},\ \bibinfo {pages} {319} (\bibinfo {year} {1999})},\ \Eprint
  {http://arxiv.org/abs/hep-ph/9901201} {hep-ph/9901201} \BibitemShut {NoStop}%
\bibitem [{\citenamefont {Bern}\ \emph
  {et~al.}(1994{\natexlab{a}})\citenamefont {Bern}, \citenamefont {Chalmers},
  \citenamefont {Dixon},\ and\ \citenamefont {Kosower}}]{Bern:1993qk}%
  \BibitemOpen
  \bibfield  {author} {\bibinfo {author} {\bibfnamefont {Z.}~\bibnamefont
  {Bern}}, \bibinfo {author} {\bibfnamefont {G.}~\bibnamefont {Chalmers}},
  \bibinfo {author} {\bibfnamefont {L.~J.}\ \bibnamefont {Dixon}}, \ and\
  \bibinfo {author} {\bibfnamefont {D.~A.}\ \bibnamefont {Kosower}},\ }\href
  {\doibase 10.1103/PhysRevLett.72.2134} {\bibfield  {journal} {\bibinfo
  {journal} {Phys. Rev. Lett.}\ }\textbf {\bibinfo {volume} {72}},\ \bibinfo
  {pages} {2134} (\bibinfo {year} {1994}{\natexlab{a}})},\ \Eprint
  {http://arxiv.org/abs/hep-ph/9312333} {hep-ph/9312333} \BibitemShut {NoStop}%
\bibitem [{\citenamefont {Bern}\ \emph
  {et~al.}(1994{\natexlab{b}})\citenamefont {Bern}, \citenamefont {Dixon},
  \citenamefont {Dunbar},\ and\ \citenamefont {Kosower}}]{Bern:1994zx}%
  \BibitemOpen
  \bibfield  {author} {\bibinfo {author} {\bibfnamefont {Z.}~\bibnamefont
  {Bern}}, \bibinfo {author} {\bibfnamefont {L.~J.}\ \bibnamefont {Dixon}},
  \bibinfo {author} {\bibfnamefont {D.~C.}\ \bibnamefont {Dunbar}}, \ and\
  \bibinfo {author} {\bibfnamefont {D.~A.}\ \bibnamefont {Kosower}},\ }\href
  {\doibase 10.1016/0550-3213(94)90179-1} {\bibfield  {journal} {\bibinfo
  {journal} {Nucl. Phys.}\ }\textbf {\bibinfo {volume} {B425}},\ \bibinfo
  {pages} {217} (\bibinfo {year} {1994}{\natexlab{b}})},\ \Eprint
  {http://arxiv.org/abs/hep-ph/9403226} {hep-ph/9403226} \BibitemShut {NoStop}%
\bibitem [{\citenamefont {Bern}\ and\ \citenamefont
  {Chalmers}(1995)}]{Bern:1995ix}%
  \BibitemOpen
  \bibfield  {author} {\bibinfo {author} {\bibfnamefont {Z.}~\bibnamefont
  {Bern}}\ and\ \bibinfo {author} {\bibfnamefont {G.}~\bibnamefont
  {Chalmers}},\ }\href {\doibase 10.1016/0550-3213(95)00226-I} {\bibfield
  {journal} {\bibinfo  {journal} {Nucl. Phys.}\ }\textbf {\bibinfo {volume}
  {B447}},\ \bibinfo {pages} {465} (\bibinfo {year} {1995})},\ \Eprint
  {http://arxiv.org/abs/hep-ph/9503236} {hep-ph/9503236} \BibitemShut {NoStop}%
\bibitem [{\citenamefont {Bern}\ \emph {et~al.}(1998)\citenamefont {Bern},
  \citenamefont {Del~Duca},\ and\ \citenamefont {Schmidt}}]{Bern:1998sc}%
  \BibitemOpen
  \bibfield  {author} {\bibinfo {author} {\bibfnamefont {Z.}~\bibnamefont
  {Bern}}, \bibinfo {author} {\bibfnamefont {V.}~\bibnamefont {Del~Duca}}, \
  and\ \bibinfo {author} {\bibfnamefont {C.~R.}\ \bibnamefont {Schmidt}},\
  }\href {\doibase 10.1016/S0370-2693(98)01495-6} {\bibfield  {journal}
  {\bibinfo  {journal} {Phys. Lett.}\ }\textbf {\bibinfo {volume} {B445}},\
  \bibinfo {pages} {168} (\bibinfo {year} {1998})},\ \Eprint
  {http://arxiv.org/abs/hep-ph/9810409} {hep-ph/9810409} \BibitemShut {NoStop}%
\bibitem [{\citenamefont {Kosower}\ and\ \citenamefont
  {Uwer}(1999)}]{Kosower:1999rx}%
  \BibitemOpen
  \bibfield  {author} {\bibinfo {author} {\bibfnamefont {D.~A.}\ \bibnamefont
  {Kosower}}\ and\ \bibinfo {author} {\bibfnamefont {P.}~\bibnamefont {Uwer}},\
  }\href {\doibase 10.1016/S0550-3213(99)00583-0} {\bibfield  {journal}
  {\bibinfo  {journal} {Nucl. Phys.}\ }\textbf {\bibinfo {volume} {B563}},\
  \bibinfo {pages} {477} (\bibinfo {year} {1999})},\ \Eprint
  {http://arxiv.org/abs/hep-ph/9903515} {hep-ph/9903515} \BibitemShut {NoStop}%
\bibitem [{\citenamefont {Bern}\ \emph {et~al.}(1999)\citenamefont {Bern},
  \citenamefont {Del~Duca}, \citenamefont {Kilgore},\ and\ \citenamefont
  {Schmidt}}]{Bern:1999ry}%
  \BibitemOpen
  \bibfield  {author} {\bibinfo {author} {\bibfnamefont {Z.}~\bibnamefont
  {Bern}}, \bibinfo {author} {\bibfnamefont {V.}~\bibnamefont {Del~Duca}},
  \bibinfo {author} {\bibfnamefont {W.~B.}\ \bibnamefont {Kilgore}}, \ and\
  \bibinfo {author} {\bibfnamefont {C.~R.}\ \bibnamefont {Schmidt}},\ }\href
  {\doibase 10.1103/PhysRevD.60.116001} {\bibfield  {journal} {\bibinfo
  {journal} {Phys. Rev.}\ }\textbf {\bibinfo {volume} {D60}},\ \bibinfo {pages}
  {116001} (\bibinfo {year} {1999})},\ \Eprint
  {http://arxiv.org/abs/hep-ph/9903516} {hep-ph/9903516} \BibitemShut {NoStop}%
\bibitem [{\citenamefont {Sborlini}\ \emph {et~al.}(2014)\citenamefont
  {Sborlini}, \citenamefont {de~Florian},\ and\ \citenamefont
  {Rodrigo}}]{Sborlini:2013jba}%
  \BibitemOpen
  \bibfield  {author} {\bibinfo {author} {\bibfnamefont {G.~F.}\ \bibnamefont
  {Sborlini}}, \bibinfo {author} {\bibfnamefont {D.}~\bibnamefont
  {de~Florian}}, \ and\ \bibinfo {author} {\bibfnamefont {G.}~\bibnamefont
  {Rodrigo}},\ }\href {\doibase 10.1007/JHEP01(2014)018} {\bibfield  {journal}
  {\bibinfo  {journal} {JHEP}\ }\textbf {\bibinfo {volume} {1401}},\ \bibinfo
  {pages} {018} (\bibinfo {year} {2014})},\ \Eprint
  {http://arxiv.org/abs/1310.6841} {arXiv:1310.6841} \BibitemShut {NoStop}%
\bibitem [{\citenamefont {Campbell}\ and\ \citenamefont
  {Glover}(1998)}]{Campbell:1997hg}%
  \BibitemOpen
  \bibfield  {author} {\bibinfo {author} {\bibfnamefont {J.~M.}\ \bibnamefont
  {Campbell}}\ and\ \bibinfo {author} {\bibfnamefont {E.~N.}\ \bibnamefont
  {Glover}},\ }\href {\doibase 10.1016/S0550-3213(98)00295-8} {\bibfield
  {journal} {\bibinfo  {journal} {Nucl. Phys.}\ }\textbf {\bibinfo {volume}
  {B527}},\ \bibinfo {pages} {264} (\bibinfo {year} {1998})},\ \Eprint
  {http://arxiv.org/abs/hep-ph/9710255} {hep-ph/9710255} \BibitemShut {NoStop}%
\bibitem [{\citenamefont {Catani}\ and\ \citenamefont
  {Grazzini}(1999)}]{Catani:1998nv}%
  \BibitemOpen
  \bibfield  {author} {\bibinfo {author} {\bibfnamefont {S.}~\bibnamefont
  {Catani}}\ and\ \bibinfo {author} {\bibfnamefont {M.}~\bibnamefont
  {Grazzini}},\ }\href {\doibase 10.1016/S0370-2693(98)01513-5} {\bibfield
  {journal} {\bibinfo  {journal} {Phys. Lett.}\ }\textbf {\bibinfo {volume}
  {B446}},\ \bibinfo {pages} {143} (\bibinfo {year} {1999})},\ \Eprint
  {http://arxiv.org/abs/hep-ph/9810389} {hep-ph/9810389} \BibitemShut {NoStop}%
\bibitem [{\citenamefont {Procura}\ and\ \citenamefont
  {Stewart}(2010)}]{Procura:2009vm}%
  \BibitemOpen
  \bibfield  {author} {\bibinfo {author} {\bibfnamefont {M.}~\bibnamefont
  {Procura}}\ and\ \bibinfo {author} {\bibfnamefont {I.~W.}\ \bibnamefont
  {Stewart}},\ }\href {\doibase 10.1103/PhysRevD.81.074009,
  10.1103/PhysRevD.83.039902} {\bibfield  {journal} {\bibinfo  {journal} {Phys.
  Rev.}\ }\textbf {\bibinfo {volume} {D81}},\ \bibinfo {pages} {074009}
  (\bibinfo {year} {2010})},\ \Eprint {http://arxiv.org/abs/0911.4980}
  {arXiv:0911.4980} \BibitemShut {NoStop}%
\bibitem [{\citenamefont {Bauer}\ and\ \citenamefont
  {Manohar}(2004)}]{Bauer:2003pi}%
  \BibitemOpen
  \bibfield  {author} {\bibinfo {author} {\bibfnamefont {C.~W.}\ \bibnamefont
  {Bauer}}\ and\ \bibinfo {author} {\bibfnamefont {A.~V.}\ \bibnamefont
  {Manohar}},\ }\href {\doibase 10.1103/PhysRevD.70.034024} {\bibfield
  {journal} {\bibinfo  {journal} {Phys. Rev.}\ }\textbf {\bibinfo {volume}
  {D70}},\ \bibinfo {pages} {034024} (\bibinfo {year} {2004})},\ \Eprint
  {http://arxiv.org/abs/hep-ph/0312109} {hep-ph/0312109} \BibitemShut {NoStop}%
\bibitem [{\citenamefont {Fleming}\ \emph {et~al.}(2003)\citenamefont
  {Fleming}, \citenamefont {Leibovich},\ and\ \citenamefont
  {Mehen}}]{Fleming:2003gt}%
  \BibitemOpen
  \bibfield  {author} {\bibinfo {author} {\bibfnamefont {S.}~\bibnamefont
  {Fleming}}, \bibinfo {author} {\bibfnamefont {A.~K.}\ \bibnamefont
  {Leibovich}}, \ and\ \bibinfo {author} {\bibfnamefont {T.}~\bibnamefont
  {Mehen}},\ }\href {\doibase 10.1103/PhysRevD.68.094011} {\bibfield  {journal}
  {\bibinfo  {journal} {Phys. Rev.}\ }\textbf {\bibinfo {volume} {D68}},\
  \bibinfo {pages} {094011} (\bibinfo {year} {2003})},\ \Eprint
  {http://arxiv.org/abs/hep-ph/0306139} {hep-ph/0306139} \BibitemShut {NoStop}%
\bibitem [{\citenamefont {Becher}\ and\ \citenamefont
  {Neubert}(2006)}]{Becher:2006qw}%
  \BibitemOpen
  \bibfield  {author} {\bibinfo {author} {\bibfnamefont {T.}~\bibnamefont
  {Becher}}\ and\ \bibinfo {author} {\bibfnamefont {M.}~\bibnamefont
  {Neubert}},\ }\href {\doibase 10.1016/j.physletb.2006.04.046} {\bibfield
  {journal} {\bibinfo  {journal} {Phys. Lett.}\ }\textbf {\bibinfo {volume}
  {B637}},\ \bibinfo {pages} {251} (\bibinfo {year} {2006})},\ \Eprint
  {http://arxiv.org/abs/hep-ph/0603140} {hep-ph/0603140} \BibitemShut {NoStop}%
\bibitem [{\citenamefont {Jain}\ \emph {et~al.}(2011)\citenamefont {Jain},
  \citenamefont {Procura},\ and\ \citenamefont {Waalewijn}}]{Jain:2011xz}%
  \BibitemOpen
  \bibfield  {author} {\bibinfo {author} {\bibfnamefont {A.}~\bibnamefont
  {Jain}}, \bibinfo {author} {\bibfnamefont {M.}~\bibnamefont {Procura}}, \
  and\ \bibinfo {author} {\bibfnamefont {W.~J.}\ \bibnamefont {Waalewijn}},\
  }\href {\doibase 10.1007/JHEP05(2011)035} {\bibfield  {journal} {\bibinfo
  {journal} {JHEP}\ }\textbf {\bibinfo {volume} {1105}},\ \bibinfo {pages}
  {035} (\bibinfo {year} {2011})},\ \Eprint {http://arxiv.org/abs/1101.4953}
  {arXiv:1101.4953} \BibitemShut {NoStop}%
\bibitem [{\citenamefont {Bauer}\ and\ \citenamefont
  {Mereghetti}(2014)}]{Bauer:2013bza}%
  \BibitemOpen
  \bibfield  {author} {\bibinfo {author} {\bibfnamefont {C.~W.}\ \bibnamefont
  {Bauer}}\ and\ \bibinfo {author} {\bibfnamefont {E.}~\bibnamefont
  {Mereghetti}},\ }\href {\doibase 10.1007/JHEP04(2014)051} {\bibfield
  {journal} {\bibinfo  {journal} {JHEP}\ }\textbf {\bibinfo {volume} {1404}},\
  \bibinfo {pages} {051} (\bibinfo {year} {2014})},\ \Eprint
  {http://arxiv.org/abs/1312.5605} {arXiv:1312.5605} \BibitemShut {NoStop}%
\bibitem [{\citenamefont {Becher}\ and\ \citenamefont
  {Neubert}(2011)}]{Becher:2010tm}%
  \BibitemOpen
  \bibfield  {author} {\bibinfo {author} {\bibfnamefont {T.}~\bibnamefont
  {Becher}}\ and\ \bibinfo {author} {\bibfnamefont {M.}~\bibnamefont
  {Neubert}},\ }\href {\doibase 10.1140/epjc/s10052-011-1665-7} {\bibfield
  {journal} {\bibinfo  {journal} {Eur. Phys. J.}\ }\textbf {\bibinfo {volume}
  {C71}},\ \bibinfo {pages} {1665} (\bibinfo {year} {2011})},\ \Eprint
  {http://arxiv.org/abs/1007.4005} {arXiv:1007.4005} \BibitemShut {NoStop}%
\bibitem [{\citenamefont {Echevarria}\ \emph {et~al.}(2012)\citenamefont
  {Echevarria}, \citenamefont {Idilbi},\ and\ \citenamefont
  {Scimemi}}]{GarciaEchevarria:2011rb}%
  \BibitemOpen
  \bibfield  {author} {\bibinfo {author} {\bibfnamefont {M.~G.}\ \bibnamefont
  {Echevarria}}, \bibinfo {author} {\bibfnamefont {A.}~\bibnamefont {Idilbi}},
  \ and\ \bibinfo {author} {\bibfnamefont {I.}~\bibnamefont {Scimemi}},\ }\href
  {\doibase 10.1007/JHEP07(2012)002} {\bibfield  {journal} {\bibinfo  {journal}
  {JHEP}\ }\textbf {\bibinfo {volume} {1207}},\ \bibinfo {pages} {002}
  (\bibinfo {year} {2012})},\ \Eprint {http://arxiv.org/abs/1111.4996}
  {arXiv:1111.4996} \BibitemShut {NoStop}%
\bibitem [{\citenamefont {Chiu}\ \emph
  {et~al.}(2012{\natexlab{a}})\citenamefont {Chiu}, \citenamefont {Jain},
  \citenamefont {Neill},\ and\ \citenamefont {Rothstein}}]{Chiu:2012ir}%
  \BibitemOpen
  \bibfield  {author} {\bibinfo {author} {\bibfnamefont {J.-Y.}\ \bibnamefont
  {Chiu}}, \bibinfo {author} {\bibfnamefont {A.}~\bibnamefont {Jain}}, \bibinfo
  {author} {\bibfnamefont {D.}~\bibnamefont {Neill}}, \ and\ \bibinfo {author}
  {\bibfnamefont {I.~Z.}\ \bibnamefont {Rothstein}},\ }\href {\doibase
  10.1007/JHEP05(2012)084} {\bibfield  {journal} {\bibinfo  {journal} {JHEP}\
  }\textbf {\bibinfo {volume} {1205}},\ \bibinfo {pages} {084} (\bibinfo {year}
  {2012}{\natexlab{a}})},\ \Eprint {http://arxiv.org/abs/1202.0814}
  {arXiv:1202.0814} \BibitemShut {NoStop}%
\bibitem [{\citenamefont {Collins}\ \emph {et~al.}(1985)\citenamefont
  {Collins}, \citenamefont {Soper},\ and\ \citenamefont
  {Sterman}}]{Collins:1984kg}%
  \BibitemOpen
  \bibfield  {author} {\bibinfo {author} {\bibfnamefont {J.~C.}\ \bibnamefont
  {Collins}}, \bibinfo {author} {\bibfnamefont {D.~E.}\ \bibnamefont {Soper}},
  \ and\ \bibinfo {author} {\bibfnamefont {G.~F.}\ \bibnamefont {Sterman}},\
  }\href {\doibase 10.1016/0550-3213(85)90479-1} {\bibfield  {journal}
  {\bibinfo  {journal} {Nucl. Phys.}\ }\textbf {\bibinfo {volume} {B250}},\
  \bibinfo {pages} {199} (\bibinfo {year} {1985})}\BibitemShut {NoStop}%
\bibitem [{\citenamefont {Collins}(2011)}]{Collins:2011zzd}%
  \BibitemOpen
  \bibfield  {author} {\bibinfo {author} {\bibfnamefont {J.}~\bibnamefont
  {Collins}},\ }\href@noop {} {\emph {\bibinfo {title} {{Foundations of
  Perturbative QCD}}}}\ (\bibinfo  {publisher} {Cambridge University Press},\
  \bibinfo {address} {Cambridge},\ \bibinfo {year} {2011})\BibitemShut
  {NoStop}%
\bibitem [{\citenamefont {Gehrmann}\ \emph {et~al.}(2012)\citenamefont
  {Gehrmann}, \citenamefont {Lubbert},\ and\ \citenamefont
  {Yang}}]{Gehrmann:2012ze}%
  \BibitemOpen
  \bibfield  {author} {\bibinfo {author} {\bibfnamefont {T.}~\bibnamefont
  {Gehrmann}}, \bibinfo {author} {\bibfnamefont {T.}~\bibnamefont {Lubbert}}, \
  and\ \bibinfo {author} {\bibfnamefont {L.~L.}\ \bibnamefont {Yang}},\ }\href
  {\doibase 10.1103/PhysRevLett.109.242003} {\bibfield  {journal} {\bibinfo
  {journal} {Phys. Rev. Lett.}\ }\textbf {\bibinfo {volume} {109}},\ \bibinfo
  {pages} {242003} (\bibinfo {year} {2012})},\ \Eprint
  {http://arxiv.org/abs/1209.0682} {arXiv:1209.0682} \BibitemShut {NoStop}%
\bibitem [{\citenamefont {Gehrmann}\ \emph {et~al.}(2014)\citenamefont
  {Gehrmann}, \citenamefont {Luebbert},\ and\ \citenamefont
  {Yang}}]{Gehrmann:2014yya}%
  \BibitemOpen
  \bibfield  {author} {\bibinfo {author} {\bibfnamefont {T.}~\bibnamefont
  {Gehrmann}}, \bibinfo {author} {\bibfnamefont {T.}~\bibnamefont {Luebbert}},
  \ and\ \bibinfo {author} {\bibfnamefont {L.~L.}\ \bibnamefont {Yang}},\
  }\href@noop {} {\  (\bibinfo {year} {2014})},\ \Eprint
  {http://arxiv.org/abs/1403.6451} {arXiv:1403.6451} \BibitemShut {NoStop}%
\bibitem [{\citenamefont {Fleming}\ \emph {et~al.}(2006)\citenamefont
  {Fleming}, \citenamefont {Leibovich},\ and\ \citenamefont
  {Mehen}}]{Fleming:2006cd}%
  \BibitemOpen
  \bibfield  {author} {\bibinfo {author} {\bibfnamefont {S.}~\bibnamefont
  {Fleming}}, \bibinfo {author} {\bibfnamefont {A.~K.}\ \bibnamefont
  {Leibovich}}, \ and\ \bibinfo {author} {\bibfnamefont {T.}~\bibnamefont
  {Mehen}},\ }\href {\doibase 10.1103/PhysRevD.74.114004} {\bibfield  {journal}
  {\bibinfo  {journal} {Phys. Rev.}\ }\textbf {\bibinfo {volume} {D74}},\
  \bibinfo {pages} {114004} (\bibinfo {year} {2006})},\ \Eprint
  {http://arxiv.org/abs/hep-ph/0607121} {hep-ph/0607121} \BibitemShut {NoStop}%
\bibitem [{\citenamefont {Stewart}\ \emph
  {et~al.}(2010{\natexlab{b}})\citenamefont {Stewart}, \citenamefont
  {Tackmann},\ and\ \citenamefont {Waalewijn}}]{Stewart:2009yx}%
  \BibitemOpen
  \bibfield  {author} {\bibinfo {author} {\bibfnamefont {I.~W.}\ \bibnamefont
  {Stewart}}, \bibinfo {author} {\bibfnamefont {F.~J.}\ \bibnamefont
  {Tackmann}}, \ and\ \bibinfo {author} {\bibfnamefont {W.~J.}\ \bibnamefont
  {Waalewijn}},\ }\href {\doibase 10.1103/PhysRevD.81.094035} {\bibfield
  {journal} {\bibinfo  {journal} {Phys. Rev.}\ }\textbf {\bibinfo {volume}
  {D81}},\ \bibinfo {pages} {094035} (\bibinfo {year} {2010}{\natexlab{b}})},\
  \Eprint {http://arxiv.org/abs/0910.0467} {arXiv:0910.0467} \BibitemShut
  {NoStop}%
\bibitem [{\citenamefont {Stewart}\ \emph
  {et~al.}(2010{\natexlab{c}})\citenamefont {Stewart}, \citenamefont
  {Tackmann},\ and\ \citenamefont {Waalewijn}}]{Stewart:2010qs}%
  \BibitemOpen
  \bibfield  {author} {\bibinfo {author} {\bibfnamefont {I.~W.}\ \bibnamefont
  {Stewart}}, \bibinfo {author} {\bibfnamefont {F.~J.}\ \bibnamefont
  {Tackmann}}, \ and\ \bibinfo {author} {\bibfnamefont {W.~J.}\ \bibnamefont
  {Waalewijn}},\ }\href {\doibase 10.1007/JHEP09(2010)005} {\bibfield
  {journal} {\bibinfo  {journal} {JHEP}\ }\textbf {\bibinfo {volume} {1009}},\
  \bibinfo {pages} {005} (\bibinfo {year} {2010}{\natexlab{c}})},\ \Eprint
  {http://arxiv.org/abs/1002.2213} {arXiv:1002.2213} \BibitemShut {NoStop}%
\bibitem [{\citenamefont {Gaunt}\ \emph
  {et~al.}(2014{\natexlab{a}})\citenamefont {Gaunt}, \citenamefont
  {Stahlhofen},\ and\ \citenamefont {Tackmann}}]{Gaunt:2014xga}%
  \BibitemOpen
  \bibfield  {author} {\bibinfo {author} {\bibfnamefont {J.~R.}\ \bibnamefont
  {Gaunt}}, \bibinfo {author} {\bibfnamefont {M.}~\bibnamefont {Stahlhofen}}, \
  and\ \bibinfo {author} {\bibfnamefont {F.~J.}\ \bibnamefont {Tackmann}},\
  }\href {\doibase 10.1007/JHEP04(2014)113} {\bibfield  {journal} {\bibinfo
  {journal} {JHEP}\ }\textbf {\bibinfo {volume} {1404}},\ \bibinfo {pages}
  {113} (\bibinfo {year} {2014}{\natexlab{a}})},\ \Eprint
  {http://arxiv.org/abs/1401.5478} {arXiv:1401.5478} \BibitemShut {NoStop}%
\bibitem [{\citenamefont {Gaunt}\ \emph
  {et~al.}(2014{\natexlab{b}})\citenamefont {Gaunt}, \citenamefont
  {Stahlhofen},\ and\ \citenamefont {Tackmann}}]{Gaunt:2014cfa}%
  \BibitemOpen
  \bibfield  {author} {\bibinfo {author} {\bibfnamefont {J.}~\bibnamefont
  {Gaunt}}, \bibinfo {author} {\bibfnamefont {M.}~\bibnamefont {Stahlhofen}}, \
  and\ \bibinfo {author} {\bibfnamefont {F.~J.}\ \bibnamefont {Tackmann}},\
  }\href@noop {} {\  (\bibinfo {year} {2014}{\natexlab{b}})},\ \Eprint
  {http://arxiv.org/abs/1405.1044} {arXiv:1405.1044} \BibitemShut {NoStop}%
\bibitem [{\citenamefont {Chiu}\ \emph
  {et~al.}(2012{\natexlab{b}})\citenamefont {Chiu}, \citenamefont {Jain},
  \citenamefont {Neill},\ and\ \citenamefont {Rothstein}}]{Chiu:2011qc}%
  \BibitemOpen
  \bibfield  {author} {\bibinfo {author} {\bibfnamefont {J.-y.}\ \bibnamefont
  {Chiu}}, \bibinfo {author} {\bibfnamefont {A.}~\bibnamefont {Jain}}, \bibinfo
  {author} {\bibfnamefont {D.}~\bibnamefont {Neill}}, \ and\ \bibinfo {author}
  {\bibfnamefont {I.~Z.}\ \bibnamefont {Rothstein}},\ }\href {\doibase
  10.1103/PhysRevLett.108.151601} {\bibfield  {journal} {\bibinfo  {journal}
  {Phys. Rev. Lett.}\ }\textbf {\bibinfo {volume} {108}},\ \bibinfo {pages}
  {151601} (\bibinfo {year} {2012}{\natexlab{b}})},\ \Eprint
  {http://arxiv.org/abs/1104.0881} {arXiv:1104.0881} \BibitemShut {NoStop}%
\bibitem [{\citenamefont {Procura}\ and\ \citenamefont
  {Waalewijn}(2012)}]{Procura:2011aq}%
  \BibitemOpen
  \bibfield  {author} {\bibinfo {author} {\bibfnamefont {M.}~\bibnamefont
  {Procura}}\ and\ \bibinfo {author} {\bibfnamefont {W.~J.}\ \bibnamefont
  {Waalewijn}},\ }\href {\doibase 10.1103/PhysRevD.85.114041} {\bibfield
  {journal} {\bibinfo  {journal} {Phys. Rev.}\ }\textbf {\bibinfo {volume}
  {D85}},\ \bibinfo {pages} {114041} (\bibinfo {year} {2012})},\ \Eprint
  {http://arxiv.org/abs/1111.6605} {arXiv:1111.6605} \BibitemShut {NoStop}%
\bibitem [{\citenamefont {Giele}\ and\ \citenamefont
  {Glover}(1992)}]{Giele:1991vf}%
  \BibitemOpen
  \bibfield  {author} {\bibinfo {author} {\bibfnamefont {W.}~\bibnamefont
  {Giele}}\ and\ \bibinfo {author} {\bibfnamefont {E.~N.}\ \bibnamefont
  {Glover}},\ }\href {\doibase 10.1103/PhysRevD.46.1980} {\bibfield  {journal}
  {\bibinfo  {journal} {Phys. Rev.}\ }\textbf {\bibinfo {volume} {D46}},\
  \bibinfo {pages} {1980} (\bibinfo {year} {1992})}\BibitemShut {NoStop}%
\bibitem [{\citenamefont {Gross}\ and\ \citenamefont
  {Wilczek}(1974)}]{Gross:1974cs}%
  \BibitemOpen
  \bibfield  {author} {\bibinfo {author} {\bibfnamefont {D.}~\bibnamefont
  {Gross}}\ and\ \bibinfo {author} {\bibfnamefont {F.}~\bibnamefont
  {Wilczek}},\ }\href {\doibase 10.1103/PhysRevD.9.980} {\bibfield  {journal}
  {\bibinfo  {journal} {Phys. Rev.}\ }\textbf {\bibinfo {volume} {D9}},\
  \bibinfo {pages} {980} (\bibinfo {year} {1974})}\BibitemShut {NoStop}%
\bibitem [{\citenamefont {Altarelli}\ and\ \citenamefont
  {Parisi}(1977)}]{Altarelli:1977zs}%
  \BibitemOpen
  \bibfield  {author} {\bibinfo {author} {\bibfnamefont {G.}~\bibnamefont
  {Altarelli}}\ and\ \bibinfo {author} {\bibfnamefont {G.}~\bibnamefont
  {Parisi}},\ }\href {\doibase 10.1016/0550-3213(77)90384-4} {\bibfield
  {journal} {\bibinfo  {journal} {Nucl. Phys.}\ }\textbf {\bibinfo {volume}
  {B126}},\ \bibinfo {pages} {298} (\bibinfo {year} {1977})}\BibitemShut
  {NoStop}%
\bibitem [{\citenamefont {Liu}(2011)}]{Liu:2010ng}%
  \BibitemOpen
  \bibfield  {author} {\bibinfo {author} {\bibfnamefont {X.}~\bibnamefont
  {Liu}},\ }\href {\doibase 10.1016/j.physletb.2011.03.055} {\bibfield
  {journal} {\bibinfo  {journal} {Phys. Lett.}\ }\textbf {\bibinfo {volume}
  {B699}},\ \bibinfo {pages} {87} (\bibinfo {year} {2011})},\ \Eprint
  {http://arxiv.org/abs/1011.3872} {arXiv:1011.3872} \BibitemShut {NoStop}%
\bibitem [{\citenamefont {Collins}\ and\ \citenamefont
  {Soper}(1982)}]{Collins:1981uw}%
  \BibitemOpen
  \bibfield  {author} {\bibinfo {author} {\bibfnamefont {J.~C.}\ \bibnamefont
  {Collins}}\ and\ \bibinfo {author} {\bibfnamefont {D.~E.}\ \bibnamefont
  {Soper}},\ }\href {\doibase 10.1016/0550-3213(82)90021-9} {\bibfield
  {journal} {\bibinfo  {journal} {Nucl. Phys.}\ }\textbf {\bibinfo {volume}
  {B194}},\ \bibinfo {pages} {445} (\bibinfo {year} {1982})}\BibitemShut
  {NoStop}%
\bibitem [{\citenamefont {Ji}\ \emph {et~al.}(2005)\citenamefont {Ji},
  \citenamefont {Ma},\ and\ \citenamefont {Yuan}}]{Ji:2004wu}%
  \BibitemOpen
  \bibfield  {author} {\bibinfo {author} {\bibfnamefont {X.-d.}\ \bibnamefont
  {Ji}}, \bibinfo {author} {\bibfnamefont {J.-p.}\ \bibnamefont {Ma}}, \ and\
  \bibinfo {author} {\bibfnamefont {F.}~\bibnamefont {Yuan}},\ }\href {\doibase
  10.1103/PhysRevD.71.034005} {\bibfield  {journal} {\bibinfo  {journal} {Phys.
  Rev.}\ }\textbf {\bibinfo {volume} {D71}},\ \bibinfo {pages} {034005}
  (\bibinfo {year} {2005})},\ \Eprint {http://arxiv.org/abs/hep-ph/0404183}
  {hep-ph/0404183} \BibitemShut {NoStop}%
\bibitem [{\citenamefont {Chiu}\ \emph {et~al.}(2009)\citenamefont {Chiu},
  \citenamefont {Fuhrer}, \citenamefont {Hoang}, \citenamefont {Kelley},\ and\
  \citenamefont {Manohar}}]{Chiu:2009yx}%
  \BibitemOpen
  \bibfield  {author} {\bibinfo {author} {\bibfnamefont {J.-y.}\ \bibnamefont
  {Chiu}}, \bibinfo {author} {\bibfnamefont {A.}~\bibnamefont {Fuhrer}},
  \bibinfo {author} {\bibfnamefont {A.~H.}\ \bibnamefont {Hoang}}, \bibinfo
  {author} {\bibfnamefont {R.}~\bibnamefont {Kelley}}, \ and\ \bibinfo {author}
  {\bibfnamefont {A.~V.}\ \bibnamefont {Manohar}},\ }\href {\doibase
  10.1103/PhysRevD.79.053007} {\bibfield  {journal} {\bibinfo  {journal} {Phys.
  Rev.}\ }\textbf {\bibinfo {volume} {D79}},\ \bibinfo {pages} {053007}
  (\bibinfo {year} {2009})},\ \Eprint {http://arxiv.org/abs/0901.1332}
  {arXiv:0901.1332} \BibitemShut {NoStop}%
\bibitem [{\citenamefont {Becher}\ and\ \citenamefont
  {Bell}(2012{\natexlab{b}})}]{Becher:2011dz}%
  \BibitemOpen
  \bibfield  {author} {\bibinfo {author} {\bibfnamefont {T.}~\bibnamefont
  {Becher}}\ and\ \bibinfo {author} {\bibfnamefont {G.}~\bibnamefont {Bell}},\
  }\href {\doibase 10.1016/j.physletb.2012.05.016} {\bibfield  {journal}
  {\bibinfo  {journal} {Phys. Lett.}\ }\textbf {\bibinfo {volume} {B713}},\
  \bibinfo {pages} {41} (\bibinfo {year} {2012}{\natexlab{b}})},\ \Eprint
  {http://arxiv.org/abs/1112.3907} {arXiv:1112.3907} \BibitemShut {NoStop}%
\bibitem [{\citenamefont {Gehrmann-De~Ridder}\ and\ \citenamefont
  {Glover}(1998)}]{GehrmannDeRidder:1997gf}%
  \BibitemOpen
  \bibfield  {author} {\bibinfo {author} {\bibfnamefont {A.}~\bibnamefont
  {Gehrmann-De~Ridder}}\ and\ \bibinfo {author} {\bibfnamefont {E.~N.}\
  \bibnamefont {Glover}},\ }\href {\doibase 10.1016/S0550-3213(97)00818-3}
  {\bibfield  {journal} {\bibinfo  {journal} {Nucl. Phys.}\ }\textbf {\bibinfo
  {volume} {B517}},\ \bibinfo {pages} {269} (\bibinfo {year} {1998})},\ \Eprint
  {http://arxiv.org/abs/hep-ph/9707224} {hep-ph/9707224} \BibitemShut {NoStop}%
\bibitem [{\citenamefont {Kosower}\ and\ \citenamefont
  {Uwer}(2003)}]{Kosower:2003np}%
  \BibitemOpen
  \bibfield  {author} {\bibinfo {author} {\bibfnamefont {D.~A.}\ \bibnamefont
  {Kosower}}\ and\ \bibinfo {author} {\bibfnamefont {P.}~\bibnamefont {Uwer}},\
  }\href {\doibase 10.1016/j.nuclphysb.2003.09.044} {\bibfield  {journal}
  {\bibinfo  {journal} {Nucl. Phys.}\ }\textbf {\bibinfo {volume} {B674}},\
  \bibinfo {pages} {365} (\bibinfo {year} {2003})},\ \Eprint
  {http://arxiv.org/abs/hep-ph/0307031} {hep-ph/0307031} \BibitemShut {NoStop}%
\bibitem [{\citenamefont {Ma\^itre}(2006)}]{Maitre:2005uu}%
  \BibitemOpen
  \bibfield  {author} {\bibinfo {author} {\bibfnamefont {D.}~\bibnamefont
  {Ma\^itre}},\ }\href {\doibase 10.1016/j.cpc.2005.10.008} {\bibfield
  {journal} {\bibinfo  {journal} {Comput. Phys. Commun.}\ }\textbf {\bibinfo
  {volume} {174}},\ \bibinfo {pages} {222} (\bibinfo {year} {2006})},\ \Eprint
  {http://arxiv.org/abs/hep-ph/0507152} {hep-ph/0507152} \BibitemShut {NoStop}%
\bibitem [{\citenamefont {Huber}\ and\ \citenamefont
  {Ma\^itre}(2006)}]{Huber:2005yg}%
  \BibitemOpen
  \bibfield  {author} {\bibinfo {author} {\bibfnamefont {T.}~\bibnamefont
  {Huber}}\ and\ \bibinfo {author} {\bibfnamefont {D.}~\bibnamefont
  {Ma\^itre}},\ }\href {\doibase 10.1016/j.cpc.2006.01.007} {\bibfield
  {journal} {\bibinfo  {journal} {Comput. Phys. Commun.}\ }\textbf {\bibinfo
  {volume} {175}},\ \bibinfo {pages} {122} (\bibinfo {year} {2006})},\ \Eprint
  {http://arxiv.org/abs/hep-ph/0507094} {hep-ph/0507094} \BibitemShut {NoStop}%
\bibitem [{\citenamefont {Anastasiou}\ and\ \citenamefont
  {Melnikov}(2002)}]{Anastasiou:2002yz}%
  \BibitemOpen
  \bibfield  {author} {\bibinfo {author} {\bibfnamefont {C.}~\bibnamefont
  {Anastasiou}}\ and\ \bibinfo {author} {\bibfnamefont {K.}~\bibnamefont
  {Melnikov}},\ }\href {\doibase 10.1016/S0550-3213(02)00837-4} {\bibfield
  {journal} {\bibinfo  {journal} {Nucl. Phys.}\ }\textbf {\bibinfo {volume}
  {B646}},\ \bibinfo {pages} {220} (\bibinfo {year} {2002})},\ \Eprint
  {http://arxiv.org/abs/hep-ph/0207004} {hep-ph/0207004} \BibitemShut {NoStop}%
\bibitem [{\citenamefont {Anastasiou}\ \emph
  {et~al.}(2003{\natexlab{a}})\citenamefont {Anastasiou}, \citenamefont
  {Dixon},\ and\ \citenamefont {Melnikov}}]{Anastasiou:2002qz}%
  \BibitemOpen
  \bibfield  {author} {\bibinfo {author} {\bibfnamefont {C.}~\bibnamefont
  {Anastasiou}}, \bibinfo {author} {\bibfnamefont {L.~J.}\ \bibnamefont
  {Dixon}}, \ and\ \bibinfo {author} {\bibfnamefont {K.}~\bibnamefont
  {Melnikov}},\ }\href {\doibase 10.1016/S0920-5632(03)80168-8} {\bibfield
  {journal} {\bibinfo  {journal} {Nucl. Phys. Proc. Suppl.}\ }\textbf {\bibinfo
  {volume} {116}},\ \bibinfo {pages} {193} (\bibinfo {year}
  {2003}{\natexlab{a}})},\ \Eprint {http://arxiv.org/abs/hep-ph/0211141}
  {hep-ph/0211141} \BibitemShut {NoStop}%
\bibitem [{\citenamefont {Anastasiou}\ \emph
  {et~al.}(2003{\natexlab{b}})\citenamefont {Anastasiou}, \citenamefont
  {Dixon}, \citenamefont {Melnikov},\ and\ \citenamefont
  {Petriello}}]{Anastasiou:2003yy}%
  \BibitemOpen
  \bibfield  {author} {\bibinfo {author} {\bibfnamefont {C.}~\bibnamefont
  {Anastasiou}}, \bibinfo {author} {\bibfnamefont {L.~J.}\ \bibnamefont
  {Dixon}}, \bibinfo {author} {\bibfnamefont {K.}~\bibnamefont {Melnikov}}, \
  and\ \bibinfo {author} {\bibfnamefont {F.}~\bibnamefont {Petriello}},\ }\href
  {\doibase 10.1103/PhysRevLett.91.182002} {\bibfield  {journal} {\bibinfo
  {journal} {Phys. Rev. Lett.}\ }\textbf {\bibinfo {volume} {91}},\ \bibinfo
  {pages} {182002} (\bibinfo {year} {2003}{\natexlab{b}})},\ \Eprint
  {http://arxiv.org/abs/hep-ph/0306192} {hep-ph/0306192} \BibitemShut {NoStop}%
\bibitem [{\citenamefont {Anastasiou}\ \emph {et~al.}(2004)\citenamefont
  {Anastasiou}, \citenamefont {Dixon}, \citenamefont {Melnikov},\ and\
  \citenamefont {Petriello}}]{Anastasiou:2003ds}%
  \BibitemOpen
  \bibfield  {author} {\bibinfo {author} {\bibfnamefont {C.}~\bibnamefont
  {Anastasiou}}, \bibinfo {author} {\bibfnamefont {L.~J.}\ \bibnamefont
  {Dixon}}, \bibinfo {author} {\bibfnamefont {K.}~\bibnamefont {Melnikov}}, \
  and\ \bibinfo {author} {\bibfnamefont {F.}~\bibnamefont {Petriello}},\ }\href
  {\doibase 10.1103/PhysRevD.69.094008} {\bibfield  {journal} {\bibinfo
  {journal} {Phys. Rev.}\ }\textbf {\bibinfo {volume} {D69}},\ \bibinfo {pages}
  {094008} (\bibinfo {year} {2004})},\ \Eprint
  {http://arxiv.org/abs/hep-ph/0312266} {hep-ph/0312266} \BibitemShut {NoStop}%
\bibitem [{\citenamefont {Smirnov}(2008)}]{Smirnov:2008iw}%
  \BibitemOpen
  \bibfield  {author} {\bibinfo {author} {\bibfnamefont {A.}~\bibnamefont
  {Smirnov}},\ }\href {\doibase 10.1088/1126-6708/2008/10/107} {\bibfield
  {journal} {\bibinfo  {journal} {JHEP}\ }\textbf {\bibinfo {volume} {0810}},\
  \bibinfo {pages} {107} (\bibinfo {year} {2008})},\ \Eprint
  {http://arxiv.org/abs/0807.3243} {arXiv:0807.3243} \BibitemShut {NoStop}%
\bibitem [{\citenamefont {Studerus}(2010)}]{Studerus:2009ye}%
  \BibitemOpen
  \bibfield  {author} {\bibinfo {author} {\bibfnamefont {C.}~\bibnamefont
  {Studerus}},\ }\href {\doibase 10.1016/j.cpc.2010.03.012} {\bibfield
  {journal} {\bibinfo  {journal} {Comput. Phys. Commun.}\ }\textbf {\bibinfo
  {volume} {181}},\ \bibinfo {pages} {1293} (\bibinfo {year} {2010})},\ \Eprint
  {http://arxiv.org/abs/0912.2546} {arXiv:0912.2546} \BibitemShut {NoStop}%
\bibitem [{\citenamefont {von Manteuffel}\ and\ \citenamefont
  {Studerus}(2012)}]{vonManteuffel:2012np}%
  \BibitemOpen
  \bibfield  {author} {\bibinfo {author} {\bibfnamefont {A.}~\bibnamefont {von
  Manteuffel}}\ and\ \bibinfo {author} {\bibfnamefont {C.}~\bibnamefont
  {Studerus}},\ }\href@noop {} {\  (\bibinfo {year} {2012})},\ \Eprint
  {http://arxiv.org/abs/1201.4330} {arXiv:1201.4330} \BibitemShut {NoStop}%
\bibitem [{\citenamefont {Kotikov}(1991{\natexlab{a}})}]{Kotikov:1990kg}%
  \BibitemOpen
  \bibfield  {author} {\bibinfo {author} {\bibfnamefont {A.}~\bibnamefont
  {Kotikov}},\ }\href {\doibase 10.1016/0370-2693(91)90413-K} {\bibfield
  {journal} {\bibinfo  {journal} {Phys. Lett.}\ }\textbf {\bibinfo {volume}
  {B254}},\ \bibinfo {pages} {158} (\bibinfo {year}
  {1991}{\natexlab{a}})}\BibitemShut {NoStop}%
\bibitem [{\citenamefont {Kotikov}(1991{\natexlab{b}})}]{Kotikov:1991hm}%
  \BibitemOpen
  \bibfield  {author} {\bibinfo {author} {\bibfnamefont {A.}~\bibnamefont
  {Kotikov}},\ }\href {\doibase 10.1016/0370-2693(91)90834-D} {\bibfield
  {journal} {\bibinfo  {journal} {Phys. Lett.}\ }\textbf {\bibinfo {volume}
  {B259}},\ \bibinfo {pages} {314} (\bibinfo {year}
  {1991}{\natexlab{b}})}\BibitemShut {NoStop}%
\bibitem [{\citenamefont {Kotikov}(1991{\natexlab{c}})}]{Kotikov:1991pm}%
  \BibitemOpen
  \bibfield  {author} {\bibinfo {author} {\bibfnamefont {A.}~\bibnamefont
  {Kotikov}},\ }\href {\doibase 10.1016/0370-2693(91)90536-Y} {\bibfield
  {journal} {\bibinfo  {journal} {Phys. Lett.}\ }\textbf {\bibinfo {volume}
  {B267}},\ \bibinfo {pages} {123} (\bibinfo {year}
  {1991}{\natexlab{c}})}\BibitemShut {NoStop}%
\bibitem [{\citenamefont {Remiddi}(1997)}]{Remiddi:1997ny}%
  \BibitemOpen
  \bibfield  {author} {\bibinfo {author} {\bibfnamefont {E.}~\bibnamefont
  {Remiddi}},\ }\href@noop {} {\bibfield  {journal} {\bibinfo  {journal} {Nuovo
  Cim.}\ }\textbf {\bibinfo {volume} {A110}},\ \bibinfo {pages} {1435}
  (\bibinfo {year} {1997})},\ \Eprint {http://arxiv.org/abs/hep-th/9711188}
  {hep-th/9711188} \BibitemShut {NoStop}%
\bibitem [{\citenamefont {Caffo}\ \emph
  {et~al.}(1998{\natexlab{a}})\citenamefont {Caffo}, \citenamefont {Czyz},
  \citenamefont {Laporta},\ and\ \citenamefont {Remiddi}}]{Caffo:1998yd}%
  \BibitemOpen
  \bibfield  {author} {\bibinfo {author} {\bibfnamefont {M.}~\bibnamefont
  {Caffo}}, \bibinfo {author} {\bibfnamefont {H.}~\bibnamefont {Czyz}},
  \bibinfo {author} {\bibfnamefont {S.}~\bibnamefont {Laporta}}, \ and\
  \bibinfo {author} {\bibfnamefont {E.}~\bibnamefont {Remiddi}},\ }\href@noop
  {} {\bibfield  {journal} {\bibinfo  {journal} {Acta Phys.Polon.}\ }\textbf
  {\bibinfo {volume} {B29}},\ \bibinfo {pages} {2627} (\bibinfo {year}
  {1998}{\natexlab{a}})},\ \Eprint {http://arxiv.org/abs/hep-th/9807119}
  {hep-th/9807119} \BibitemShut {NoStop}%
\bibitem [{\citenamefont {Caffo}\ \emph
  {et~al.}(1998{\natexlab{b}})\citenamefont {Caffo}, \citenamefont {Czyz},
  \citenamefont {Laporta},\ and\ \citenamefont {Remiddi}}]{Caffo:1998du}%
  \BibitemOpen
  \bibfield  {author} {\bibinfo {author} {\bibfnamefont {M.}~\bibnamefont
  {Caffo}}, \bibinfo {author} {\bibfnamefont {H.}~\bibnamefont {Czyz}},
  \bibinfo {author} {\bibfnamefont {S.}~\bibnamefont {Laporta}}, \ and\
  \bibinfo {author} {\bibfnamefont {E.}~\bibnamefont {Remiddi}},\ }\href@noop
  {} {\bibfield  {journal} {\bibinfo  {journal} {Nuovo Cim.}\ }\textbf
  {\bibinfo {volume} {A111}},\ \bibinfo {pages} {365} (\bibinfo {year}
  {1998}{\natexlab{b}})},\ \Eprint {http://arxiv.org/abs/hep-th/9805118}
  {hep-th/9805118} \BibitemShut {NoStop}%
\bibitem [{\citenamefont {Gehrmann}\ and\ \citenamefont
  {Remiddi}(2000)}]{Gehrmann:1999as}%
  \BibitemOpen
  \bibfield  {author} {\bibinfo {author} {\bibfnamefont {T.}~\bibnamefont
  {Gehrmann}}\ and\ \bibinfo {author} {\bibfnamefont {E.}~\bibnamefont
  {Remiddi}},\ }\href {\doibase 10.1016/S0550-3213(00)00223-6} {\bibfield
  {journal} {\bibinfo  {journal} {Nucl. Phys.}\ }\textbf {\bibinfo {volume}
  {B580}},\ \bibinfo {pages} {485} (\bibinfo {year} {2000})},\ \Eprint
  {http://arxiv.org/abs/hep-ph/9912329} {hep-ph/9912329} \BibitemShut {NoStop}%
\bibitem [{\citenamefont {Ligeti}\ \emph {et~al.}(2008)\citenamefont {Ligeti},
  \citenamefont {Stewart},\ and\ \citenamefont {Tackmann}}]{Ligeti:2008ac}%
  \BibitemOpen
  \bibfield  {author} {\bibinfo {author} {\bibfnamefont {Z.}~\bibnamefont
  {Ligeti}}, \bibinfo {author} {\bibfnamefont {I.~W.}\ \bibnamefont {Stewart}},
  \ and\ \bibinfo {author} {\bibfnamefont {F.~J.}\ \bibnamefont {Tackmann}},\
  }\href {\doibase 10.1103/PhysRevD.78.114014} {\bibfield  {journal} {\bibinfo
  {journal} {Phys. Rev.}\ }\textbf {\bibinfo {volume} {D78}},\ \bibinfo {pages}
  {114014} (\bibinfo {year} {2008})},\ \Eprint {http://arxiv.org/abs/0807.1926}
  {arXiv:0807.1926} \BibitemShut {NoStop}%
\bibitem [{\citenamefont {Neubert}(2005)}]{Neubert:2004dd}%
  \BibitemOpen
  \bibfield  {author} {\bibinfo {author} {\bibfnamefont {M.}~\bibnamefont
  {Neubert}},\ }\href {\doibase 10.1140/epjc/s2005-02141-1} {\bibfield
  {journal} {\bibinfo  {journal} {Eur. Phys. J.}\ }\textbf {\bibinfo {volume}
  {C40}},\ \bibinfo {pages} {165} (\bibinfo {year} {2005})},\ \Eprint
  {http://arxiv.org/abs/hep-ph/0408179} {hep-ph/0408179} \BibitemShut {NoStop}%
\bibitem [{\citenamefont {Mantry}\ and\ \citenamefont
  {Petriello}(2010)}]{Mantry:2009qz}%
  \BibitemOpen
  \bibfield  {author} {\bibinfo {author} {\bibfnamefont {S.}~\bibnamefont
  {Mantry}}\ and\ \bibinfo {author} {\bibfnamefont {F.}~\bibnamefont
  {Petriello}},\ }\href {\doibase 10.1103/PhysRevD.81.093007} {\bibfield
  {journal} {\bibinfo  {journal} {Phys. Rev.}\ }\textbf {\bibinfo {volume}
  {D81}},\ \bibinfo {pages} {093007} (\bibinfo {year} {2010})},\ \Eprint
  {http://arxiv.org/abs/0911.4135} {arXiv:0911.4135} \BibitemShut {NoStop}%
\bibitem [{\citenamefont {Jain}\ \emph {et~al.}(2012)\citenamefont {Jain},
  \citenamefont {Procura},\ and\ \citenamefont {Waalewijn}}]{Jain:2011iu}%
  \BibitemOpen
  \bibfield  {author} {\bibinfo {author} {\bibfnamefont {A.}~\bibnamefont
  {Jain}}, \bibinfo {author} {\bibfnamefont {M.}~\bibnamefont {Procura}}, \
  and\ \bibinfo {author} {\bibfnamefont {W.~J.}\ \bibnamefont {Waalewijn}},\
  }\href {\doibase 10.1007/JHEP04(2012)132} {\bibfield  {journal} {\bibinfo
  {journal} {JHEP}\ }\textbf {\bibinfo {volume} {1204}},\ \bibinfo {pages}
  {132} (\bibinfo {year} {2012})},\ \Eprint {http://arxiv.org/abs/1110.0839}
  {arXiv:1110.0839} \BibitemShut {NoStop}%
\bibitem [{\citenamefont {Larkoski}\ \emph {et~al.}(2014)\citenamefont
  {Larkoski}, \citenamefont {Moult},\ and\ \citenamefont
  {Neill}}]{Larkoski:2014tva}%
  \BibitemOpen
  \bibfield  {author} {\bibinfo {author} {\bibfnamefont {A.~J.}\ \bibnamefont
  {Larkoski}}, \bibinfo {author} {\bibfnamefont {I.}~\bibnamefont {Moult}}, \
  and\ \bibinfo {author} {\bibfnamefont {D.}~\bibnamefont {Neill}},\
  }\href@noop {} {\  (\bibinfo {year} {2014})},\ \Eprint
  {http://arxiv.org/abs/1401.4458} {arXiv:1401.4458} \BibitemShut {NoStop}%
\bibitem [{\citenamefont {Kang}\ \emph {et~al.}(2013)\citenamefont {Kang},
  \citenamefont {Lee},\ and\ \citenamefont {Stewart}}]{Kang:2013nha}%
  \BibitemOpen
  \bibfield  {author} {\bibinfo {author} {\bibfnamefont {D.}~\bibnamefont
  {Kang}}, \bibinfo {author} {\bibfnamefont {C.}~\bibnamefont {Lee}}, \ and\
  \bibinfo {author} {\bibfnamefont {I.~W.}\ \bibnamefont {Stewart}},\ }\href
  {\doibase 10.1103/PhysRevD.88.054004} {\bibfield  {journal} {\bibinfo
  {journal} {Phys. Rev.}\ }\textbf {\bibinfo {volume} {D88}},\ \bibinfo {pages}
  {054004} (\bibinfo {year} {2013})},\ \Eprint {http://arxiv.org/abs/1303.6952}
  {arXiv:1303.6952} \BibitemShut {NoStop}%
\bibitem [{\citenamefont {Krohn}\ \emph {et~al.}(2013)\citenamefont {Krohn},
  \citenamefont {Schwartz}, \citenamefont {Lin},\ and\ \citenamefont
  {Waalewijn}}]{Krohn:2012fg}%
  \BibitemOpen
  \bibfield  {author} {\bibinfo {author} {\bibfnamefont {D.}~\bibnamefont
  {Krohn}}, \bibinfo {author} {\bibfnamefont {M.~D.}\ \bibnamefont {Schwartz}},
  \bibinfo {author} {\bibfnamefont {T.}~\bibnamefont {Lin}}, \ and\ \bibinfo
  {author} {\bibfnamefont {W.~J.}\ \bibnamefont {Waalewijn}},\ }\href {\doibase
  10.1103/PhysRevLett.110.212001} {\bibfield  {journal} {\bibinfo  {journal}
  {Phys. Rev. Lett.}\ }\textbf {\bibinfo {volume} {110}},\ \bibinfo {pages}
  {212001} (\bibinfo {year} {2013})},\ \Eprint {http://arxiv.org/abs/1209.2421}
  {arXiv:1209.2421} \BibitemShut {NoStop}%
\bibitem [{\citenamefont {Waalewijn}(2012)}]{Waalewijn:2012sv}%
  \BibitemOpen
  \bibfield  {author} {\bibinfo {author} {\bibfnamefont {W.~J.}\ \bibnamefont
  {Waalewijn}},\ }\href {\doibase 10.1103/PhysRevD.86.094030} {\bibfield
  {journal} {\bibinfo  {journal} {Phys. Rev.}\ }\textbf {\bibinfo {volume}
  {D86}},\ \bibinfo {pages} {094030} (\bibinfo {year} {2012})},\ \Eprint
  {http://arxiv.org/abs/1209.3019} {arXiv:1209.3019} \BibitemShut {NoStop}%
\bibitem [{\citenamefont {Ellis}\ \emph {et~al.}(2010)\citenamefont {Ellis},
  \citenamefont {Vermilion}, \citenamefont {Walsh}, \citenamefont {Hornig},\
  and\ \citenamefont {Lee}}]{Ellis:2010rwa}%
  \BibitemOpen
  \bibfield  {author} {\bibinfo {author} {\bibfnamefont {S.~D.}\ \bibnamefont
  {Ellis}}, \bibinfo {author} {\bibfnamefont {C.~K.}\ \bibnamefont
  {Vermilion}}, \bibinfo {author} {\bibfnamefont {J.~R.}\ \bibnamefont
  {Walsh}}, \bibinfo {author} {\bibfnamefont {A.}~\bibnamefont {Hornig}}, \
  and\ \bibinfo {author} {\bibfnamefont {C.}~\bibnamefont {Lee}},\ }\href
  {\doibase 10.1007/JHEP11(2010)101} {\bibfield  {journal} {\bibinfo  {journal}
  {JHEP}\ }\textbf {\bibinfo {volume} {1011}},\ \bibinfo {pages} {101}
  (\bibinfo {year} {2010})},\ \Eprint {http://arxiv.org/abs/1001.0014}
  {arXiv:1001.0014} \BibitemShut {NoStop}%
\bibitem [{\citenamefont {Becher}\ \emph {et~al.}(2013)\citenamefont {Becher},
  \citenamefont {Neubert},\ and\ \citenamefont {Rothen}}]{Becher:2013xia}%
  \BibitemOpen
  \bibfield  {author} {\bibinfo {author} {\bibfnamefont {T.}~\bibnamefont
  {Becher}}, \bibinfo {author} {\bibfnamefont {M.}~\bibnamefont {Neubert}}, \
  and\ \bibinfo {author} {\bibfnamefont {L.}~\bibnamefont {Rothen}},\ }\href
  {\doibase 10.1007/JHEP10(2013)125} {\bibfield  {journal} {\bibinfo  {journal}
  {JHEP}\ }\textbf {\bibinfo {volume} {1310}},\ \bibinfo {pages} {125}
  (\bibinfo {year} {2013})},\ \Eprint {http://arxiv.org/abs/1307.0025}
  {arXiv:1307.0025} \BibitemShut {NoStop}%
\bibitem [{\citenamefont {Stewart}\ \emph
  {et~al.}(2014{\natexlab{b}})\citenamefont {Stewart}, \citenamefont
  {Tackmann}, \citenamefont {Walsh},\ and\ \citenamefont
  {Zuberi}}]{Stewart:2013faa}%
  \BibitemOpen
  \bibfield  {author} {\bibinfo {author} {\bibfnamefont {I.~W.}\ \bibnamefont
  {Stewart}}, \bibinfo {author} {\bibfnamefont {F.~J.}\ \bibnamefont
  {Tackmann}}, \bibinfo {author} {\bibfnamefont {J.~R.}\ \bibnamefont {Walsh}},
  \ and\ \bibinfo {author} {\bibfnamefont {S.}~\bibnamefont {Zuberi}},\
  }\href@noop {} {\bibfield  {journal} {\bibinfo  {journal} {Phys. Rev.}\
  }\textbf {\bibinfo {volume} {D89}},\ \bibinfo {pages} {054001} (\bibinfo
  {year} {2014}{\natexlab{b}})},\ \Eprint {http://arxiv.org/abs/1307.1808}
  {arXiv:1307.1808} \BibitemShut {NoStop}%
\bibitem [{\citenamefont {Fickinger}\ \emph {et~al.}(2013)\citenamefont
  {Fickinger}, \citenamefont {Ovanesyan},\ and\ \citenamefont
  {Vitev}}]{Fickinger:2013xwa}%
  \BibitemOpen
  \bibfield  {author} {\bibinfo {author} {\bibfnamefont {M.}~\bibnamefont
  {Fickinger}}, \bibinfo {author} {\bibfnamefont {G.}~\bibnamefont
  {Ovanesyan}}, \ and\ \bibinfo {author} {\bibfnamefont {I.}~\bibnamefont
  {Vitev}},\ }\href {\doibase 10.1007/JHEP07(2013)059} {\bibfield  {journal}
  {\bibinfo  {journal} {JHEP}\ }\textbf {\bibinfo {volume} {1307}},\ \bibinfo
  {pages} {059} (\bibinfo {year} {2013})},\ \Eprint
  {http://arxiv.org/abs/1304.3497} {arXiv:1304.3497} \BibitemShut {NoStop}%
\bibitem [{\citenamefont {Remiddi}\ and\ \citenamefont
  {Vermaseren}(2000)}]{Remiddi:1999ew}%
  \BibitemOpen
  \bibfield  {author} {\bibinfo {author} {\bibfnamefont {E.}~\bibnamefont
  {Remiddi}}\ and\ \bibinfo {author} {\bibfnamefont {J.}~\bibnamefont
  {Vermaseren}},\ }\href {\doibase 10.1142/S0217751X00000367} {\bibfield
  {journal} {\bibinfo  {journal} {Int. J. Mod. Phys.}\ }\textbf {\bibinfo
  {volume} {A15}},\ \bibinfo {pages} {725} (\bibinfo {year} {2000})},\ \Eprint
  {http://arxiv.org/abs/hep-ph/9905237} {hep-ph/9905237} \BibitemShut {NoStop}%
\bibitem [{\citenamefont {Gehrmann-De~Ridder}\ \emph
  {et~al.}(2004)\citenamefont {Gehrmann-De~Ridder}, \citenamefont {Gehrmann},\
  and\ \citenamefont {Heinrich}}]{GehrmannDeRidder:2003bm}%
  \BibitemOpen
  \bibfield  {author} {\bibinfo {author} {\bibfnamefont {A.}~\bibnamefont
  {Gehrmann-De~Ridder}}, \bibinfo {author} {\bibfnamefont {T.}~\bibnamefont
  {Gehrmann}}, \ and\ \bibinfo {author} {\bibfnamefont {G.}~\bibnamefont
  {Heinrich}},\ }\href {\doibase 10.1016/j.nuclphysb.2004.01.023} {\bibfield
  {journal} {\bibinfo  {journal} {Nucl. Phys.}\ }\textbf {\bibinfo {volume}
  {B682}},\ \bibinfo {pages} {265} (\bibinfo {year} {2004})},\ \Eprint
  {http://arxiv.org/abs/hep-ph/0311276} {hep-ph/0311276} \BibitemShut {NoStop}%
\bibitem [{\citenamefont {Furmanski}\ and\ \citenamefont
  {Petronzio}(1980)}]{Furmanski:1980cm}%
  \BibitemOpen
  \bibfield  {author} {\bibinfo {author} {\bibfnamefont {W.}~\bibnamefont
  {Furmanski}}\ and\ \bibinfo {author} {\bibfnamefont {R.}~\bibnamefont
  {Petronzio}},\ }\href {\doibase 10.1016/0370-2693(80)90636-X} {\bibfield
  {journal} {\bibinfo  {journal} {Phys.Lett.}\ }\textbf {\bibinfo {volume}
  {B97}},\ \bibinfo {pages} {437} (\bibinfo {year} {1980})}\BibitemShut
  {NoStop}%
\bibitem [{\citenamefont {Ellis}\ \emph {et~al.}(1996)\citenamefont {Ellis},
  \citenamefont {Stirling},\ and\ \citenamefont {Webber}}]{Ellis:1991qj}%
  \BibitemOpen
  \bibfield  {author} {\bibinfo {author} {\bibfnamefont {R.~K.}\ \bibnamefont
  {Ellis}}, \bibinfo {author} {\bibfnamefont {W.~J.}\ \bibnamefont {Stirling}},
  \ and\ \bibinfo {author} {\bibfnamefont {B.}~\bibnamefont {Webber}},\
  }\href@noop {} {\emph {\bibinfo {title} {QCD and Collider Physics}}}\
  (\bibinfo  {publisher} {Cambridge University Press},\ \bibinfo {address}
  {Cambridge},\ \bibinfo {year} {1996})\BibitemShut {NoStop}%
\bibitem [{\citenamefont {Curci}\ \emph {et~al.}(1980)\citenamefont {Curci},
  \citenamefont {Furmanski},\ and\ \citenamefont {Petronzio}}]{Curci:1980uw}%
  \BibitemOpen
  \bibfield  {author} {\bibinfo {author} {\bibfnamefont {G.}~\bibnamefont
  {Curci}}, \bibinfo {author} {\bibfnamefont {W.}~\bibnamefont {Furmanski}}, \
  and\ \bibinfo {author} {\bibfnamefont {R.}~\bibnamefont {Petronzio}},\ }\href
  {\doibase 10.1016/0550-3213(80)90003-6} {\bibfield  {journal} {\bibinfo
  {journal} {Nucl. Phys.}\ }\textbf {\bibinfo {volume} {B175}},\ \bibinfo
  {pages} {27} (\bibinfo {year} {1980})}\BibitemShut {NoStop}%
\bibitem [{\citenamefont {Mitov}\ and\ \citenamefont
  {Moch}(2006)}]{Mitov:2006wy}%
  \BibitemOpen
  \bibfield  {author} {\bibinfo {author} {\bibfnamefont {A.}~\bibnamefont
  {Mitov}}\ and\ \bibinfo {author} {\bibfnamefont {S.-O.}\ \bibnamefont
  {Moch}},\ }\href {\doibase 10.1016/j.nuclphysb.2006.05.018} {\bibfield
  {journal} {\bibinfo  {journal} {Nucl. Phys.}\ }\textbf {\bibinfo {volume}
  {B751}},\ \bibinfo {pages} {18} (\bibinfo {year} {2006})},\ \Eprint
  {http://arxiv.org/abs/hep-ph/0604160} {hep-ph/0604160} \BibitemShut {NoStop}%
\bibitem [{\citenamefont {H{\"o}schele}\ \emph {et~al.}(2014)\citenamefont
  {H{\"o}schele}, \citenamefont {Hoff}, \citenamefont {Pak}, \citenamefont
  {Steinhauser},\ and\ \citenamefont {Ueda}}]{Hoeschele:2013gga}%
  \BibitemOpen
  \bibfield  {author} {\bibinfo {author} {\bibfnamefont {M.}~\bibnamefont
  {H{\"o}schele}}, \bibinfo {author} {\bibfnamefont {J.}~\bibnamefont {Hoff}},
  \bibinfo {author} {\bibfnamefont {A.}~\bibnamefont {Pak}}, \bibinfo {author}
  {\bibfnamefont {M.}~\bibnamefont {Steinhauser}}, \ and\ \bibinfo {author}
  {\bibfnamefont {T.}~\bibnamefont {Ueda}},\ }\href {\doibase
  10.1016/j.cpc.2013.10.007} {\bibfield  {journal} {\bibinfo  {journal}
  {Comput. Phys. Commun.}\ }\textbf {\bibinfo {volume} {185}},\ \bibinfo
  {pages} {528} (\bibinfo {year} {2014})},\ \Eprint
  {http://arxiv.org/abs/1307.6925} {arXiv:1307.6925} \BibitemShut {NoStop}%
\end{thebibliography}%


\end{document}